\documentclass[tradiabstract]{aa} 

\usepackage{graphicx}
\usepackage{txfonts}
\usepackage{amsmath}
\usepackage{bm}
\usepackage{braket}
\usepackage{ulem}
\usepackage{widetext}
\usepackage{fancybox}
\begin{document}

\thisfancyput(16cm,1.8cm){YITP-21-134}

\title{Cold dark matter protohalo structure around collapse: Lagrangian cosmological perturbation theory versus Vlasov simulations}
\titlerunning{Cold dark matter structure around collapse}

\author{
Shohei Saga\inst{1,4}
\and
Atsushi Taruya\inst{2,3}
\and
St{\'e}phane Colombi\inst{4}
}
\authorrunning{S. Saga et al.}

\institute{Laboratoire Univers et Th{\'e}ories, Observatoire de Paris, Universit{\'e} PSL, Universit{\'e} de Paris, CNRS, F-92190 Meudon, France\\
\email{shohei.saga@obspm.fr}
\and
Center for Gravitational Physics and Quantum Information, Yukawa Institute for Theoretical Physics, Kyoto University, Kyoto 606-8502, Japan
\and
Kavli Institute for the Physics and Mathematics of the Universe (WPI), Todai institute
for Advanced Study, University of Tokyo, Kashiwa, Chiba 277-8568, Japan
\and
Sorbonne Universti\'e, CNRS, UMR7095, Institut d'Astrophysique de Paris, 98bis boulevard Arago, F-75014 Paris, France
}

\date{Received --; accepted --}

\abstract{
We explore the structure around shell-crossing time of cold dark matter protohaloes seeded by two or three crossed sine waves of various relative initial amplitudes, by comparing Lagrangian perturbation theory~(LPT) up to 10th order to high-resolution cosmological simulations performed with the public Vlasov code {\tt ColDICE}. Accurate analyses of the density, the velocity, and related quantities such as the vorticity are performed by exploiting the fact that {\tt ColDICE} can follow locally the phase-space sheet at the quadratic level. To test LPT predictions beyond shell-crossing, we employ a ballistic approximation, which assumes that the velocity field is frozen just after shell-crossing. 
 
In the generic case, where the amplitudes of the sine waves are all different, high-order LPT predictions match very well the exact solution, even beyond collapse. As expected, convergence slows down when going from quasi-1D dynamics where one wave dominates over the two others, to the axial-symmetric configuration, where all the amplitudes of the waves are equal. It is also noticed that LPT convergence is slower when considering velocity related quantities. Additionally, the structure of the system at and beyond collapse given by LPT and the simulations agrees very well with singularity theory predictions, in particular with respect to the caustic and vorticity patterns that develop beyond collapse. Again, this does not apply to axial-symmetric configurations, that are still correct from the qualitative point of view, but where multiple foldings of the phase-space sheet produce very high density contrasts, hence a strong back-reaction of the gravitational force. 
}

\keywords{gravitation - methods: numerical - galaxies: kinematics and dynamics - dark matter}

\maketitle

%%%%%%%%%%%%%%%%%%%%%%%%%%%%%%%%%%%%%%%%%%%%%%%%%%
%%%%%%%%%%%%%%%%%%%%%%%%%%%%%%%%%%%%%%%%%%%%%%%%%%
%%%%%%%%%%%%%%%%%%%%%%%%%%%%%%%%%%%%%%%%%%%%%%%%%%
\section{Introduction}
%%%%%%%%%%%%%%%%%%%%%%%%%%%%%%%%%%%%%%%%%%%%%%%%%%
%%%%%%%%%%%%%%%%%%%%%%%%%%%%%%%%%%%%%%%%%%%%%%%%%%
%%%%%%%%%%%%%%%%%%%%%%%%%%%%%%%%%%%%%%%%%%%%%%%%%%
In the concordance model of large-scale structure formation, the matter content of the Universe is dominated by collisionless cold dark matter (CDM) following Vlasov-Poisson equations \citep{1982ApJ...263L...1P,1984ApJ...277..470P,1984Natur.311..517B}. The cold nature of initial conditions implies that the dark matter distribution is concentrated on a three-dimensional sheet evolving in six-dimensional phase-space. At shell crossing, that is in places where the phase-space sheet first self-intersects, the seeds of first dark matter haloes are created. In these regions, the fluid enters a multi-stream regime during which violent relaxation takes place \citep{1967MNRAS.136..101L} to form primordial CDM haloes. Numerical investigations suggest that first dark matter haloes formed during this process have a power-law density profile $\rho\propto r^{\alpha}$ with $\alpha \approx -1.5$ \citep{1991ApJ...382..377M,2005Natur.433..389D,2014ApJ...788...27I,2017MNRAS.471.4687A,2018MNRAS.473.4339O,2018PhRvD..97d1303D,2021A&A...647A..66C}. During their subsequent evolution, which includes successive mergers, dark matter haloes relax to the well known universal Navarro-Frenk-White profile~\citep[hereafter NFW,][]{1996ApJ...462..563N,1997ApJ...490..493N}. 

From an analytical point of view, many approaches have been proposed to describe the results of simulations and the different steps of dark matter haloes history, relying, for example, on entropy maximisation~\citep{1967MNRAS.136..101L,2010ApJ...722..851H,2013MNRAS.432.3161C,2013MNRAS.430..121P}, self-similarity~\citep{1984ApJ...281....1F,1985ApJS...58...39B,1995MNRAS.276..679H,1997PhRvD..56.1863S,2010PhRvD..82j4045Z,2010PhRvD..82j4044Z,2013MNRAS.428..340A}, or more recently, a post-collapse perturbative treatment \citep{2015MNRAS.446.2902C,2017MNRAS.470.4858T,2019arXiv191200868R}.
However, because of the highly nonlinear complex processes taking place during the post-collapse phase, the formation and evolution of dark matter haloes remain a subject of debate, and a consistent framework explaining the different phases of halo history remains yet to be proposed.

The early growth of large scale structures up to first shell-crossing, on the other hand, is well understood thanks to perturbation theory. Indeed, restricting to the early phase of structure formation, i.e., to the single-stream regime before shell-crossing, we can employ perturbation theory as long as fluctuations in the density field remain small~\citep[see, e.g.,][for a review and references therein]{2002PhR...367....1B}. Lagrangian perturbation theory (LPT)~\citep[e.g.,][]{1989RvMP...61..185S,1992ApJ...394L...5B,1992MNRAS.254..729B,1993MNRAS.264..375B,1995A&A...296..575B,1994ApJ...427...51B} uses the displacement field as a small quantity in the expansion of the equations of motion.
First-order LPT corresponds to the classic Zel'dovich approximation~\citep{1970A&A.....5...84Z}, and in one-dimensional space, it is an exact solution until shell-crossing~\citep{1969JETP...30..512N}.
Because the Lagrangian description follows elements of fluid along the motion, Zel'dovich approximation and higher-order LPT provide us with a rather accurate description of the large scale matter distribution, even in the nonlinear regime, shortly after shell crossing. 
The families of singularities that form at shell-crossing and after have been examined in detail in the Lagrangian dynamics framework~\citep{1982GApFD..20..111A,2014MNRAS.437.3442H,2018JCAP...05..027F}, and the structure of cosmological systems at shell-crossing has been investigated for specific initial conditions~\citep{1969JETP...30..512N,2017MNRAS.471..671R,2018PhRvL.121x1302S,2019MNRAS.484.5223R} and for random initial conditions~\citep{2021MNRAS.501L..71R}.

As described above, investigations into the early stages of large-scale structure formation represent a key to understand the bottom-up scenario underlying the CDM model. Practically, perturbation theory under the single-stream assumption, which is strictly valid only during the early stages of the dynamical evolution, provides the basis for predicting statistics of the large-scale structure distribution, and has been successfully confronted with $N$-body simulations and observations~\citep[see, e.g.,][for a review]{2002PhR...367....1B}. Recently, in order to incorporate the effects of multi-streaming at small scales, it has been proposed to introduce effective fluid equations with a non-vanishing stress tensor in the dark matter fluids, the so-called effective field theory of large-scale structure~\citep{2012JCAP...07..051B,2012JHEP...09..082C,2014PhRvD..89d3521H,2015PhRvD..92l3007B}. Although this approach needs parameters in the stress tensor to be calibrated with $N$-body simulations, it has attracted much attention, and has been recently applied to real observational datasets to derive cosmological parameter constraints \citep{2020JCAP...05..042I,2020JCAP...05..005D}.
Accordingly, the understanding of multi-streaming effects from first principles, even at an early stage, would provide significant insights into the precision theoretical modelling of large-scale structure. 

The aim of this paper is to extend the investigations of \citet[][]{2018PhRvL.121x1302S}, hereafter STC18.
In this work, we studied the phase-space structure of protohaloes at shell-crossing with simplified initial conditions composed of three crossed sine waves, following the footsteps of \citet{1991ApJ...382..377M,1995ApJ...441...10M}. Thanks to a detailed comparison between LPT predictions up to 10th order to numerical simulations performed with the state-of-the-art Vlasov-Poisson solver {\tt ColDICE}~\citep{2016JCoPh.321..644S}, we explicitly showed that the convergence of the LPT series slows down when going from quasi-one dimensional to triaxial-symmetric initial conditions, where a spiky structure in phase-space appears. Using a global fitting form, we were able to formally extrapolate the LPT solution to infinite order, obtaining a remarkable agreement with the simulations, even at shell-crossing. At present, we further thoroughly examine the structure both at and shortly after shell-crossing of these systems by considering as well the two-dimensional case.  Although sine wave initial conditions are restrictive, they are to a large extent representative of high peaks of a smooth random Gaussian field~\citep[see, e.g.,][]{1986ApJ...304...15B} and should provide fruitful guidance on the general case. Furthermore, initial conditions expressed only in terms of sine waves considerably simplify analytical calculations while still allowing one an insightful exploration of the pre- and post-collapse dynamics.

The prominent features shortly after shell-crossing are, for instance, the appearance of caustics, and in the multi-stream region delimited by these latter, non-trivial vorticity patterns~\citep{1973ApL....14...11D,1993A&A...267..315C,1999A&A...343..663P,2015MNRAS.454.3920H}.
Thanks to a description of the phase-space sheet at the quadratic level in {\tt ColDICE}, we can measure these quantities in high-resolution Vlasov simulations, in particular the vorticity, with unprecedented accuracy. To perform theoretical predictions shortly after shell-crossing, we use a simple dynamical approximation based on ballistic motion applied to the state of the system described by high-order LPT solutions at shell-crossing.
While convergence with perturbation order of LPT remains a complex subject of investigation \citep{2014JFM...749..404Z,2015MNRAS.452.1421R, 2021MNRAS.500..663M}, it seems to take place at least up to shell crossing, not only for the three sine waves configurations we aim to examine (STC18), but also for more general, random initial conditions \citep{2021MNRAS.501L..71R}, although the effects of cut-offs on power-spectra remain to be investigated furthermore in the latter case. 
Therefore, as long as the back-reaction from multi-stream flows on post-collapse dynamics remains negligible, shortly after shell-crossing, the approximation of the dynamics we propose here should work. This is the first step toward a proper analytical description of post-collapse dynamics in 6D phase-space.

This paper is organized as follows. In Sec.~\ref{sec: LPT}, we begin by introducing the basics of LPT and its recurrence relations, as well as its applications to initial conditions given by linear superpositions of trigonometric functions, such as our sine waves initial conditions. In Sec.~\ref{sec: vlasov}, the important features of the Vlasov solver {\tt ColDICE} are briefly summarised. We also address some aspects of measurements uncertainties in simulation data. In Sec.~\ref{sec: result shell crossing}, we examine the phase-space structure and radial profiles at shell-crossing with a comparison of analytic predictions to simulations, in the framework of singularity theory. In Sec.~\ref{sec: shortly after sc}, the structure shortly after shell-crossing is explored using the ballistic approximation by examining phase-space diagrams, caustics, density and vorticity fields. Again, analytical predictions are compared to the Vlasov runs. Finally, Sec.~\ref{sec: summary} summaries the main results of this article. To complement the main text, Appendices~\ref{app: solutions}, \ref{app: coldice}, \ref{app: Q1DPT} and \ref{sec: analytic log slope} provide details on the explicit form of high-order LPT solutions for the sine waves initial conditions up to 5th order, the measurements in the Vlasov simulations, the predictions from quasi-1D LPT, and the expected structure of singularities at collapse time, respectively. 

Throughout the paper, we will use the following units: a box size $L=1$ and an inverse of the Hubble parameter at present time $H^{-1}_{0}=1$ for the dimensions of length and time, respectively. In particular, the Lagrangian/Eulerian coordinates, velocity, and vorticity will be explicitly expressed as $\bm{q}/L$, $\bm{x}/L$, $\bm{v}/(L\,H_{0})$, and $\bm{\omega}/H_{0} = \left(L\,\bm{\nabla}\right) \times \left( \bm{v}/(L\,H_{0}) \right)$, respectively.

%%%%%%%%%%%%%%%%%%%%%%%%%%%%%%%%%%%%%%%%%%%%%%%%%%
%%%%%%%%%%%%%%%%%%%%%%%%%%%%%%%%%%%%%%%%%%%%%%%%%%
%%%%%%%%%%%%%%%%%%%%%%%%%%%%%%%%%%%%%%%%%%%%%%%%%%
\section{Lagrangian Perturbation Theory}
\label{sec: LPT}
%%%%%%%%%%%%%%%%%%%%%%%%%%%%%%%%%%%%%%%%%%%%%%%%%%
%%%%%%%%%%%%%%%%%%%%%%%%%%%%%%%%%%%%%%%%%%%%%%%%%%
%%%%%%%%%%%%%%%%%%%%%%%%%%%%%%%%%%%%%%%%%%%%%%%%%%
The Lagrangian equation of motion of a fluid element is given by \citep[e.g.,][]{1980lssu.book.....P}
%-------------------------------------------------
\begin{align}
\frac{{\rm d}^2\bm{x}}{{\rm d}t^2} + 2H \frac{{\rm d}\bm{x}}{{\rm d}t} &= -\frac{1}{a^{2}}\bm{\nabla}_{x} \phi(\bm{x}), 
\label{eq:basic_EoM}
\end{align}
%-------------------------------------------------
where the quantities $\bm{x}$, $a$, and $H(t)=a^{-1} {\rm d}a/{\rm d}t$ are the Eulerian comoving position, the scale factor of the Universe, and Hubble parameter, respectively.
The derivative operator $\bm{\nabla}_{x} = \partial/\partial\bm{x}$ is a spatial derivative in Eulerian space.
The Newton gravitational potential $\phi(\bm{x})$ is related to the matter density contrast $\delta(\bm{x}) = \rho(\bm{x})/\bar{\rho}-1$ with $\bar{\rho}$ being the background mass density, through the Poisson equation:
%-------------------------------------------------
\begin{align}
\bm{\nabla}^{2}_{x}\phi(\bm{x}) &= 4\pi G\bar{\rho}\, a^{2} \,\delta(\bm{x}). 
\label{eq:Poisson}
\end{align}
%-------------------------------------------------
In this framework, the velocity of each mass element is given by $\bm{v} = a\,{\rm d}\bm{x}/{\rm d}t$.

Taking the divergence and curl of Eq.~(\ref{eq:basic_EoM}) with respect to Eulerian coordinates, Eqs.~(\ref{eq:basic_EoM}) and (\ref{eq:Poisson}) can be expressed by the equivalent set of equations:
\begin{align}
\bm{\nabla}_{x}\cdot\left( \ddot{\bm{x}} + 2H\dot{\bm{x}}\right) &= 4\pi G\bar{\rho} \,\delta(\bm{x}), \label{eq: div} \\
\bm{\nabla}_{x}\times \left( \ddot{\bm{x}} + 2H \dot{\bm{x}}\right) &= 0, \label{eq: rot}
\end{align}
where the dot represents the Lagrangian derivative of time, ${\rm d}/{\rm d}t$.

In the Lagrangian approach, for each mass element, the Eulerian position $\bm{x}$ at the time of interest $t$ is related to the Lagrangian coordinate (initial position) $\bm{q}$ through the displacement field $\bm{\Psi}(\bm{q}, t)$ by
\begin{align}
\bm{x}(\bm{q},t)=\bm{q}+ \bm{\Psi}(\bm{q},t) . \label{eq: def Psi}
\end{align}
The velocity field is expressed as $\bm{v}(\bm{q},t)=a\,{\rm d}\bm{\Psi}/{\rm d}t$.
In the single flow regime, i.e., before first shell-crossing time $t_{\rm sc}$, mass conservation implies $\bar{\rho}\,{\rm d}^3\bm{q}=\rho(\bm{x})\,{\rm d}^3\bm{x}$, which leads to
\begin{align}
1 + \delta(\bm{x}) = \frac{\rho(\bm{x})}{\bar{\rho}} = \frac{1}{J} , \label{eq: delta-J}
\end{align}
where the quantity $J = \det{J_{ij}}$ is the Jacobian of the matrix $J_{ij}$ defined by
\begin{equation}
 J_{ij}(\bm{q}, t) =\frac{\partial x_{i}(\bm{q}, t)}{\partial q_{j}} = \delta_{ij} + \Psi_{i,j}(\bm{q}, t) ,
 \label{eq: def Jij}
\end{equation}
and its inverse is given as
\begin{equation}
J^{-1}_{ij} = \frac{\partial q_{i}}{\partial x_{j}} . \label{eq:jm1}
\end{equation}
While equation (\ref{eq: delta-J}) is well defined only until the first shell-crossing time $t_{\rm sc}$, that is the first occurrence of $J=0$, Eqs.~(\ref{eq: def Jij}) and (\ref{eq:jm1}) can be formally used beyond $t_{\rm sc}$, as long as they are expressed in terms of Lagrangian coordinates (except that Eq.~(\ref{eq:jm1}) might become singular).

%%%%%%%%%%%%%%%%%%%%%%%%%%%%%%%%%%%%%%%%%%%%%%%%%%
%%%%%%%%%%%%%%%%%%%%%%%%%%%%%%%%%%%%%%%%%%%%%%%%%%
\subsection{Recursion relation}
\label{sec:recur}
%%%%%%%%%%%%%%%%%%%%%%%%%%%%%%%%%%%%%%%%%%%%%%%%%%
%%%%%%%%%%%%%%%%%%%%%%%%%%%%%%%%%%%%%%%%%%%%%%%%%%
In Lagrangian perturbation theory, the displacement field, $\bm{\Psi}$, is the fundamental building block which is considered as a small quantity. It can be systematically formally expanded as follows,
\begin{align}
\bm{\Psi}(\bm{q},t) = \sum_{n = 1}^{\infty}\bm{\Psi}^{(n)}(\bm{q},t) . \label{eq: def LPT}
\end{align}
In what follows, we shall assume that the fastest growing modes dominate. They are well known to be given to a very good approximation by \citep[see, e.g.,][and references therein]{2002PhR...367....1B}
\begin{align}
\bm{\Psi}^{(n)}(\bm{q},t) = D^{n}_{+}(t)\, \bm{\Psi}^{(n)}(\bm{q}) , \label{eq: fastest growing}
\end{align}
where the purely time-dependent function $D_{+}(t)$ corresponds to the linear growth factor.
The velocity field is then given by
\begin{align}
\bm{v}(\bm{q},t) = a\, H\, f\,\sum_{n = 1}^{\infty}n\,D^{n}_{+}(t) \,\bm{\Psi}^{(n)}(\bm{q}) ,
\end{align}
where function $f(t)\equiv {\rm d}\ln{D_{+}}/{\rm d}\ln{a}$ corresponds to the logarithmic derivative of the growth factor. Note that the analyses performed in subsequent sections will consider the Einstein-de Sitter cosmology, that is a total matter density parameter $\Omega_{\rm m}=1$ and a cosmological constant density parameter $\Omega_{\rm \Lambda} =0$. In this case, one simply has $D_{+} \propto a$ and $f = 1$.

During early phases of structure formation, the Universe approaches Einstein-de Sitter cosmology and $f$ is approximately given by $f \approx \Omega_{\rm m}^{3/5}$. With the further approximation $f\approx \Omega_{m}^{1/2}$, as implicitly assumed in all the subsequent calculations \citep[see, e.g.,][]{1980lssu.book.....P,2002PhR...367....1B,2015PhRvD..92b3534M}, by substituting Eqs.~(\ref{eq: def Psi}), (\ref{eq: delta-J}), (\ref{eq: def Jij}), and (\ref{eq: def LPT}) into Eqs.~(\ref{eq: div}) and (\ref{eq: rot}), one obtains simple recurrence formulae for the longitudinal and transverse parts of the motion \citep{2012JCAP...12..004R,2014JFM...749..404Z,2015MNRAS.452.1421R,2015PhRvD..92b3534M}:
\begin{align}
\left( \hat{\mathcal{T}} - \frac{3}{2}\right)\Psi^{(n)}_{k,k}
 &=
 -\varepsilon_{im}\,\varepsilon_{jk}
 \sum_{n_{1}+n_{2}=n}
 \Psi^{(n_{1})}_{m,k}\left( \hat{\mathcal{T}} - \frac{3}{4}\right) \Psi^{(n_{2})}_{i,j}, \label{eq:2Dlongitudinal2} \\
\varepsilon_{ij}\,\hat{\mathcal{T}} \Psi^{(n)}_{j,i}
 &= \varepsilon_{ij}\sum_{n_{1} + n_{2}=n}\Psi^{(n_{1})}_{k,j} \,\hat{\mathcal{T}}\Psi^{(n_{2})}_{k,i}, 
\label{eq:2Dtransverse2}
\end{align}
in the two-dimensional case, and 
\begin{align}
\left( \hat{\mathcal{T}} - \frac{3}{2}\right)\Psi^{(n)}_{k,k}
&=
-\varepsilon_{ijk}\,\varepsilon_{ipq}
\sum_{n_{1}+n_{2}=n}
\Psi^{(n_{1})}_{j,p}\left( \hat{\mathcal{T}} - \frac{3}{4}\right) \Psi^{(n_{2})}_{k,q} \notag \\
&
 -\frac{1}{2}\varepsilon_{ijk}\,\varepsilon_{pqr} \notag \\
 & \quad \quad \times \sum_{n_{1}+n_{2}+n_{3}=n}\Psi^{(n_{1})}_{i,p}\Psi^{(n_{2})}_{j,q}\left( \hat{\mathcal{T}} - \frac{1}{2}\right)\Psi^{(n_{3})}_{k, r}, \label{eq:longitudinal2} \\
\varepsilon_{ijk}\,\hat{\mathcal{T}} \Psi^{(n)}_{j,k}
&= -\varepsilon_{ijk}\sum_{n_{1} + n_{2}=n}\Psi^{(n_{1})}_{p,j}\, \hat{\mathcal{T}} \Psi^{(n_{2})}_{p,k}, 
\label{eq:transverse2}
\end{align}
in the three-dimensional case. Here and in what follows, we adopt the Einstein summation convention when the equation includes repeated lowercase Latin letters, and the subsrcipts $i,j,..$ take $1$, $2$  or $3$ ($1$ or $2$) in the three-dimensional (two-dimensional) case. In the above, we defined $\Psi_{i,j}\equiv\partial\Psi_i/\partial q_j$, and the tensors $\varepsilon_{ij}$ and $\varepsilon_{ijk}$ are, respectively, the two-dimensional and three-dimensional Levi-Civita symbols. The symbol $\hat{\mathcal{T}}$ stands for a differential operator:
\begin{align}
\hat{\mathcal{T}}\equiv \frac{\partial^2}{\partial \ln{D_{+}}^2}+\frac{1}{2}\frac{\partial}{\partial \ln{D_{+}}} .
\end{align}
In obtaining the recursion relations, we used the following formulae of linear algebra:
\begin{align}
J = \frac{1}{2}\varepsilon_{ij}\,\varepsilon_{kr}\,J_{ik}\,J_{jr} ,~~~
J^{-1}_{ij} = \frac{1}{J}\varepsilon_{ik}\,\varepsilon_{jr}\,J_{rk} ,
\end{align}
for the two-dimensional case, and 
\begin{align}
J = \frac{1}{6}\varepsilon_{ijk}\,\varepsilon_{pqr}\,J_{ip}\,J_{jq}\,J_{kr},~~~
J^{-1}_{ij} = \frac{1}{2J}\varepsilon_{jkp}\,\varepsilon_{iqr}\,J_{kq}\,J_{pr} ,
\end{align}
for the three-dimensional case.

In the fastest-growing mode regime (\ref{eq: fastest growing}), which will be assumed in the LPT calculations of this work, the time dependence in the recursion relations simplifies and one obtains
\begin{align}
\Psi^{(n)}_{k,k}
&=
- \varepsilon_{im}\,\varepsilon_{jk}
\sum_{n_{1}+n_{2}=n}
\frac{4n^{2}_{2} + 2n_{2} - 3}{2(n-1)(2n+3)}
\Psi^{(n_{1})}_{m,k} \,\Psi^{(n_{2})}_{i,j}
, \label{eq:2Dlongitudinal3} \\
\varepsilon_{ij} \Psi^{(n)}_{j,i}
&= \varepsilon_{ij}\sum_{n_{1} + n_{2}=n}
\frac{n_{2}(2n_{2} + 1)}{n(2n + 1)}
\Psi^{(n_{1})}_{k,j} \,\Psi^{(n_{2})}_{k,i}, 
\label{eq:2Dtransverse3}
\end{align}
for the two-dimensional case, and
\begin{align}
\Psi^{(n)}_{k,k}
&=
- \varepsilon_{ijk}\,\varepsilon_{ipq}
\sum_{n_{1}+n_{2}=n}
\frac{4n^{2}_{2} + 2n_{2} - 3}{2(n-1)(2n+3)}
\Psi^{(n_{1})}_{j,p} \,\Psi^{(n_{2})}_{k,q} \notag \\
&
 - \varepsilon_{ijk}\,\varepsilon_{pqr} \notag \\
 & \quad \quad \times \sum_{n_{1}+n_{2}+n_{3}=n}
\frac{(n_{3}+1)(2n_{3}-1)}{2(n-1)(2n+3)}
\Psi^{(n_{1})}_{i,p}\,\Psi^{(n_{2})}_{j,q}\,\Psi^{(n_{3})}_{k, r}, \label{eq:longitudinal3} \\
\varepsilon_{ijk} \,\Psi^{(n)}_{j,k}
&= -\varepsilon_{ijk}\sum_{n_{1} + n_{2}=n}
\frac{n_{2}(2n_{2} + 1)}{n(2n + 1)}
\Psi^{(n_{1})}_{p,j}\, \Psi^{(n_{2})}_{p,k}, 
\label{eq:transverse3}
\end{align}
for the three-dimensional case.

%%%%%%%%%%%%%%%%%%%%%%%%%%%%%%%%%%%%%%%%%%%%%%%%%%
%%%%%%%%%%%%%%%%%%%%%%%%%%%%%%%%%%%%%%%%%%%%%%%%%%
\subsection{Two sine and three sine waves initial conditions}
\label{sec:sine_ini}
%%%%%%%%%%%%%%%%%%%%%%%%%%%%%%%%%%%%%%%%%%%%%%%%%%
%%%%%%%%%%%%%%%%%%%%%%%%%%%%%%%%%%%%%%%%%%%%%%%%%%
Throughout this paper, we focus on specific initial conditions: two or three sine waves in a periodic box covering the interval $[-L/2, L/2[$:
\begin{align}
\Psi_{i}^{\rm ini}(\bm{q}, t_{\rm ini}) =\frac{L}{2\pi}\,D_+(t_{\rm ini})\,\epsilon_{i} \sin\left( \frac{2\pi}{L}q_{i}\right).
\label{eq:init_psi}
\end{align}
The parameters $\epsilon_{i} < 0$ with $|\epsilon_x| \geq |\epsilon_y| \geq |\epsilon_z|$ quantify the linear amplitudes of the sine waves in each direction.
The initial time, $t_{\rm ini}$, is chosen such that $D_+(t_{\rm ini})|\epsilon_{i}|\le 0.012 \ll1$, which makes the fastest growing mode approximation very accurate, as shown by STC18. In this framework, the dependence on $\epsilon_i$ of the dynamics is reduced to a function of the ratios $\epsilon_{\rm 2D} = \epsilon_{y}/\epsilon_{x}$ and $\bm{\epsilon}_{\rm 3D}=(\epsilon_{y}/\epsilon_{x}, \epsilon_{z}/\epsilon_{x})$, respectively for two and three sine waves initial conditions. These ratios will therefore be the quantities of relevance to define our initial conditions.
In this setting, the initial density field, given by $\delta^{\rm ini} \simeq -\bm{\nabla}_{q}\cdot\bm{\Psi}^{\rm ini} = D_+(t_{\rm ini}) \sum_{\rm i}|\epsilon_{i}| \cos\left( 2\pi/L\,q_{i}\right)$,
presents a small peak at the origin, and mass elements subsequently infall toward the central overdense region. With the proper choice of $\epsilon_i $, this initial overdensity can asymptotically match any peak of a smooth random Gaussian field \cite[see, e.g.,][]{1986ApJ...304...15B}, which actually makes the three sine waves initial conditions quite generic, hence providing many insights into the dynamics during the early stages of protohalo formation.

Naturally, this very symmetrical set-up remains unrealistic, with a tidal environment restricted to periodic replica, but has the advantage of belonging to the family of linear superpositions of trigonometric functions, here simple sine functions,  which considerably simplifies LPT calculations, as described below.
Initial conditions that only involve linear superpositions of trigonometric functions include exact Fourier transforms, so are in principle very general. Furthermore, with only a few Fourier modes, one can theoretically account for more realistic initial conditions with proper tidal environments and mergers, while keeping the analytical description still very tractable. However, as far as we are concerned, applications beyond two or three sine waves are left for future work. 

For initial conditions given by
linear superpositions of trigonometric functions, it is trivial to see that all the terms of the perturbation series are also expressed in the same way:
\begin{align}
\nabla_{q}\cdot\bm{\Psi}^{(n)}_{\rm L} &= \sum_{\bm{m}}\alpha^{(n)}_{\bm{m}}{\rm e}^{{\rm i}\, \bm{m}\cdot\bm{q}} ,\\
\nabla_{q}\times\bm{\Psi}^{(n)}_{\rm T} &= \sum_{\bm{m}}\bm{\beta}^{(n)}_{\bm{m}}{\rm e}^{{\rm i}\, \bm{m}\cdot\bm{q}} ,
\end{align}
where the $n$th-order scalar coefficients $\alpha^{(n)}_{\bm{m}}$ and the $n$th-order vector coefficients $\bm{\beta}^{(n)}_{\bm{m}}$ are obtained recursively by calculating the right-hand-side of Eqs.~(\ref{eq:longitudinal3}) and (\ref{eq:transverse3}), which depends on lower order terms, starting from the $n=1$ coefficients determined by equation (\ref{eq:init_psi}).

By imposing the conditions $\nabla_{q}\times\bm{\Psi}^{(n)}_{\rm L} = \bm{0}$ and $\nabla_{q}\cdot\bm{\Psi}^{(n)}_{\rm T} =0$, one can build up the perturbative solutions $\bm{\Psi}^{(n)}_{\rm L}$ and $\bm{\Psi}^{(n)}_{\rm T}$ using simple algebraic manipulations involving coefficients $\alpha^{(n)}_{\bm{m}}$ and $\bm{\beta}^{(n)}_{\bm{m}}$:
\begin{align}
\bm{\Psi}^{(n)}_{\rm L} &= \sum_{\bm{m}}(-{\rm i})\,\alpha^{(n)}_{\bm{m}}\frac{\bm{m}}{|\bm{m}|^{2}}\, {\rm e}^{{\rm i}\,\bm{m}\cdot\bm{q}} , \label{eq: rec psi L}\\
\bm{\Psi}^{(n)}_{\rm T} &= \sum_{\bm{m}}{\rm i}\, \frac{\bm{m}\times \bm{\beta}^{(n)}_{\bm{m}}}{|\bm{m}|^{2}}\, {\rm e}^{{\rm i}\,\bm{m}\cdot\bm{q}} . \label{eq: rec psi T}
\end{align}
These solutions lead to the $n$th-order displacement field given by $\bm{\Psi}^{(n)} = \bm{\Psi}^{(n)}_{\rm L} + \bm{\Psi}^{(n)}_{\rm T}$.
Using this solution for $\bm{\Psi}^{(n)}$ as well as Eqs.~(\ref{eq:longitudinal3}) and (\ref{eq:transverse3}), we subsequently construct the source of the $(n+1)$th-order derivatives $\nabla_{q}\cdot\bm{\Psi}^{(n+1)}_{\rm L}$ and $\nabla_{q}\times\bm{\Psi}^{(n+1)}_{\rm T}$.
By repeating the above operation together with the recursive relations, one can derive, in principle, arbitrary high-order LPT solutions.
In Appendix~\ref{app: solutions}, we present the explicit forms of the LPT solutions up to 5th order derived in this way\footnote{The LPT solutions up to 10th order can be provided upon request as a {\tt Mathematica} notebook.}.

The above prescription is valid in three-dimensional space, and can be applied to the 2D case by cancelling all fluctuations along $z$-axis, that is by performing 3D calculations with $\epsilon_z=0$, i.e., $\bm{\epsilon}_{\rm 3D}=(\epsilon_y/\epsilon_x,0)$ for the three sine waves case. However, one can also realize that in two dimensions, vector coefficients $\bm{\beta}^{(n)}_{\bm{m}}$ become scalars $\beta^{(n)}_{\bm m}$, and that the solutions take the following form:
\begin{align}
\bm{\Psi}^{(n)}_{\rm L} &= \sum_{\bm{m}}(-{\rm i})\frac{\alpha^{(n)}_{\bm{m}}}{|\bm{m}|^{2}}
\left(
 \begin{array}{c}
 m_{x}\\
 m_{y}
 \end{array}
\right)\, {\rm e}^{{\rm i}\, \bm{m}\cdot\bm{q}} , \\
\bm{\Psi}^{(n)}_{\rm T} &= \sum_{\bm{m}}{\rm i}\,\frac{\beta^{(n)}_{\bm{m}}}{|\bm{m}|^{2}}
\left(
 \begin{array}{c}
 m_{y}\\
 -m_{x}
 \end{array}
\right)
\, {\rm e}^{{\rm i}\,\bm{m}\cdot\bm{q}} .
\end{align}
Our analytical investigations can easily cover a large range of values of $\epsilon_{\rm 2D}$ and $\bm{\epsilon}_{\rm 3D}$, while the simulations, much more costly, will only focus on three configurations, as detailed in Table~\ref{tab: initial conditions}, reflecting various regimes in the dynamics: quasi one-dimensional with $|\epsilon_x| \gg |\epsilon_{y,z}|$, anisotropic with $|\epsilon_x| > |\epsilon_y| > |\epsilon_z|$, and what we design by axial-symmetric, with $|\epsilon_x|=|\epsilon_y|(=|\epsilon_z|)$, denoted by Q1D, ANI and SYM, respectively\footnote{The designations named here, Q1D-3SIN, ANI-3SIN, and SYM-3SIN, are the same as Q1D-S, ASY-Sb, and SYM-S, used in \citet{2018PhRvL.121x1302S}, respectively.}. 
%-------------------------------------------------
\begin{table*}
\centering
\begin{tabular}{ccccccccc}
\hline
Designation & $\epsilon_{\rm 2D}$ or $\bm{\epsilon}_{\rm 3D}$ & $\epsilon_x$ & $n_{\rm g}$ & $n_{\rm s}$ & $a_{\rm sc}^{\infty}$ & $a_{\rm sc}$ & ${\hat a}_{\rm sc}$ & $a_{\rm sc}+\Delta a$ \\
\hline
{\it Quasi 1D}\\
Q1D-2SIN & 1/6 & -18 & 2048 & 2048 & 0.05279 & 0.05285 & 0.05281 & 0.05402 \\
Q1D-3SIN & (1/6, 1/8) & -24 & 512 & 256 & 0.03815 & 0.03832 & 0.03819 & 0.03907 \\
\hline
{\it Anisotropic}\\
ANI-2SIN & 2/3 & -18 & 2048 & 2048 & 0.04531 & 0.04545 & 0.04534 & 0.04601 \\
ANI-3SIN & (3/4, 1/2) & -24 & 512 & 512 & 0.02911 & 0.02919 & 0.02915 & 0.03003 \\
\hline
{\it Axial-symmetric}\\
SYM-2SIN & 1 & -18 & 2048 & 2048 & 0.04076 & 0.04090 & 0.04078 & 0.04101 \\
SYM-3SIN & (1, 1) & -18 & 512 & 512 & 0.03190 & 0.03155 & 0.03190 & 0.03201 \\
\hline
\end{tabular}
\caption{Parameters of the runs performed with \texttt{ColDICE}~\citep{2016JCoPh.321..644S}.
The first column indicates the designation of the run.
The second column corresponds to the relative amplitudes of the initial sine waves, namely, $\epsilon_{\rm 2D} = \epsilon_{y}/\epsilon_{x}$ and $\bm{\epsilon}_{\rm 3D} = (\epsilon_{y}/\epsilon_{x}, \epsilon_{z}/\epsilon_{x})$ for two and three sine waves, respectively.
The third column gives the value of $\epsilon_x$.
The fourth and fith columns, respectively, indicate the spatial resolution $n_{\rm g}$ of the grid used to solve the Poisson equation, and the spatial resolution $n_{\rm s}$ of the mesh of vertices used to construct the initial tessellation (see Sec.~\ref{sec: vlasov} for details).
The sixth column indicates the scale factor $a_{\rm sc}^{\infty}$ at shell-crossing estimated by LPT extrapolated to infinite order~\citep{2018PhRvL.121x1302S}, while the seventh one provides the value $a_{\rm sc}$ measured in the Vlasov runs (see Appendix~\ref{app:A1}) and which is actually used in Sec.~\ref{sec: phase space shell-crossing}. The eighth column indicates the value ${\hat a}_{\rm sc}$ of the expansion factor of the closest available simulation snapshot to collapse time for the comparisons performed in section~\ref{sec: result shell crossing}. Finally, the last column indicates the value of the expansion factor used for the analyses performed beyond collapse time in Sec~\ref{sec: shortly after sc}. Note: with the normalisation of the scale factor used here, the value of $a_{\rm sc}$ has to be multiplied by $|\epsilon_x|$ to obtain a more intuitive collapse time that would be equal to unity in the purely one dimensional case. }
\label{tab: initial conditions}
\end{table*}
%-------------------------------------------------

%%%%%%%%%%%%%%%%%%%%%%%%%%%%%%%%%%%%%%%%%%%%%%%%%%
%%%%%%%%%%%%%%%%%%%%%%%%%%%%%%%%%%%%%%%%%%%%%%%%%%
%%%%%%%%%%%%%%%%%%%%%%%%%%%%%%%%%%%%%%%%%%%%%%%%%%
\section{Vlasov-Poisson simulations}
\label{sec: vlasov}
%%%%%%%%%%%%%%%%%%%%%%%%%%%%%%%%%%%%%%%%%%%%%%%%%%
%%%%%%%%%%%%%%%%%%%%%%%%%%%%%%%%%%%%%%%%%%%%%%%%%%
%%%%%%%%%%%%%%%%%%%%%%%%%%%%%%%%%%%%%%%%%%%%%%%%%%
To perform the numerical experiments, we use the public parallel Vlasov solver {\tt ColDICE} \citep{2016JCoPh.321..644S}. {\tt ColDICE} follows the phase-space sheet with an adaptive tessellation of simplices, composed, in 2 and 3 dimensions, of connected triangles and connected tetrahedra, respectively. The phase-space coordinates of the vertices of the tessellation, $[\bm{X}(t),\bm{V}(t)]$, follow the standard Lagrangian equations of motion, similarly as in an $N$-body code, but matter is distributed linearly inside each simplex instead of being transported by the vertices. In what follows, after providing a few additional technical details on the algorithm (Sec.~\ref{sec:coldice}), we address measurement uncertainties issues (Sec.~\ref{sec:accuracy}).

%%%%%%%%%%%%%%%%%%%%%%%%%%%%%%%%%%%%%%%%%%%%%%%%%%
\subsection{The Vlasov code {\tt ColDICE}}
\label{sec:coldice}
%%%%%%%%%%%%%%%%%%%%%%%%%%%%%%%%%%%%%%%%%%%%%%%%%%
The Lagrangian coordinates defined in Sec.~\ref{sec:recur} correspond to the following unperturbed initial set-up,
\begin{align}
 \bm{X}(\bm{Q},t_{0 })&\equiv \bm{Q},\\
 \bm{V}(\bm{Q},t_{0}) &\equiv 0,
\end{align}
where $t_0$ is a virtual time corresponding to $a=0$, $\bm{Q}$ is the Lagrangian coordinate of each vertex. Note that we use capital letters to distinguish between vertices coordinates and actual coordinates of fluid elements of the phase-space sheet that they are supposed to trace. These notations are used in Appendix~\ref{app: coldice}.

Vertices phase-space coordinates are then perturbed using Zel'dovich motion to set actual initial conditions defined in Sec.~\ref{sec:sine_ini}:
\begin{align}
 \bm{X}(\bm{Q},t_{\rm ini })&= \bm{Q} +\bm{\Psi}^{\rm ini}(\bm{Q}, t_{\rm ini}),\\
 \bm{V}(\bm{Q},t_{\rm ini}) &= a\,H\,f\,\bm{\Psi}^{\rm ini}(\bm{Q}, t_{\rm ini}),
\end{align}
with $\bm{\Psi}^{\rm ini}$ given by equation (\ref{eq:init_psi}). 

To update the position and the velocity of each vertex, a standard second order predictor-corrector scheme with slowly varying time step is employed. Constraints on the value of the time step combine bounds on the relative variations of the expansion factor, classical Courant-Friedrichs-Lewy criterion and a harmonic condition related to the local projected density. More details about the parameters used for the time step constraints in the 3D simulations are given in \citet[][hereafter C21]{2021A&A...647A..66C}, so we do not repeat these details here. Additionally, we chose the same constraints on the time step for the 2D runs as for the 3D simulations with $n_{\rm g}=n_{\rm s}=512$ (see below for the definitions of $n_{\rm g}$ and $n_{\rm s}$). 

Poisson equation is solved using the Fast-Fourier-Technique in a mesh of fixed resolution $n_{\rm g}$. To estimate the projected density $\rho(\bm{x})$ on this mesh, the intersection of each simplex of the phase-space sheet with each voxel/pixel of the mesh is computed exactly up to linear order with a special ray-tracing technique. Once the gravitational potential is obtained on the mesh, the gravitational force is computed from the gradient of the potential using a standard 4 points stencil, and then it is interpolated to the vertices using second-order triangular shape cloud interpolation \citep[see, e.g.,][]{1988csup.book.....H}, in order to update their velocities.

Initially, the tessellation is constructed from a regular network of $n_{\rm s}^D$ vertices corresponding, respectively to $2 n_{\rm s}^2$ and $6 n_{\rm s}^3$ simplices in 2 and 3 dimensions. {\tt ColDICE} allows for local refinement of the simplices following criteria based on local Poincar\'e invariant conservation as explained more in detail in \citet[][]{2016JCoPh.321..644S}. Values of the refinement criterion parameter $I$ used for our 3D runs are listed in detail in C21. For completeness, following the notations of C21, we used $I=10^{-8}$ for the 2D runs.

To perform local refinement, the phase-space sheet is locally described at quadratic order inside each simplex with the help of $3$ and $6$ additional tracers per simplex in 2D and 3D, respectively. At the dynamical times considered in this work, which correspond at most to short periods after collapse, refinement is not triggered, except for a small number of simplices in the 3D axial-symmetric simulation, SYM-3SIN, so we do not deem it necessary to discuss more about refinement. However, the ability to describe the phase-space sheet at the quadratic level is important to have correct estimates of derivatives of the velocity field, in particular of the local vorticity of the mean flow. 

The parameters used for all the simulations are listed in Table~\ref{tab: initial conditions}, in particular the resolution $n_{\rm g}$ of the mesh used to solve Poisson equation and the initial number of vertices of the tessellation, $n_{\rm s}^3$. In Appendix~\ref{app: coldice}, we explain how measurements of various quantities are performed, such as the set of curves corresponding to the intersection of the phase-space sheet with the hyperplane $y(=z)=0$ used in Secs.~\ref{sec: phase space shell-crossing} and \ref{sec: postcollapse phase space}, the collapse time $t_{\rm sc}$ shown in Table~\ref{tab: initial conditions}, the radial profiles used in Sec.~\ref{sec:radialprof}, as well as the caustic network, the projected density and the vorticity analysed in Sec.~\ref{sec: postcollapse caustics}.

%%%%%%%%%%%%%%%%%%%%%%%%%%%%%%%%%%%%%%%%%%%%%%%%%%
\subsection{Accuracy and related considerations}
\label{sec:accuracy}
%%%%%%%%%%%%%%%%%%%%%%%%%%%%%%%%%%%%%%%%%%%%%%%%%%
Accuracy and possible defects of {\tt ColDICE} are discussed in \citet[][]{2016JCoPh.321..644S} and in C21 in the 3D case. In our work, precise determination of collapse time is critical, since we consider, in the analyses performed below, either collapse time or a moment very shortly after it. The uncertainty on collapse time depends on the nature of initial conditions and the parameters controlling the accuracy of the simulations, in particular the number of simplices used to represent the phase-space sheet and the spatial resolution $n_{\rm g}$ of the mesh used to solve Poisson equation. This latter turns out to be a very important parameter. Indeed, decreasing $n_{\rm g}$ augments collapse time, as discussed in C21. In Sec.~\ref{app:colla}, relying on measurements in ANI-3SIN configurations with various spatial resolutions, we estimate that the quantity $a_{\rm sc}$ shown in Table~\ref{tab: initial conditions}  should be accurate at the fourth significant digit level or better for all configurations, e.g. $a_{\rm sc}=0.02919 \pm 10^{-4}$ for ANI-3SIN, except for SYM-3SIN, where the uncertainty on the measured $a_{\rm sc}$ could be of a few $10^{-4}$. Indeed, as just mentioned above, accuracy on the determination of collapse time also depends on the nature of initial conditions, the quasi-1D case and the triaxial-symmetric configuration being the easiest and the most difficult to deal with, respectively, as expected.

Note that accuracy on collapse time also depends on the ability to determine the exact position of the shell-crossing point. Using the tessellation technique, shell crossing coincides with a temporal change of sign of the orientation of the simplices. In our symmetrical set-up, the position of the shell-crossing point is simply the center of the system and the determination of collapse time can be simply performed using the intersection of the phase-space sheet with the $y=z=0$ hyperplane (see Appendix~\ref{app:colla}). However, even in a more complex case where the center of the forming halo would be moving, one expects to be able to compute collapse time accurately thanks to the tessellation representation as long as sampling of the phase-space sheet is sufficiently dense and, of course, if the quadratic representation inside each simplex is fully exploited. Sampling of the phase-space sheet is controlled by $n_{\rm s}$, as well as the refinement parameter $I$ limiting deviations from symplectic motion. As discussed in Sec.~\ref{sec:coldice}, for our simulations, parameter $I$ does not influence the results except (probably marginally) for SYM-3SIN. Effects related to the choice of $n_{\rm s}$ are discussed in \citet[][]{2016JCoPh.321..644S} and in C21. They should be negligible for our sine wave simulations compared to other sources of uncertainty on collapse time for the values of $n_{\rm s}$ we adopted. However, we did not test the accuracy of {\tt ColDICE} in more complex cases, because we do not need to.

An additional advantage of our setting is that the axes of the sine waves are aligned with the mesh used to solve Poisson equation, which helps making the simulations more accurate: the same configurations in an oblique fashion would not achieve the same level of accuracy. This is illustrated by appendix H of \citet[][]{2016JCoPh.321..644S} in the single sine wave case.  As shown in this Appendix, anisotropies due to misalignments between preferred directions in the representation and preferred directions in the dynamics are the source of significant accuracy loss. However, these cumulative effects should remain small for {\tt ColDICE} at the early times considered in this work.

Another critical issue is to determine accurately the shape of the caustics and then the density and vorticity fields inside the multi-stream region. The technique employed to extract the caustics curves and surfaces from the 2D and the 3D simulations, respectively, is rather simple and robust, as briefly sketched in Appendix~\ref{sec:appcaus}. Based on a detection of a spatial sign change on the simplices orientation, it consists in extracting a subset of segments (in 2D) or triangles (in 3D) from the tessellation. These segments and triangles correspond to intersections between adjacent simplices. Obviously this is a rather crude way to draw the caustics, since, in Lagrangian space, it ends up into a subset of segments/triangles of a regular pattern. Therefore, the accuracy in the determination of the caustics location in Lagrangian space is at best of the order of the distance between two neighbouring vertices, i.e. $L/n_{\rm s}$ where $L$ is the size of the simulation cube. This uncertainty, which adds up to the effects discussed above, is difficult to transpose to Eulerian space due to the variations in the strain tensor. It is however probably the main source of inaccuracy on the determination of the caustics in {\tt ColDICE}. A way to improve greatly on this would consist in exploiting the quadratic representation inside each simplex, which we do not do here.

Turning to the fields, the vorticity, because it is composed of derivatives, is particularly challenging to measure even with the considerable gain brought by the tessellation technique compared to more traditional representations based on particles. Appendix~\ref{app:quadfields} explains in details how we measure density and vorticity, making use, this time, of the locally quadratic representation of the phase-space sheet inside each simplex. Accuracy on the fields measurements is not discussed quantitatively, as it was not deemed necessary for the analyses performed in this work, but we notice that vorticity becomes very noisy when very close to the caustic curves/surfaces, which we will keep in mind for the analyses carried out in Sec.~\ref{sec: postcollapse caustics}.

%%%%%%%%%%%%%%%%%%%%%%%%%%%%%%%%%%%%%%%%%%%%%%%%%%
%%%%%%%%%%%%%%%%%%%%%%%%%%%%%%%%%%%%%%%%%%%%%%%%%%
%%%%%%%%%%%%%%%%%%%%%%%%%%%%%%%%%%%%%%%%%%%%%%%%%%
\section{Shell crossing structure}
\label{sec: result shell crossing}
%%%%%%%%%%%%%%%%%%%%%%%%%%%%%%%%%%%%%%%%%%%%%%%%%%
%%%%%%%%%%%%%%%%%%%%%%%%%%%%%%%%%%%%%%%%%%%%%%%%%%
%%%%%%%%%%%%%%%%%%%%%%%%%%%%%%%%%%%%%%%%%%%%%%%%%%
We are now in a position to study the structure of our protohaloes at collapse time, $t_{\rm sc}$, and concentrate our investigations on phase-space diagrams and radial profiles, with comparisons of LPT pushed up to 10th order to the Vlasov runs. This section is organized as follows. First, Sec.~\ref{sec: shell-crossing time} presents the calculation of collapse time itself. Indeed, this quantity depends on initial conditions and perturbation order, and the ability of LPT to provide an accurate determination of $t_{\rm sc}$ is of prime importance. We discuss the extrapolation to infinite order of the LPT series proposed by STC18 and generalize it to the 2D case. Second, Sec.~\ref{sec: phase space shell-crossing} examines the convergence of LPT at collapse with phase-space diagrams, extending the earlier investigation of STC18 to the 2D case. For comparison, we also test the formal extension of LPT to infinite order and the predictions of the quasi one-dimensional approach proposed by \citet[][hereafter RF17]{2017MNRAS.471..671R}. Finally, in Sec.~\ref{sec:radialprof}, LPT predictions and their convergence are studied in terms of radial profiles of the density, the velocity as well as the pseudo phase-space density, and put into perspective in relation to singularity theory.
%%%%%%%%%%%%%%%%%%%%%%%%%%%%%%%%%%%%%%%%%%%%%%%%%%
%%%%%%%%%%%%%%%%%%%%%%%%%%%%%%%%%%%%%%%%%%%%%%%%%%
\subsection{Shell-crossing time}
\label{sec: shell-crossing time}
%%%%%%%%%%%%%%%%%%%%%%%%%%%%%%%%%%%%%%%%%%%%%%%%%%
%%%%%%%%%%%%%%%%%%%%%%%%%%%%%%%%%%%%%%%%%%%%%%%%%%
This subsection presents estimates of the expansion factor at first shell-crossing to which we refer to as collapse time. In the following, the expansion factor will be formally identified to a time variable, still denoted by $a$ to contrast with actual physical time $t$. We compare the value $a_{\rm sc}^{(n)}$ of collapse time obtained from $n$th-order LPT and its extrapolation to infinite order $a^{\infty}_{\rm sc}$, as described in STC18, to the value $a_{\rm sc}$ measured in the Vlasov runs as explained in Appendix~\ref{app:colla}. Our approach follows closely that of STC18. It is basically intended to repeat its main steps and to supplement it with additional discussions and comparisons to Vlasov runs in the 2D case.

\begin{figure}[!htbp]
\centering
\includegraphics[width=0.45\textwidth]{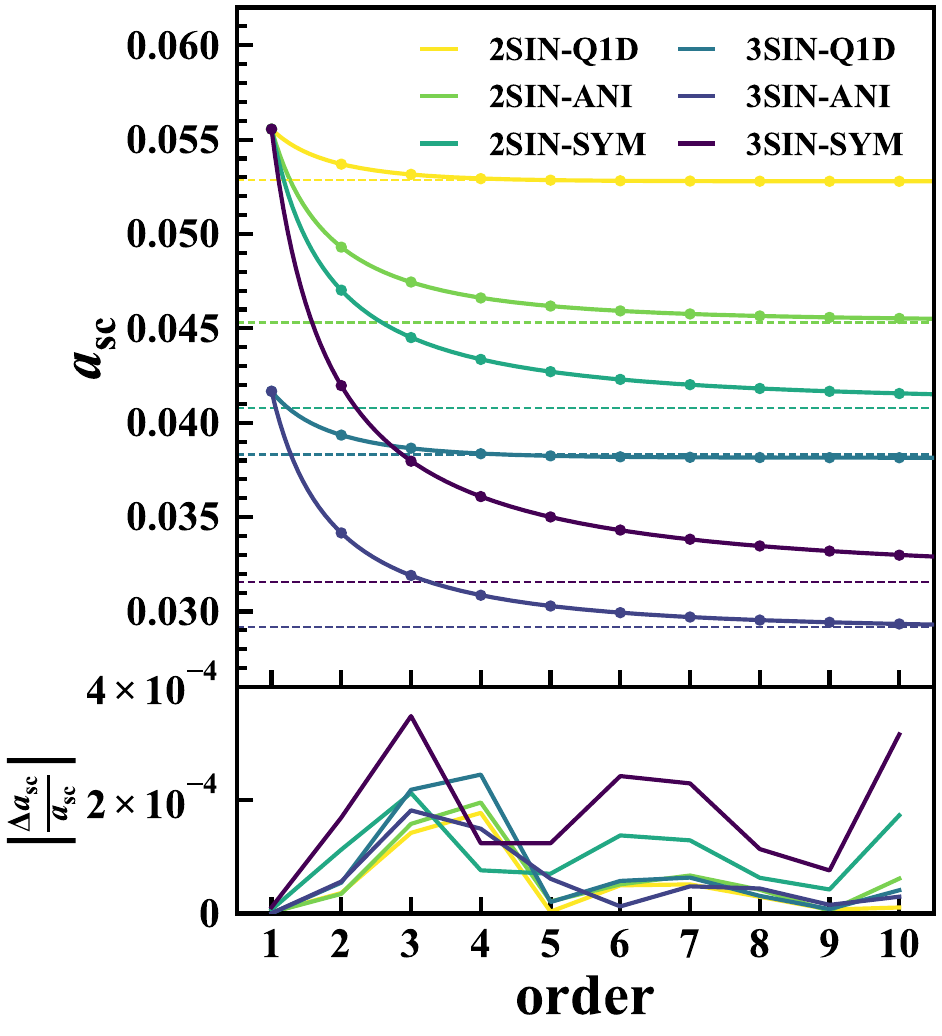}
\caption{
{\it Top:} the shell-crossing time calculated at $n$th-order LPT~(dots), together with the fitting function given by equation (\ref{eq:fitting}) (solid lines), as a function of the perturbation order, for various initial conditions as indicated on the panel. The dashed horizontal lines presents the values estimated from the simulations (see Appendix~\ref{app:A1}), corresponding to the sixth column of Table~\ref{tab: initial conditions}.
{\it Bottom:} The relative error between the shell-crossing time calculated with $n$th-order LPT and the one obtained with the fitting formula.
}
\label{fig: shell-crossing time}
\end{figure}

Using $n$th-order LPT predictions, we explore the sequence of shell-crossing times, $a_{\rm sc}^{(n)}$, as a function of order $n$ up to $n=10$, by solving $J^{(n)} = 0$ at the origin, where $J^{(n)}$ is the Jacobian of the $n$th-order LPT solution.
Generally, except in the pure one-dimensional case where first-order LPT is exact before collapse, $a_{\rm sc}^{(n)}$ becomes smaller with increasing perturbation order $n$ \citep[see, e.g., RF17,][hereafter RH21, for recent works]{2021MNRAS.501L..71R}. As illustrated by Fig.~\ref{fig: shell-crossing time}, the shell-crossing time calculated at $n$th-order with LPT is very accurately described by the following fitting form (STC18): 
%-------------------------------------------------
\begin{align}
a_{\rm sc}^{(n)} = a^{\infty}_{\rm sc} + \left( b + c \exp\left[ d \,n^{e}\right] \right)^{-1} ,
\label{eq:fitting}
\end{align}
% -------------------------------------------------
with $e > 0$. This fitting form, also used for each coordinate of the positions and velocities in Fig.~\ref{fig: phase precollapse} below, does not necessarily represent the sole choice for approximating the $n$ dependence of collapse time, but using the exponential of a power-law might be the only way to match the convergence speed of LPT at large $n$, when considering quantities computed at collapse time $a_{\rm sc}^{(n)}$ of each respective order. Equation (\ref{eq:fitting}) also implies $a_{\rm sc}^{(n)}-a_{\rm sc}^{(n-1)} \sim \exp(-|d|\, n^e)$ when $n \gg 1$, which might, at first sight, seem incompatible with the findings of RH21, who examined the LPT series in the slightly different context of a Gaussian random field in a periodic box. RH21 concluded from their analyses that convergence speed was asymptotically compatible with a power-law times an exponential of $n$. This would naively correspond to $a_{\rm sc}^{(n)}-a_{\rm sc}^{(n-1)} \sim n^{\kappa} \exp(-\eta\, n)$, i.e. $e=1$, while our measurements suggest values of $e$ different from unity, ranging from e.g. $e \sim 0.003$ (SYM-3SIN) to $e \sim 0.8$ (Q1D-3SIN) for fitting $a_{\rm sc}^{(n)}$ and from e.g. $e \sim 0.6 \mbox{--} 0.7$ (SYM-3SIN) to $e \sim 1.3 \mbox{--} 1.5$ (Q1D-3SIN) for fitting the Eulerian position. However, in RH21, the LPT series is studied in terms of the Taylor coefficients of the displacement field expanded as a function of the linear growing mode $D_+$, while we consider the sequence of shell-crossing times $a_{\rm sc}^{(n)}$, or equivalently $D_+^{(n)}$. This additional $n$ dependence introduced in the time variable fundamentally changes the nature of the LPT series as a function of $n$, which makes a direct comparison of equation (\ref{eq:fitting}) to the findings of RH21 inappropriate. Yet, it would be interesting to investigate the convergence properties of our systems seeded with sine waves using the approach of RH21.

One important thing to note is that convergence of collapse time with order is rather slow, except in the quasi-1D case, which still requires at least third order for reaching a $\sim$percent level of accuracy for approximate convergence, while much higher order is required for other configurations, especially the 3D axial-symmetric case, $\bm{\epsilon}_{\rm 3D}=(1,1)$, for which convergence does not seem to be achieved even at 10th order. Fig.~\ref{fig: shell-crossing time} thus demonstrates that low-order perturbation theory cannot be used to estimate accurately collapse times (see also STC18, in particular, Fig.~2 of this article, and RH21).

Despite these convergence issues, the extrapolated values of $a^{\infty}_{\rm sc}$ obtained with equation (\ref{eq:fitting}) are very accurate, as illustrated by Table~\ref{tab: initial conditions}. Indeed, the relative difference between $a^{\infty}_{\rm sc}$ and the value $a_{\rm sc}$ measured in the Vlasov simulations remains of the order of the percent or below, which suggests that the error on the extrapolated estimate of collapse time remains small and should not affect significantly the analytical predictions of the phase-space structure and the radial profiles at shell-crossing. Therefore, we will use $a^{\infty}_{\rm sc}$ for the analytical predictions of ``exact'' collapse time in the remaining section \ref{sec: result shell crossing}. As for the simulations data, we will use the snapshot with the closest possible available value ${\hat a}_{\rm sc}$ of the expansion factor to $a_{\rm sc}$, as indicated in Table~\ref{tab: initial conditions}.

%-------------------------------------------------
%-------------------------------------------------

%%%%%%%%%%%%%%%%%%%%%%%%%%%%%%%%%%%%%%%%%%%%%%%%%%
%%%%%%%%%%%%%%%%%%%%%%%%%%%%%%%%%%%%%%%%%%%%%%%%%%
\subsection{Phase-space structure}
\label{sec: phase space shell-crossing}
%%%%%%%%%%%%%%%%%%%%%%%%%%%%%%%%%%%%%%%%%%%%%%%%%%
%%%%%%%%%%%%%%%%%%%%%%%%%%%%%%%%%%%%%%%%%%%%%%%%%%

%-------------------------------------------------
\begin{figure*}[!htbp]
\centering
\includegraphics[width=0.75\textwidth]{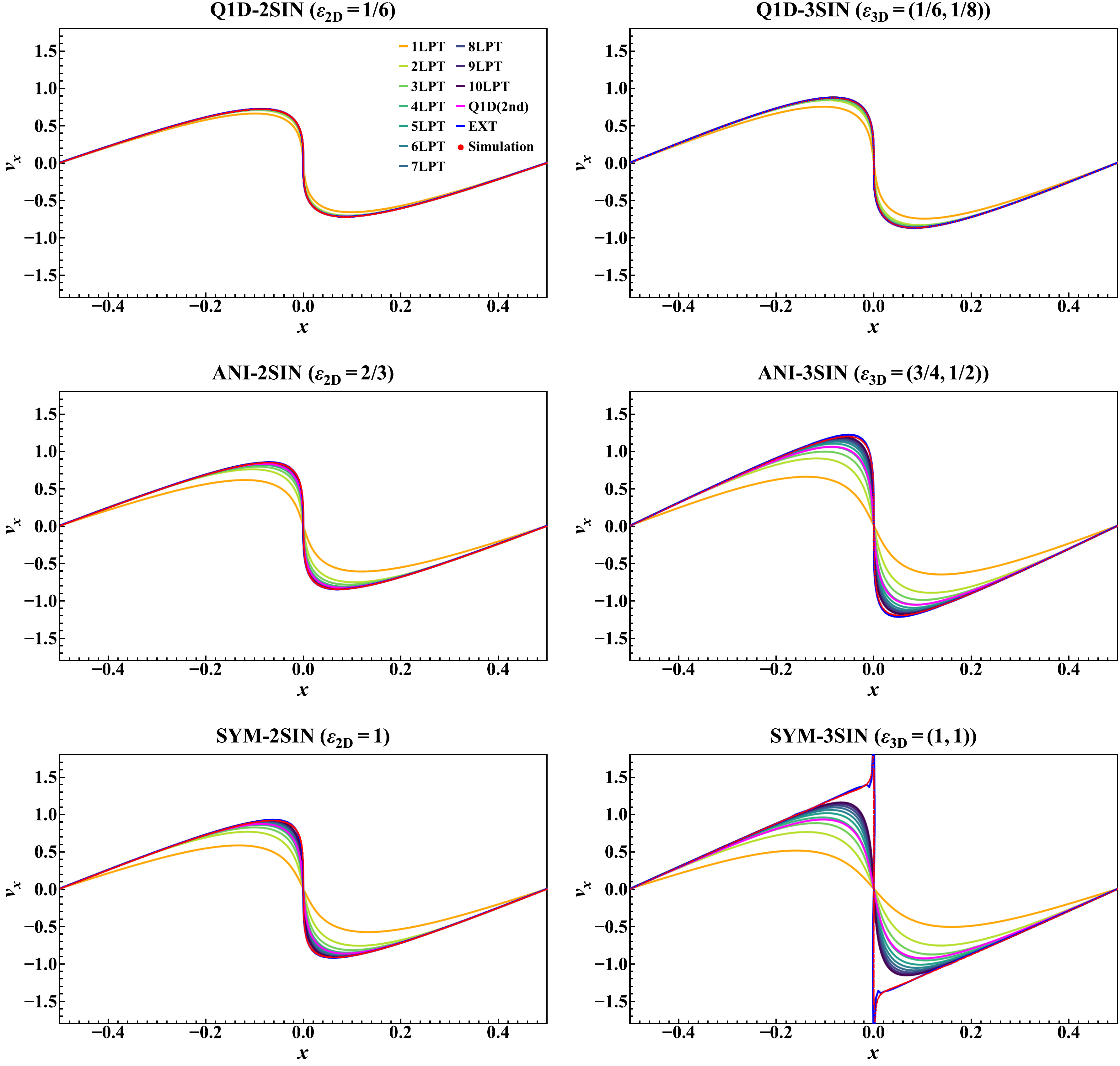}
\caption{Phase-space structure for two and three sine waves initial conditions at collapse time: Q1D-2SIN~({\it top left}), ANI-2SIN~({\it center left}), SYM-2SIN~({\it bottom left}), Q1D-3SIN~({\it top right}), ANI-3SIN~({\it center right}), and SYM-3SIN~({\it bottom right}).
The intersection of the phase-space sheet with the $y = 0$ plane for two sine waves and $y = z = 0$ hyper-plane for three sine waves is displayed in $(x,v_{x})$ subspace. Simulation results are compared to standard LPT predictions, that are supplemented with the blue line, denoted by ``EXT'', which corresponds to the formal extrapolation to infinite order proposed by STC18 and sketched in Sec.~\ref{sec: shell-crossing time} for the collapse time. For completeness, the quasi one-dimensional approach \citep{2017MNRAS.471..671R}, denoted by Q1D, is also presented (see Appendix~\ref{app: Q1DPT} for details).}
\label{fig: phase precollapse}
\end{figure*}
%-------------------------------------------------
Fig.~\ref{fig: phase precollapse} shows the $(x,v_{x})$ subspace of the phase-space structure at shell-crossing by considering the intersection of the phase-space sheet with the hyperplane $y=0$ for two sine waves and $y=z=0$ for three sine waves. For comparison, in addition to standard LPT results, we also present the formal extension of LPT to infinite order as developed by STC18 and described in Sec.~\ref{sec: shell-crossing time} for the shell-crossing time, as well as the analytical prediction obtained in the context of the quasi one-dimensional approach developed by RF17, that we extended to second order in the transverse direction (see Appendix~\ref{app: Q1DPT} for details). 

For quasi-1D (Q1D-2SIN and Q1D-3SIN) and anisotropic (ANI-2SIN and ANI-3SIN) initial conditions, the phase-space sheet first self-intersects along $x$-axis, and, until then, the dynamics is similar to the pure one-dimensional case, in which 1LPT (the Zel'dovich solution) is exact until shell-crossing. As illustrated by Fig.~\ref{fig: phase precollapse}, the LPT prediction quickly converges with perturbation order $n$ and describes the simulation results well even at shell-crossing, especially in the Q1D case.  From visual inspection of Fig.~\ref{fig: phase precollapse}, it appears that the agreement between the LPT predictions and the simulations is the worse in the vicinity of the extrema of the $x$-coordinate of the velocity field as a function of position $x$. Note that, due to the symmetry of the initial conditions, these extrema are global and not specific to the phase-space slice represented in Fig.~\ref{fig: phase precollapse}.

In the axial-symmetric cases, SYM-2SIN and SYM-3SIN, the phase-space sheet self-intersects simultaneously along all the axes of the dynamics. Interestingly, the phase-space structure for $\epsilon_{\rm 2D}=1$~(SYM-2SIN) is still qualitatively similar to one-dimensional collapse and LPT again quickly converges with perturbation order to the exact solution. This is clearly not the case for the three-dimensional axial-symmetric configuration, $\bm{\epsilon}_{\rm 3D}=(1,1)$, where LPT convergence is very slow. This set-up is indeed qualitatively different from other initial conditions, with the appearance of a spiky structure in phase-space (SCT18) similarly as in spherical collapse. Interestingly, in the spherical case and in the Einstein-de Sitter cosmology that we consider in this paper, convergence of the LPT is likely to be lost at collapse time \citep[e.g.,][]{2019MNRAS.484.5223R}, and the analyses of RH21 suggest that the velocity blows up when convergence is lost. The exact properties of the spike we observe in our numerical data, in particular, whether the velocity diverges or remains finite, and, whether if finite, the velocity is actually smooth at the fine level, remain unknown. While this spike is not present in the LPT predictions at finite order, it is well reproduced by the formal extrapolation to infinite order. This is a hint that convergence of the LPT series might be lost at collapse for SYM-3SIN.

A question might arise whether, because of such a spike and because of their highly contrasted nature, close to axial-symmetric 3D configurations correspond to a potentially different population of protohaloes. A partial answer can be found in C21, who followed numerically the evolution of the three sine-wave configurations further in the non-linear regime. As described in C21, collapse of our protohaloes is followed by a violent relaxation phase leading to a power-law profile $\rho(r) \propto r^{-\alpha}$, with $\alpha \in [1.5, 1.7]$ and then by relaxation to an NFW like universal profile. After violent relaxation, C21 did not find specific signatures in the density profile nor the pseudo phase-space density for the axial-symmetric case compared to the non-axial-symmetric ones, except that $\alpha$ tends to augment when going from Q1D to axial-symmetric, and that the axial-symmetric configuration is subject to significant (and expected) radial orbit instabilities. 

To complete the analyses, note that the formal extrapolation of LPT to infinite order matches well (but not perfectly) the simulation results for all the configurations, as already found by SCT18 in the three-dimensional case. We also notice that the quasi one-dimensional approach of RF17 can describe very well the quasi-1D configurations. Interestingly, at the second order in the transverse fluctuations considered here, the predictions are rather similar to those of standard 4th-order LPT, irrespective of initial conditions. Thus, as expected, the quasi one-dimensional approach becomes less accurate when the ratio $\epsilon_{\rm 2D}$ or $\epsilon_{\rm 3D}$ approaches unity, particularly in 3D as already shown by STC18.

%%%%%%%%%%%%%%%%%%%%%%%%%%%%%%%%%%%%%%%%%%%%%%%%%%
%%%%%%%%%%%%%%%%%%%%%%%%%%%%%%%%%%%%%%%%%%%%%%%%%%
\subsection{Radial profiles}
\label{sec:radialprof}
%%%%%%%%%%%%%%%%%%%%%%%%%%%%%%%%%%%%%%%%%%%%%%%%%%
%%%%%%%%%%%%%%%%%%%%%%%%%%%%%%%%%%%%%%%%%%%%%%%%%%
We now focus on radial profiles, and define Eulerian polar and spherical coordinates for two and three sine waves initial conditions by $(x,y) = (r\cos{\theta}, r\sin{\theta})$ and $(x,y,z) = (r\sin{\theta}\cos{\phi}, r\sin{\theta}\sin{\phi},r\cos{\theta})$, respectively.
Then the angular averaged radial profiles of the density $\rho/\bar{\rho}$, velocity dispersion $v^{2}$, radial velocity dispersion $v^{2}_{r}$, and infall velocity $-v_{r}$ are given by
\begin{align}
\rho(r,t)/\bar{\rho} &= \Braket{J^{-1}(\bm{q}, t)}_{\Omega} , \\
v^{2}(r,t) &= \frac{\Braket{J^{-1}(\bm{q}, t) \,v^{2}(\bm{q}, t)}_{\Omega}}{\Braket{J^{-1}(\bm{q}, t)}_{\Omega}} , \\
v^{2}_{r}(r,t) &= \frac{\Braket{J^{-1}(\bm{q}, t) \,\left( \bm{v}(\bm{q}, t)\cdot\hat{\bm{x}} \right)^{2}}_{\Omega}}{\Braket{J^{-1}(\bm{q}, t)}_{\Omega}}, \\
v_{r}(r,t) &= \frac{\Braket{J^{-1}(\bm{q}, t) \,\left( \bm{v}(\bm{q}, t)\cdot\hat{\bm{x}} \right)}}{\Braket{J^{-1}(\bm{q}, t)}_{\Omega}} ,
\end{align}
where $\hat{\bm{x}} = \bm{x}/r$ with $r = |\bm{x}|$ being the radial coordinate. In these equations, we used the angle average defined by
\begin{align}
\Braket{f(\bm{q})}_{\Omega} = \int\frac{{\rm d}\theta}{2\pi}\, \left. f(\bm{q})\right|_{\bm{x} = \bm{x}(\bm{q},t)}  ,
\end{align}
for two sine waves initial conditions, and
\begin{align}
\Braket{f(\bm{q})}_{\Omega} = \int\frac{\sin{\theta}{\rm d}\theta{\rm d}\phi}{4\pi}\, \left.f(\bm{q})\right|_{\bm{x} = \bm{x}(\bm{q},t)},
\end{align}
for three sine waves initial conditions.
It is important to note that the angular coordinates $\theta$ and $\phi$ in the integrands are the Eulerian coordinates, and the integrands should be evaluated in terms of the Eulerian coordinate by solving the equation $\bm{x} = \bm{x}(\bm{q},t)$.

To complete the analyses, we also study the pseudo phase-space density $Q(r,t)$, defined by
\begin{align}
Q(r,t) = \frac{\rho(r,t)/\bar{\rho}}{v^{3}(r,t)} . \label{eq: pseudo Q}
\end{align}
Note in addition that we present not only the radial profiles but also their logarithmic slopes defined by
\begin{align}
n \equiv \frac{{\rm d}\ln{X}}{{\rm d}\ln{r}} ,
\end{align}
where the quantity $X$ stands for the radial profile under consideration.

%-------------------------------------------------
\begin{figure}[!htbp]
\centering
\includegraphics[width=0.45\textwidth]{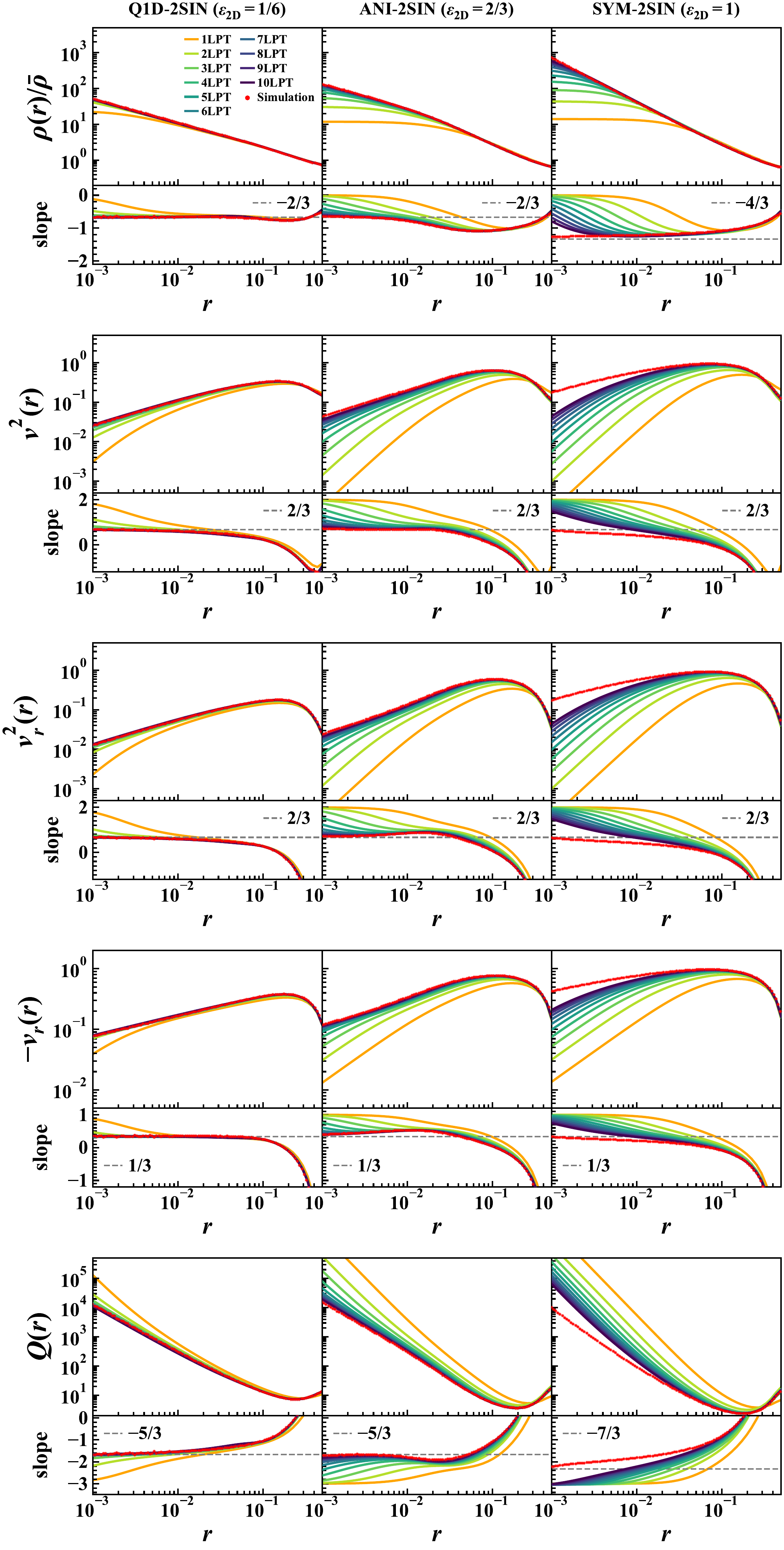}
\caption{Radial profiles and their logarithmic slopes for the two sine waves initial conditions, at shell-crossing time. As indicated on {\it top left panel}, LPT predictions are given by solid lines of various colours while Vlasov simulations results are represented by red dots. 
From {\it left} to {\it right}, the initial conditions are given by Q1D-2SIN~($\epsilon_{\rm 2D}=1/6$), ANI-2SIN~($\epsilon_{\rm 2D}=2/3$), and SYM-2SIN~($\epsilon_{\rm 2D}=1$), respectively.
From {\it top} to {\it bottom}, we present the radial profiles of the normalised density $\rho/\bar{\rho}$, the velocity dispersion $v^{2}$, the radial velocity dispersion $v^{2}_{r}$, the infall velocity $-v_{r}$, and the pseudo pseudo phase-space density $Q(r)$, respectively.
Note that when plotting the logarithmic slopes in Vlasov simulations, we used the Savitzky-Golay filter implemented in {\tt savgol\_filter} of {\tt SciPy}~\citep{2020SciPy} to smooth the data, for presentation purposes. In the logarithmic slopes panels, the horizontal dashed lines correspond to the theoretical predictions at small radii of Appendix~\ref{sec: analytic log slope} as listed in Table~\ref{tab: summary at shell-crossing}.
}
\label{fig: radial profile inft 2SIN}
\end{figure}
%-------------------------------------------------

%-------------------------------------------------
\begin{figure}[!htbp]
\centering
\includegraphics[width=0.45\textwidth]{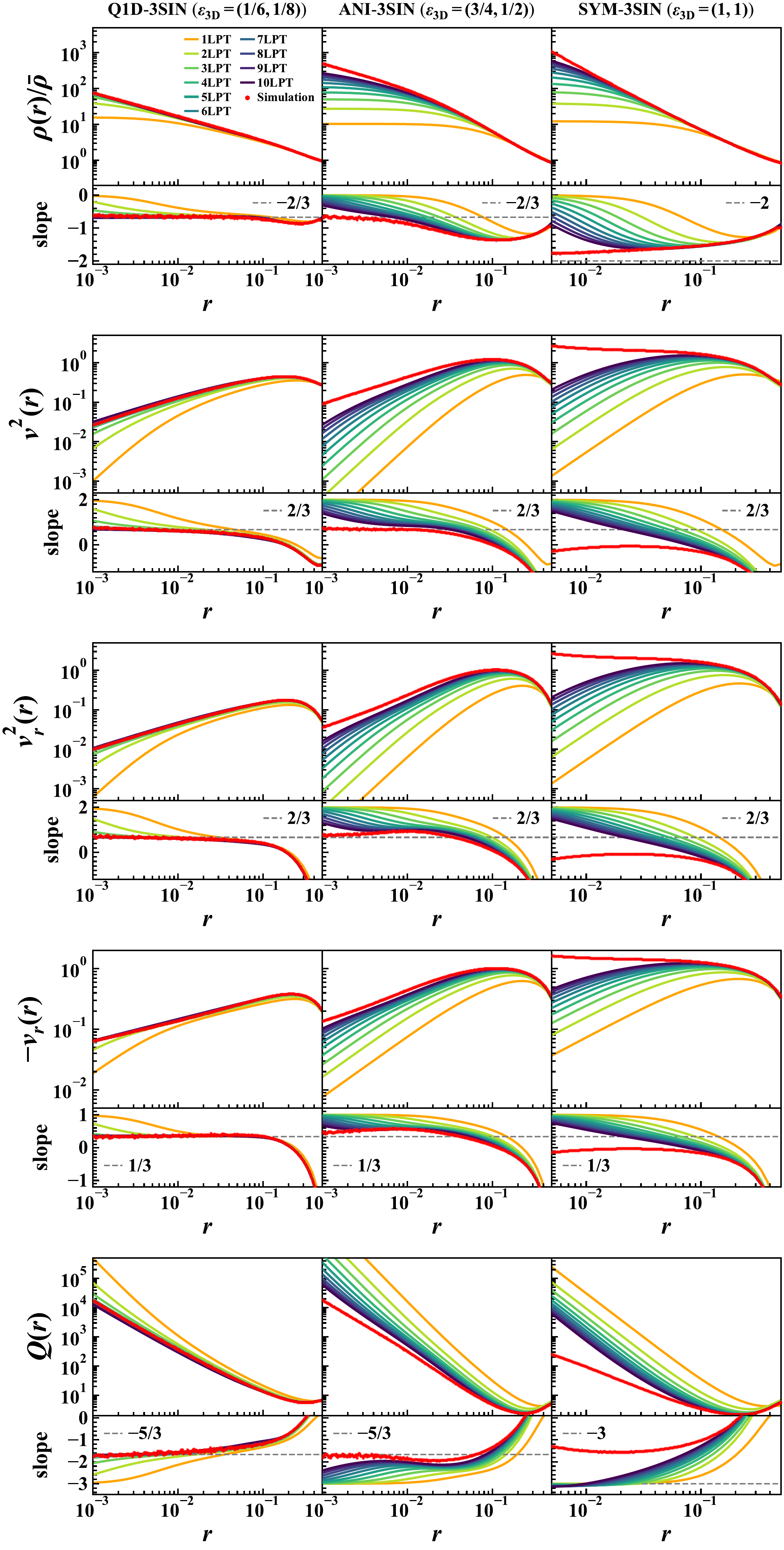}
\caption{Same as Fig.~\ref{fig: radial profile inft 2SIN} but for the three sine waves initial conditions, from {\it left} to {\it right}, Q1D-3SIN [$\bm{\epsilon}_{\rm 3D}=(1/6,1/8)$], ANI-3SIN [$\bm{\epsilon}_{\rm 3D}=(3/4,1/2)$], and SYM-3SIN~[$\bm{\epsilon}_{\rm 3D}=(1,1)$], respectively.
Note that in SYM-3SIN, the closest snapshot from collapse we had at disposal from our Vlasov runs, ${\hat a}_{\rm sc}=0.03190$, is significantly beyond actual shell-crossing time estimated by the method described in Sec.~\ref{app:colla}, $a_{\rm sc}=0.03155$, which explains some discrepancies between the theory and the simulation at small radii.
}
\label{fig: radial profile inft 3SIN}
\end{figure}
%-------------------------------------------------

%-------------------------------------------------
\begin{figure}[!htbp]
\centering
\includegraphics[width=0.45\textwidth]{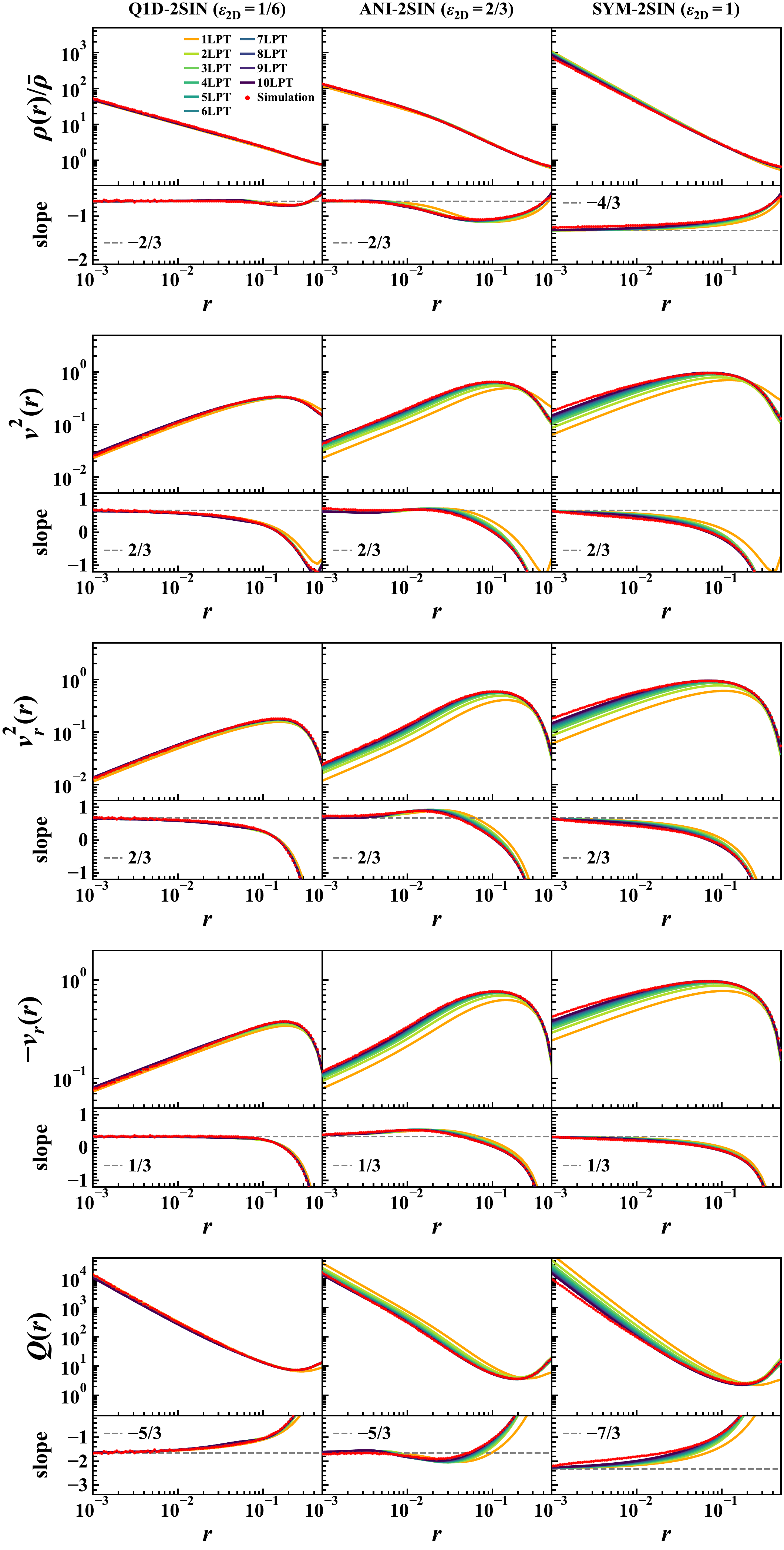}
\caption{Same as Fig.~\ref{fig: radial profile inft 2SIN} but all analytical predictions are evaluated at individual shell-crossing times computed at each perturbation order.}
\label{fig: radial profile sync 2SIN}
\end{figure}
%-------------------------------------------------

%-------------------------------------------------
\begin{figure}[!htbp]
\centering
\includegraphics[width=0.45\textwidth]{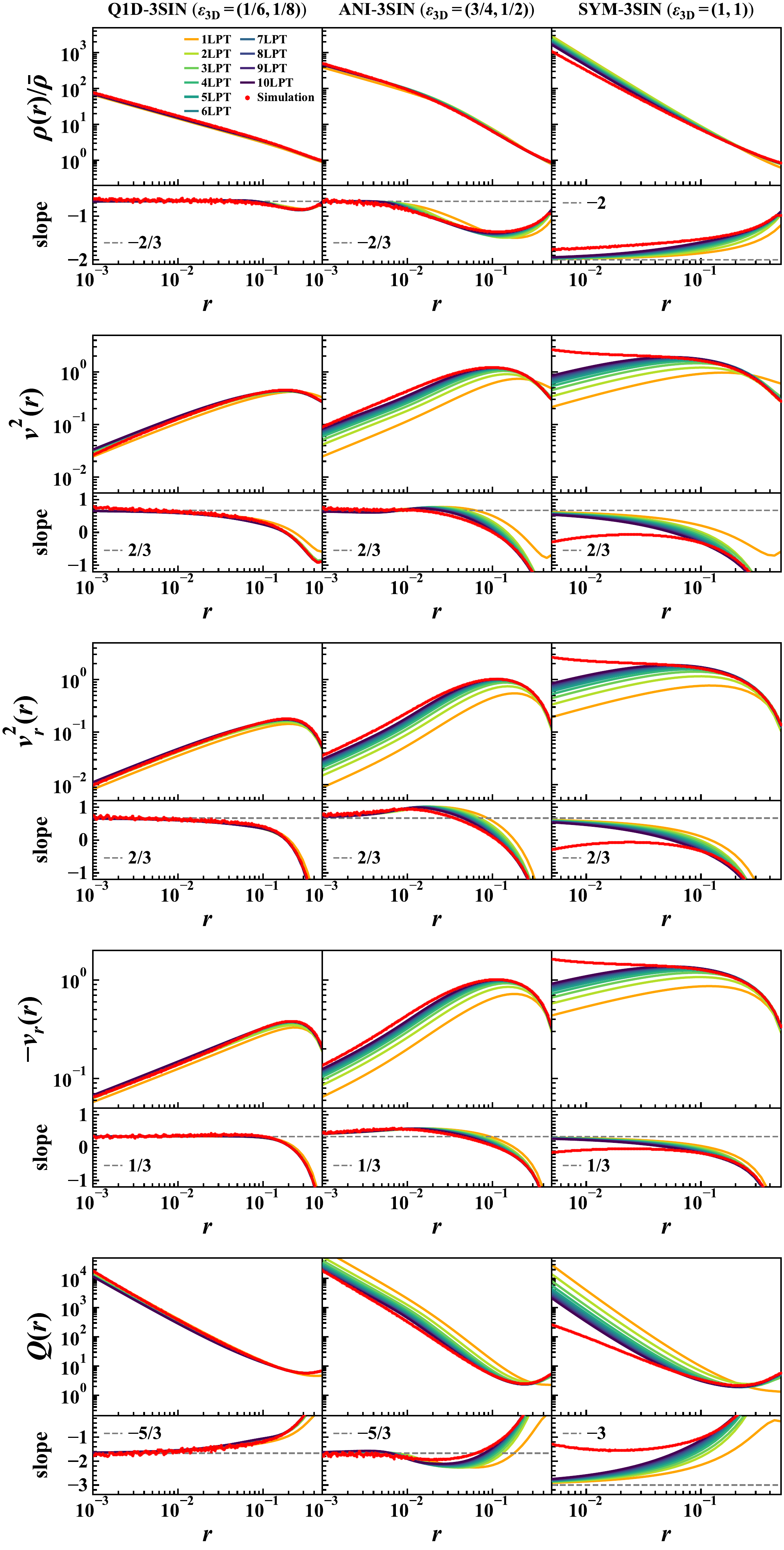}
\caption{Same as Fig.~\ref{fig: radial profile inft 3SIN} but all analytical predictions are evaluated at individual shell-crossing times computed at each perturbation order.}
\label{fig: radial profile sync 3SIN}
\end{figure}
%-------------------------------------------------

Radial profiles at shell-crossing are presented in Figs.~\ref{fig: radial profile inft 2SIN}--\ref{fig: radial profile sync 3SIN}. In these figures, measurements in the Vlasov code are performed by replacing each simplex of the phase-space sheet with a large ensemble of particles as explained in detail in C21 (see also Appendix~\ref{app:radp}). Two approaches of collapse are considered. In Figs.~\ref{fig: radial profile inft 2SIN} and \ref{fig: radial profile inft 3SIN}, which correspond respectively to two and three sine waves initial conditions, calculations are performed at ``exact'' shell-crossing time, that is the extrapolated value $a_{\rm sc}^{\infty}$ for LPT and the approximate value ${\hat a}_{\rm sc}$ for the simulations. Figs.~\ref{fig: radial profile sync 2SIN} and \ref{fig: radial profile sync 3SIN} are analogous, but examine LPT predictions of $n$th-order synchronised to their own shell-crossing time $a_{\rm sc}^{(n)}$, which allows us to examine in detail the structure of the singularity at collapse produced at each perturbation order. It is important to note that from now on, we do not examine further the quasi one-dimensional approach proposed by RF17. We also do not perform the formal extension to infinite order (Eq.~\ref{eq:fitting}), because calculating radial profiles in this framework was found to be too costly with the computational methods we are currently using. 

The examination of these figures confirms the conclusions of the phase-space diagram analysis in Sec.~\ref{sec: phase space shell-crossing}. In particular, quasi-1D profiles (Q1D-2SIN and Q1D-3SIN) are well described by LPT predictions even at low order. At ``exact'' collapse, considered in Figs.~\ref{fig: radial profile inft 2SIN} and \ref{fig: radial profile inft 3SIN}, LPT predictions generally accurately describe the outer part of the halo, where density contrasts are lower, and then deviate from the exact solution when the radius becomes smaller than some value $r_{\rm min}(n)$ decreasing with increasing $n$, where $n$ is the perturbation order. LPT can describe arbitrarily large densities, as long as $n$ is large enough. In the results presented here, density contrasts as large as 100 or larger can be probed accurately by LPT, depending on the order considered and on the nature of initial conditions. Again, convergence worsens when approaching three-dimensional axial-symmetry, the worse case being, as expected, SYM-3SIN with $\bm{\epsilon}_{\rm 3D} = (1,1)$. It is worth noticing that velocity profiles require significantly higher LPT order than density to achieve a comparable visual match between theory and measurements. In particular, deviations between LPT predictions and simulations arise at quite larger radii $r_{\rm min}(n)$ for the velocities than for the density, with differences as large as an order of magnitude. With synchronisation, performances of LPT predictions greatly improve, as expected and as illustrated by Figs.~\ref{fig: radial profile sync 2SIN} and \ref{fig: radial profile sync 3SIN}. All perturbation orders predict strikingly similar density profiles, in close to perfect match with the simulations, except for SYM-3SIN discussed further below, while the velocity profiles still diverge slightly from each other except for quasi one-dimensional initial conditions, Q1D-2SIN and Q1D-3SIN. 

We now turn to the logarithmic slope of various profiles shown in the bottom insert of each panel of Figs.~\ref{fig: radial profile inft 2SIN}--\ref{fig: radial profile sync 3SIN}. The nature of the singularities of cold dark matter structures at collapse time and beyond has been widely studied in the literature, particularly in the framework of Zel'dovich motion~\citep[e.g.,][]{1970A&A.....5...84Z,1982GApFD..20..111A,2014MNRAS.437.3442H,2018JCAP...05..027F}. In Appendix~\ref{sec: analytic log slope}, we rederive the asymptotic properties of our protohaloes at small radii by Taylor expanding the equations of motion up to third order in the Lagrangian coordinate $\bm{q}$. For our sine-wave configurations, whatever the order $n$ of the perturbation order,\footnote{Strictly speaking, our calculations are performed for $n \leq 10$, but it is reasonable to expect that the result applies to arbitrary order.} three kinds of singularities are expected at shell-crossing, as summarised in Table~\ref{tab: summary at shell-crossing}: the classic one-dimensional pancake with a power-law profile at small radii of the form (S1) $\rho(r) \propto r^{-2/3}$, valid for all the configurations expect SYM-2SIN and SYM-3SIN; then (S2) $\rho(r) \propto r^{-4/3}$ and (S3) $\rho(r) \propto r^{-2}$, for SYM-2SIN and SYM-3SIN, respectively. On the other hand, velocities are expected to follow the same power-law pattern whatever initial conditions or dimensionality, with logarithmic slopes equal to $2/3$, $2/3$, and $1/3$ respectively, for $v^{2}$, $v^{2}_{r}$, and $-v_{r}$, which in turn implies $Q(r) \propto r^{-7/3}$ for SYM-2SIN, $r^{-3}$ for SYM-3SIN, and $r^{-5/3}$ for other configurations.
%-------------------------------------------------
\begin{table*}
\centering
\begin{tabular}{lcccccc}
\hline
Sine wave amplitudes & Designation & $\frac{{\rm d}\ln{\rho/\bar{\rho}}}{{\rm d}\ln{r}}$ & $\frac{{\rm d}\ln{v^{2}}}{{\rm d}\ln{r}}$ & $\frac{{\rm d}\ln{v^{2}_{r}}}{{\rm d}\ln{r}}$ & $\frac{{\rm d}\ln{(-v_{r})}}{{\rm d}\ln{r}}$ & $\frac{{\rm d}\ln{Q}}{{\rm d}\ln{r}}$ \\
\hline
$0 \leq -\epsilon_z \leq -\epsilon_y < -\epsilon_x$ & Q1D-2SIN, ANI-2SIN, Q1D-3SIN, ANI-3SIN & -2/3 & 2/3 & 2/3 & 1/3 & -5/3\\
\hline
 $ \epsilon_z=0$, $0 <- \epsilon_y=-\epsilon_x$ & SYM-2SIN & -4/3 & 2/3 & 2/3 & 1/3 & -7/3\\
 \hline
$0< -\epsilon_z=-\epsilon_y=-\epsilon_x$ & SYM-3SIN & -2 & 2/3 & 2/3 & 1/3 & -3\\
\hline
\end{tabular}
\caption{Summary of the logarithmic slopes at shell-crossing obtained by singularity theory applied to LPT predictions at various orders (see Appendix~\ref{sec: analytic log slope} for details). The first and second columns indicate the relative amplitudes of the initial sine waves and the designation of the runs, respectively. The third through seventh columns show the logarithmic slopes of the normalised density, velocity dispersion, radial velocity dispersion, infall velocity, and pseudo phase-space density, respectively. Note thus that the logarithmic slopes do not depend, as expected, on the perturbation order of LPT.}
\label{tab: summary at shell-crossing}
\end{table*}
%-------------------------------------------------

Figs.~\ref{fig: radial profile inft 2SIN} and \ref{fig: radial profile inft 3SIN} consider LPT predictions for perturbation order $n$ calculated at ``exact'' theoretical collapse time $a_{\rm sc}^{\infty}$ and not at individual collapse time $a_{\rm sc}^{(n)}$ at this perturbation order, so shell-crossing is not reached exactly, but gets nearer as $n$ increases. Consequently, the asymptotic slope is only approached approximately, and better so with larger $n$. Note that convergence is slower for velocities than for the density, especially for axial-symmetric configurations, in particular SYM-3SIN. With synchronisation, as illustrated by Figs.~\ref{fig: radial profile sync 2SIN} and \ref{fig: radial profile sync 3SIN}, the convergence at a small radius to the prediction of singularity theory becomes clear.

Except for SYM-3SIN for which the best available snapshot is still too far off collapse with ${\hat a}_{\rm sc} > a_{\rm sc}$, one notices that all the simulated data show a close to perfect agreement between the measured slope at small radii and the predicted ones for all the radial profiles. This result is non-trivial in the sense that the Taylor expansion at small $|\bm{q}|$ underlying singularity theory is not necessarily valid in the actual fully nonlinear framework. Indeed, while singularity theory seems to apply to each order $n$ for an arbitrarily large value of $n$ in LPT framework, this does not mean that the limit to $n \rightarrow \infty$ should converge to the same singularity. The fact that the simulation agrees with singularity theory predictions beyond the well known, but nonetheless somewhat trivial, 1D case, is remarkable even though naturally expected. 

\begin{figure*}[!htbp]
\centering
\includegraphics[width=\textwidth]{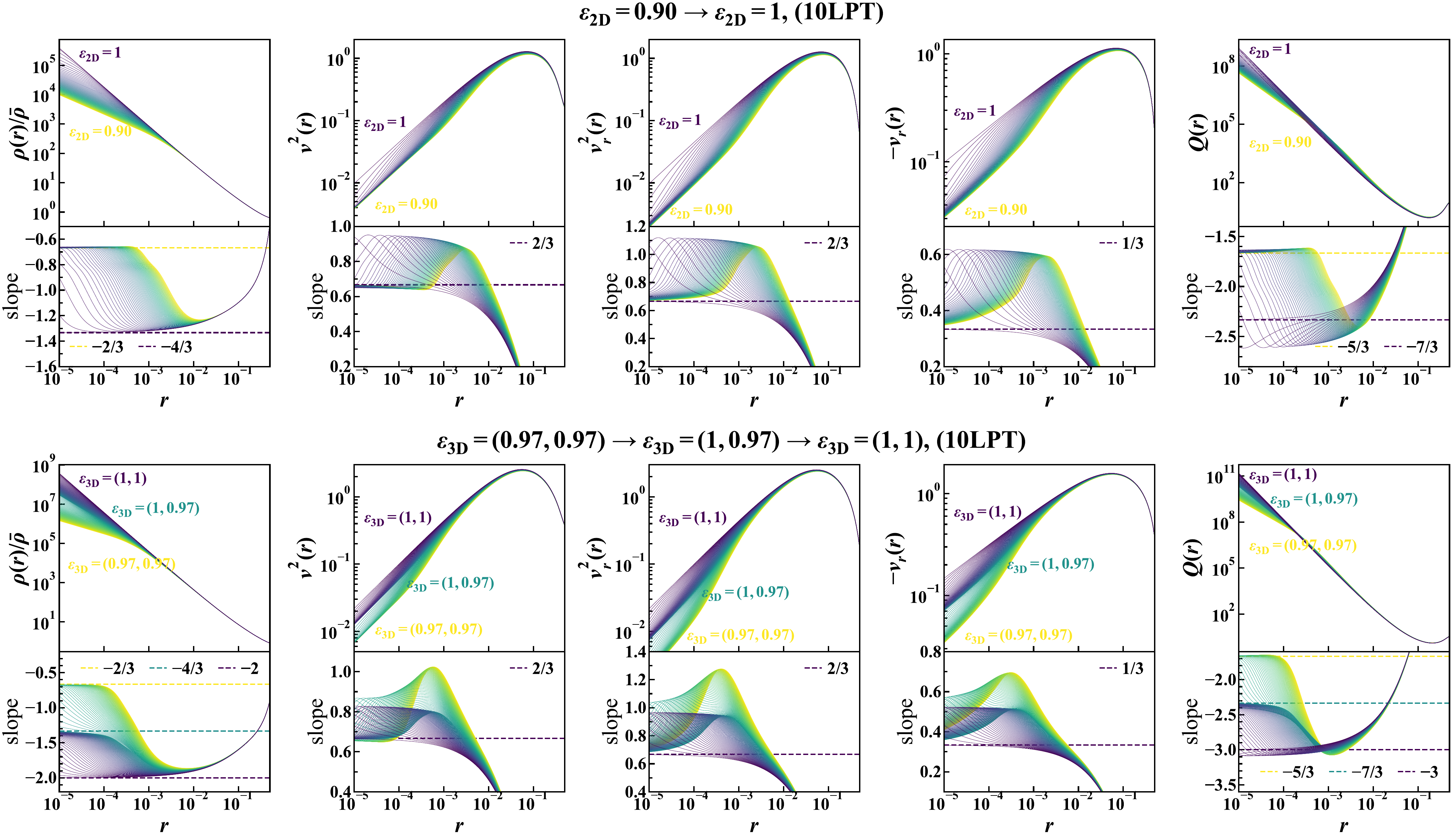}
\caption{Radial profiles at shell-crossing using 10th-order LPT in the vicinity of $\epsilon_{\rm 2D} \simeq \epsilon_{\rm 2D}^{\rm SYM} \equiv 1$ ({\it top panels}) and $\bm{\epsilon}_{\rm 3D}\simeq \bm{\epsilon}_{\rm 3D}^{\rm SYM} \equiv (1,1)$ ({\it bottom panels}). On {\it top panels}, $\epsilon_{\rm 2D}$ is progressively varying in the range $[0.9,1]$ with linear intervals of $\Delta\epsilon = 0.1/64$. On {\it Bottom panels}, $\bm{\epsilon}_{\rm 3D}$ progressively changes from $(0.97, 0.97)$ to $(1,1)$ with linear intervals of $\Delta\epsilon = 0.03/64$. From {\it left} to {\it right}, we present radial profiles of the normalised density $\rho/\bar{\rho}$, the velocity dispersion $v^{2}$, the radial velocity dispersion $v^{2}_{r}$, the infall velocity $-v_{r}$, and the pseudo phase-space density $Q$, respectively.
In the lower inserts corresponding to logarithmic slopes, the horizontal dashed lines correspond to the values expected from singularity theory as computed in Appendix \ref{sec: analytic log slope} and listed in Table~\ref{tab: summary at shell-crossing}.
}
\label{fig: 1D to 3D}
\end{figure*}

Several additional remarks are in order. First, if crossed sine waves are considered as generic approximations of local peaks of a smooth random Gaussian field, we also point out in this framework that the probability of having exactly $\epsilon_{\rm 2D}=\epsilon_{\rm 2D}^{\rm SYM} \equiv 1$ or $\bm{\epsilon}_{\rm 3D}=\bm{\epsilon}_{\rm 3D}^{\rm SYM} \equiv (1,1)$ is null, hence one expects the classic one-dimensional pancake to be exclusively dominant, from the pure mathematical view. However, this can be unrealistic at the coarse level, where high-density peaks can be close to spherical or filaments locally close to cylindrically symmetric. This is illustrated by Fig.~\ref{fig: 1D to 3D}, which examines the transition between various regimes in the vicinity of $\epsilon_{\rm 2D}^{\rm SYM}$ (top panels) and $\bm{\epsilon}_{\rm 3D}^{\rm SYM}$ (bottom panels) for $10$th-order LPT. Clearly, at sufficiently large radii, values of $\epsilon_{\rm 2D}$ or $\bm{\epsilon}_{\rm 3D}$ close to $\epsilon_{\rm 2D}^{\rm SYM}$ and $\bm{\epsilon}_{\rm 3D}^{\rm SYM}$ give similar results for the profiles, even for the logarithmic slope which can approach that of the axial-symmetric case. The exact mathematical asymptotic behaviour is indeed reached only at very small radii. In practice, the axial-symmetric configurations cannot be ignored, which shows the limits of singularity theory if applied blindly. 
%-------------------------------------------------
%-------------------------------------------------

Second, we note that the slope of the density profile predicted by LPT at collapse in the axial-symmetric 3D case, SYM-3SIN, is different from the prediction of pure spherical collapse, $\rho(r) \propto r^{-12/7}$, for an initial profile with the same asymptotic behaviour at a small radius of the form $\rho^{\rm ini} \propto 1 - \alpha r^2$ with $\alpha > 0$ \citep[see, e.g.,][]{1985PThPh..73..657N,1995ApJ...441...10M}. Hence the anisotropic nature of the axial-symmetric case (contained in terms beyond leading order $r^2$ in the Taylor expansion of trigonometric functions) provides significantly different results from pure spherical collapse. Note, though, that this state of fact is not fully proved by our simulations measurements, which as mentioned above, are slightly beyond actual collapse time.

Third, we can compare, in the three-dimensional case, the logarithmic slope of the pseudo phase-space density to that seen after violent relaxation and at late stages of the evolution of dark matter haloes, $Q(r) \sim r^{-\beta}$ \citep[see, e.g.,][]{2001ApJ...563..483T,2010MNRAS.402...21N,2010MNRAS.406..137L}, which has been found to be close to the prediction of secondary infall model, $\beta_{\rm sph} =1.875$ \citep[][]{1985ApJS...58...39B}. This is also the case of our three sine waves haloes, as analysed in detail by C21, including SYM-3SIN, which shows again that this particularly symmetric set-up, while being significantly more singular at collapse, with $\beta=3$, relaxes to a state similar to more generic three-dimensional configurations. The slope prior to collapse in the ``generic'' cases, $\beta=-5/3$, although of the same order of $\beta_{\rm sph}$, remains still different. Concerning the 2D case, which is different from the dynamical point of view, further analyses of the evolution of the haloes beyond collapse need to be done.

%%%%%%%%%%%%%%%%%%%%%%%%%%%%%%%%%%%%%%%%%%%%%%%%%%
%%%%%%%%%%%%%%%%%%%%%%%%%%%%%%%%%%%%%%%%%%%%%%%%%%
%%%%%%%%%%%%%%%%%%%%%%%%%%%%%%%%%%%%%%%%%%%%%%%%%%
\section{Structure of the system shortly after shell crossing}
\label{sec: shortly after sc}
%%%%%%%%%%%%%%%%%%%%%%%%%%%%%%%%%%%%%%%%%%%%%%%%%%
%%%%%%%%%%%%%%%%%%%%%%%%%%%%%%%%%%%%%%%%%%%%%%%%%%
%%%%%%%%%%%%%%%%%%%%%%%%%%%%%%%%%%%%%%%%%%%%%%%%%%
Once shell-crossing time is passed, the system enters into a highly non-linear phase of violent relaxation. In this regime, standard LPT is not applicable anymore, but it is still possible, shortly after collapse, to take into account the effects of the multi-streaming flow on the force field using an LPT approach \cite[see, e.g.,][for the introduction of ``post-collapse'' perturbation theory]{2015MNRAS.446.2902C,2017MNRAS.470.4858T}. While the motion can still be described again for a short while in a perturbative way from shell-crossing, it is not, overall, a smooth function of time anymore due to the formation of the singularity at collapse time \citep[][]{2019arXiv191200868R}. Still, it is reasonable to assume that very shortly after shell crossing, the back-reaction corrections due to the multi-streaming flow can be neglected, as a first approximation. However, it is important to notice that the convergence radius in time of the LPT series is finite in the generic case \citep[e.g.,][RH21]{2014JFM...749..404Z,2015MNRAS.452.1421R}. As illustrated by the measurement of the previous section, the convergence of the perturbation series is expected (although not proven) at least up to collapse time, but not necessarily far beyond collapse (see also RH21). Therefore, in what follows, we shall use the LPT solution of $n$th-order at collapse as a starting point, and from there, the standard ballistic approximation, where velocity field is frozen, to study the structure of the system shortly after collapse.

In practice, one would like to use sufficiently large perturbation order so that convergence of LPT is achieved, and ideally the formal extrapolation to infinite order proposed by STC18 and reintroduced in Sec.~\ref{sec: shell-crossing time}. However, this extrapolation is insufficiently accurate for the purpose of the calculations performed next, where very small time intervals after collapse are considered. Also, it is very costly from the computational point of view to exploit this extrapolation when solving the multivalued problem intrinsic to the multi-streaming solutions. So from now on, we shall consider only higher-order LPT calculations and not their extrapolation to infinite order (nor the quasi-1D approximation proposed by RF17). 

This section is organized as follows. In Sec.~\ref{sec: ballistic approx}, we introduce the ballistic approximation. We also relate, in the multi-stream regime, the Eulerian density and velocity fields to their Lagrangian counterparts, and introduce, in this framework, the vorticity field. Indeed, after shell-crossing, caustics form and non-zero vorticity is generated in the multi-stream region delimited by the caustics. In this section, we also discuss the caustic network created up to second order by our systems seeded by sine waves. Sec.~\ref{sec: postcollapse phase space} turns to actual comparisons of analytical predictions to Vlasov simulations using, similarly as in Sec.~\ref{sec: phase space shell-crossing}, phase-space diagrams, but shortly after collapse. In particular, we explore the limits of the ballistic approximation. Finally, Sec.~\ref{sec: postcollapse caustics} focuses on the overall structure of the multi-stream region in configuration space, by successively testing LPT predictions against simulations for the caustic pattern, the projected density and the vorticity fields. 

%%%%%%%%%%%%%%%%%%%%%%%%%%%%%%%%%%%%%%%%%%%%%%%%%%
%%%%%%%%%%%%%%%%%%%%%%%%%%%%%%%%%%%%%%%%%%%%%%%%%%
\subsection{Multi-stream regime and ballistic approximation}
\label{sec: ballistic approx}
%%%%%%%%%%%%%%%%%%%%%%%%%%%%%%%%%%%%%%%%%%%%%%%%%%
%%%%%%%%%%%%%%%%%%%%%%%%%%%%%%%%%%%%%%%%%%%%%%%%%%
Because collapse time $a_{\rm sc}^{(n)}$ depends on perturbation order, it only makes sense to test various perturbation orders shortly after shell-crossing and compare them to simulations only if collapse times are synchronised. In other words, in what follows, we consider the time $a_{\rm pc}^{(n)}=a_{\rm sc}^{(n)}+\Delta a$ for LPT of $n$th-order and $a_{\rm pc}^{\rm SIM}=a_{\rm sc}+\Delta a$ for the simulation, as listed in last column of Table~\ref{tab: initial conditions}. The value of $\Delta a$ used for each sine waves configuration is given by the difference between the values of the last and $6$th columns of the table. Remind that the quantity $a_{\rm sc}$ corresponds to the ``true'' collapse time measured in the Vlasov runs as explained in Appendix~\ref{app:colla}.

From the Eulerian coordinate and velocity fields given by $n$th-order LPT solutions at shell-crossing time $a_{\rm sc}^{(n)}$ of each perturbation order, we model the Eulerian coordinate after collapse as follows:
\begin{align}
\bm{x}(\bm{q}, a) &= \bm{x}^{(n)}(\bm{q}, a^{(n)}_{\rm sc}) + \left.\frac{\partial \bm{x}^{(n)}}{\partial a}\right|_{a=a_{\rm sc}^{(n)}}\Delta a , \\
&= 
\bm{x}^{(n)}(\bm{q}, a^{(n)}_{\rm sc}) + \left.\frac{\bm{v}^{(n)}}{a^{2}H(a)}\right|_{a=a^{(n)}_{\rm sc}}\Delta a ,
\label{eq: ballistic approx}
\end{align}
while the $n$th-order velocity field is frozen to its value at shell-crossing time of each perturbation order, $\bm{v} (\bm{q},a)= \bm{v}^{(n)}(\bm{q},a^{(n)}_{\rm sc})$.

In the multi-stream region, given the Eulerian coordinate $\bm{x}$, the solution of the equation $\bm{x} = \bm{x}(\bm{q})$ has an odd number of solutions for the Lagrangian coordinate, that we denote $\bm{q}_F$ labelled by the subscript $F=[1,\cdots,n_{\rm F}]$. Shortly after collapse, if $\epsilon_i\neq \epsilon_j$ for $i\neq j$, there are at most three streams, $n_{\rm F} \leq 3$, in a given point of space. For symmetric cases, e.g. $\epsilon_{\rm 2D}=1$ in 2D or $\epsilon_x=\epsilon_y$ in 3D, due to simultaneous shell-crossing along several axes, the number of streams can reach 9 and 27, respectively in two and three dimensions, as discussed further in Sec.~\ref{sec: postcollapse phase space}.

In what follows, we omit the time dependence for brevity. After defining the Lagrangian density and velocity fields: 
\begin{align}
\rho_{\rm L}(\bm{q}) &= |\det{J(\bm{q})}|^{-1} , \label{eq:rhoLdef} \\
\bm{v}_{\rm L}(\bm{q}) &= a \frac{{\rm d}\bm{\Psi}(\bm{q})}{{\rm d}t} ,
\end{align}
the Eulerian density and velocity fields are expressed by the superpositions of each flow:
\begin{align}
\rho(\bm{x}) &= \sum_{F} \rho_{\rm L}(\bm{q}_{F}) , \label{eq:rhoE} \\
\bm{v}(\bm{x}) &= \frac{\sum_{F}\rho_{\rm L}(\bm{q}_{F})\,\bm{v}_{\rm L}(\bm{q}_{F})}{\sum_{F}\rho_{\rm L}(\bm{q}_{F})} .\label{eq:vE}
\end{align}
where the summation $\sum_{F}$ is performed over all the solutions $\bm{q}_{F}$ of the equation $\bm{x} = \bm{x}(\bm{q}_{F})$.

While vorticity is not expected in single-stream regions, unless already present in the initial conditions (and in the latter case, it corresponds to a decaying mode), its generation is one of the prominent features of shell-crossing \citep[e.g.,][]{1999A&A...343..663P,2009PhRvD..80d3504P}.
By taking the curl of the velocity field with respect to the Eulerian coordinate, the vorticity is given, respectively in 2 and 3 dimensions, by
\begin{align}
\omega^{\rm 2D}(\bm{x})
&=\varepsilon_{ij}\frac{\partial v_{j}(\bm{x})}{\partial x_{i}}
\notag \\
&=
\varepsilon_{ij}
\frac{1}{\rho(\bm{x})}
\sum_{F}
J^{-1}_{mi}(\bm{q}_{F})
\frac{\partial \rho_{\rm L}(\bm{q}_{F})}{\partial q_{{F}, m}}
\Bigl[
v_{{\rm L},j}(\bm{q}_{F}) - v_{j}(\bm{x})
\Bigr]
\notag \\
& \qquad+
\varepsilon_{ij}
\frac{1}{\rho(\bm{x})}
\sum_{F}
J^{-1}_{mi}(\bm{q}_{F})
\frac{\partial v_{{\rm L},j}(\bm{q}_{F})}{\partial q_{{F}, m}}
\rho_{\rm L}(\bm{q}_{F})
, \label{eq: vorticity 2D} \\
\omega^{\rm 3D}_{i}(\bm{x})
&=
\varepsilon_{ijk}\frac{\partial v_{k}(\bm{x})}{\partial x_{j}} ,
\notag \\
&=
\varepsilon_{ijk}
\frac{1}{\rho(\bm{x})}
\sum_{F}
J^{-1}_{mj}(\bm{q}_{F})
\frac{\partial \rho_{\rm L}(\bm{q}_{F})}{\partial q_{{F}, m}}
\Bigl[
v_{{\rm L},k}(\bm{q}_{F}) - v_{k}(\bm{x})
\Bigr]
\notag \\
& \qquad+
\varepsilon_{ijk}
\frac{1}{\rho(\bm{x})}
\sum_{F}
J^{-1}_{mj}(\bm{q}_{F})
\frac{\partial v_{{\rm L},k}(\bm{q}_{F})}{\partial q_{{F}, m}}
\rho_{\rm L}(\bm{q}_{F}) .
\label{eq: vorticity 3D}
\end{align}
The vorticity fields in two- and three-dimensional space are scalar and vector quantities, respectively.
Because the acceleration derives from a gravitational potential, local vorticity on each individual fold of the phase-space sheet cancels.
Only the nonlinear superposition of folds in the first term of equations (\ref{eq: vorticity 2D}) and (\ref{eq: vorticity 3D}) induces non-zero vorticity, while the second term should not contribute.
Using LPT however generates spurious vorticity in Eulerian space due to the truncation at finite perturbation order (see, e.g., \citealt{1993MNRAS.264..375B,1994MNRAS.267..811B} and also \citealt{2019PhRvD..99h3524U} for a recent numerical investigation).
This spurious vorticity is mainly coming from the second term in Eqs.~(\ref{eq: vorticity 2D}) and (\ref{eq: vorticity 3D}), and can be especially noticeable in the single-stream region. We neglected it in our predictions by simply forcing this term to zero.

Before studying the structure of the system beyond collapse, we take the time to analytically investigate the properties of the ballistic solution for the sine waves case in the first- and second-order LPT, in order to understand better the measurements performed in the next paragraphs. In particular, it is important to know how the solution behaves in the multi-stream region, of which the boundaries are given by the caustics, where $J(\bm{q}) = 0$.

First, we start with 1LPT (Zel'dovich approximation). In the ballistic approximation, the Eulerian coordinate can be easily calculated shortly after shell-crossing:
\begin{align}
A^{(\rm 1LPT)}(\bm{q},a) = q_A + \frac{L}{2\pi}D^{(\rm 1LPT)}_{+, {\rm sc}}\epsilon_{A}\left( 1 + f\frac{\Delta a}{a_{\rm sc}}\right)\sin\left( \frac{2\pi}{L}q_{A} \right) ,\label{eq:zelbal}
\end{align}
with $A=x$, $y$, or $z$ and $D^{(\rm 1LPT)}_{+, {\rm sc}}=-1/\epsilon_{x}$ stands for the growth factor at shell-crossing time evaluated by using 1LPT. Note that in equation (\ref{eq:zelbal}), the growth rate $f$ and the expansion factor $a_{\rm sc}$ are evaluated at the same time as $D^{(\rm 1LPT)}_{+, {\rm sc}}$ (remind that $f=1$ in the Einstein-de Sitter cosmology considered in this work). The value of $D^{(\rm 1LPT)}_{+, {\rm sc}} $ is obtained in practice by solving the equation $\partial x^{(\rm 1LPT)}/\partial q_x=0$ at the origin, keeping in mind that $|\epsilon_x| \geq |\epsilon_{y,z}|$. Remind that equation (\ref{eq:zelbal}) is exact in the pure 1D case, that is when $\epsilon_y=\epsilon_z=0$. The condition for the caustics, $J^{(\rm 1LPT)}(\bm{q},a) = 0$, is reduced to the relation:
\begin{align}
\cos{\left( \frac{2\pi}{L}q_{i}\right)} = \left( 1 + f\frac{\Delta a}{a_{\rm sc}} \right)^{-1} . \label{eq: caustics 1LPT}
\end{align}
Eq.~(\ref{eq: caustics 1LPT}) implies that the equations of the caustics are given by $\bm{q} = \bm{q}_{0}$ with $\bm{q}_{0}$ being a constant vector depending on time, i.e. on $\Delta a$.
Therefore the 1LPT caustics consist of one-dimensional lines in 2D and two-dimensional planes in 3D, even at collapse time. This configuration is actually degenerate and is not expected in realistic cases, for instance in the framework of a smooth Gaussian random field \citep[see, e.g.,][]{1999A&A...343..663P}. Indeed, the regions where first shell-crossing occurs should be composed, in the non-degenerate case (that is non-vanishing $\epsilon_i$), of a set of points. In the vicinity of such points and shortly after shell-crossing, even in 1LPT, the equation of the caustics should correspond, at leading order in $\bm{q}$ and in the non-axial-symmetric case, $\epsilon_i \neq \epsilon_j$, $i \neq j$, to the equation of an ellipse in 2D and an ellipsoid in 3D \citep[see, e.g.,][]{2014MNRAS.437.3442H,2018JCAP...05..027F}. The reason for this not being the case here is due to the extremely restrictive class of initial conditions we have chosen.
%-------------------------------------------------
\begin{figure}[!tbp]
\centering
\includegraphics[width=0.45\textwidth]{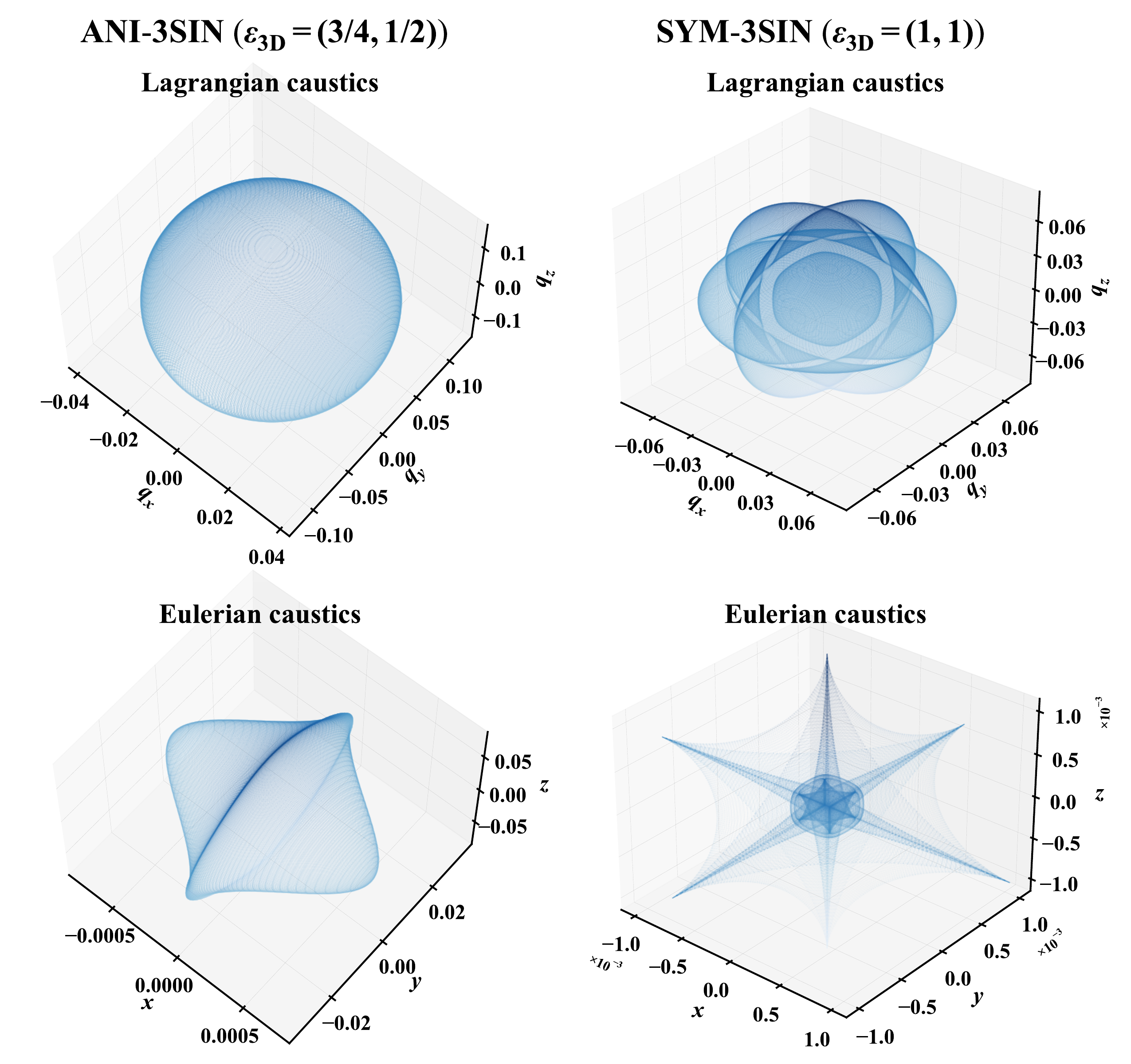}
\caption{Three-dimensional view of the expected caustic pattern shortly after shell-crossing for three sine waves initial conditions. The {\it lefts panels} correspond to the most typical pancake, here embodied by ANI-3SIN, but Q1D-3SIN would look alike, while the {\it right ones} show the axial-symmetric case, that is SYM-3SIN. The calculations are performed using 2LPT along with ballistic approximation (Eq.~\ref{eq:bal2LPT}), but higher-order LPT would provide the same topology from the qualitative point of view, except that the position of the caustic surfaces would change, especially in the axial-symmetric case.}
\label{fig: 3SIN caustics}
\end{figure}
%-------------------------------------------------
%-------------------------------------------------
\begin{figure*}[!tbp]
\centering
\includegraphics[width=0.75\textwidth]{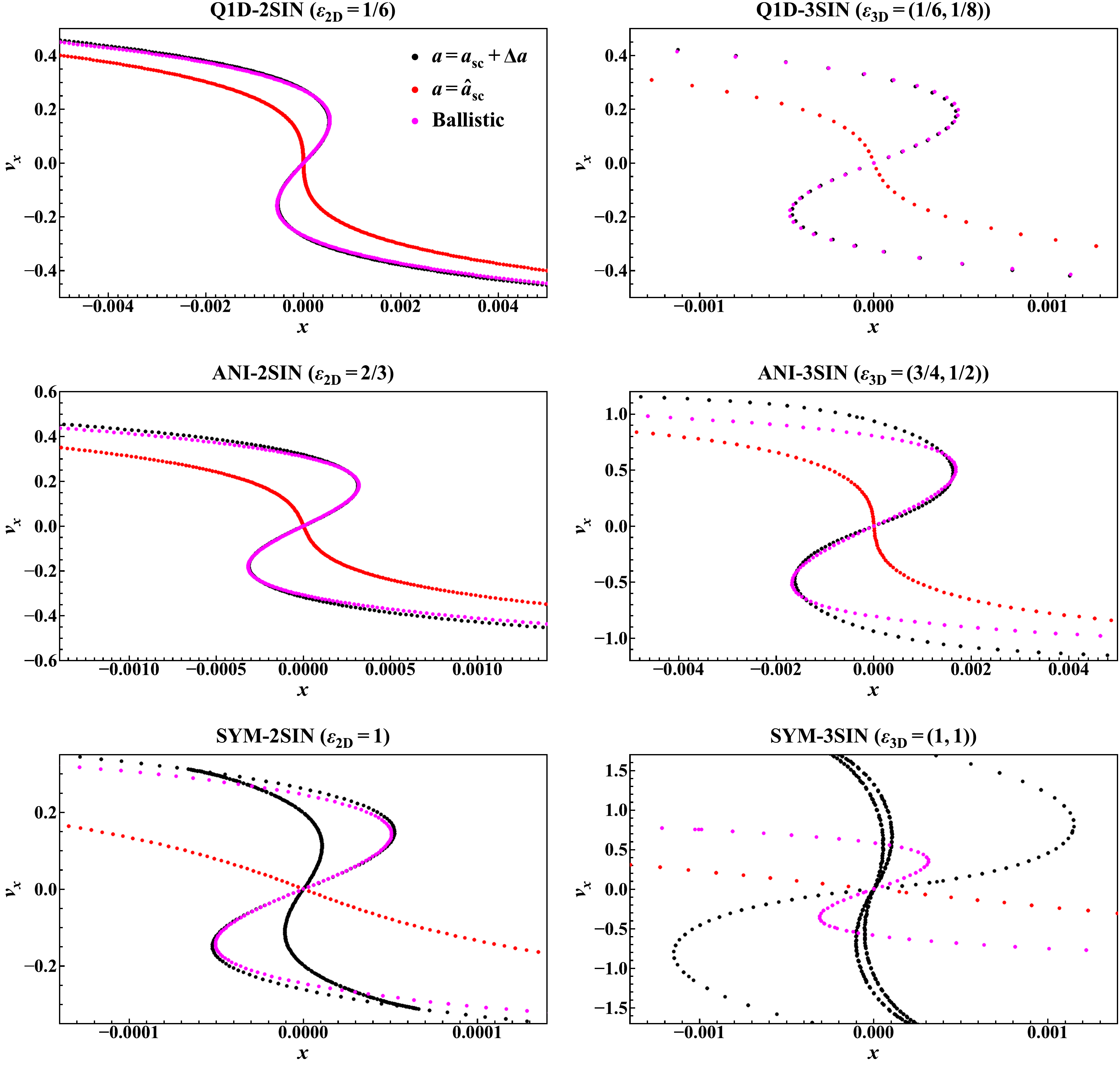}
\caption{
Tests of the ballistic approximation in Vlasov simulations: measured phase-space structure for two and three sine waves initial conditions. This figure is analogous to Fig.~\ref{fig: phase precollapse}, except that it shows a zoom on the central part of the system shortly after collapse. We test the validity of the ballistic approximation, by using two simulation snapshots slightly before (red curves, with $a={\hat a}_{\rm sc}$ as shown in Table~\ref{tab: initial conditions} except for $\bm{\epsilon}_{3{\rm D}}=(1,1)$, for which $a=0.03103$) and slightly after collapse time (black, with $a=a_{\rm sc}+\Delta a$). The ballistic approximation, as described in the main text, is applied to the red curves to obtain the magenta ones, to be compared directly to the ``exact'' solution, in black, given by the simulation.
}
\label{fig:sim_balval}
\end{figure*}
%-------------------------------------------------
%-------------------------------------------------
\begin{figure*}[!htbp]
\centering
\includegraphics[width=0.7\textwidth]{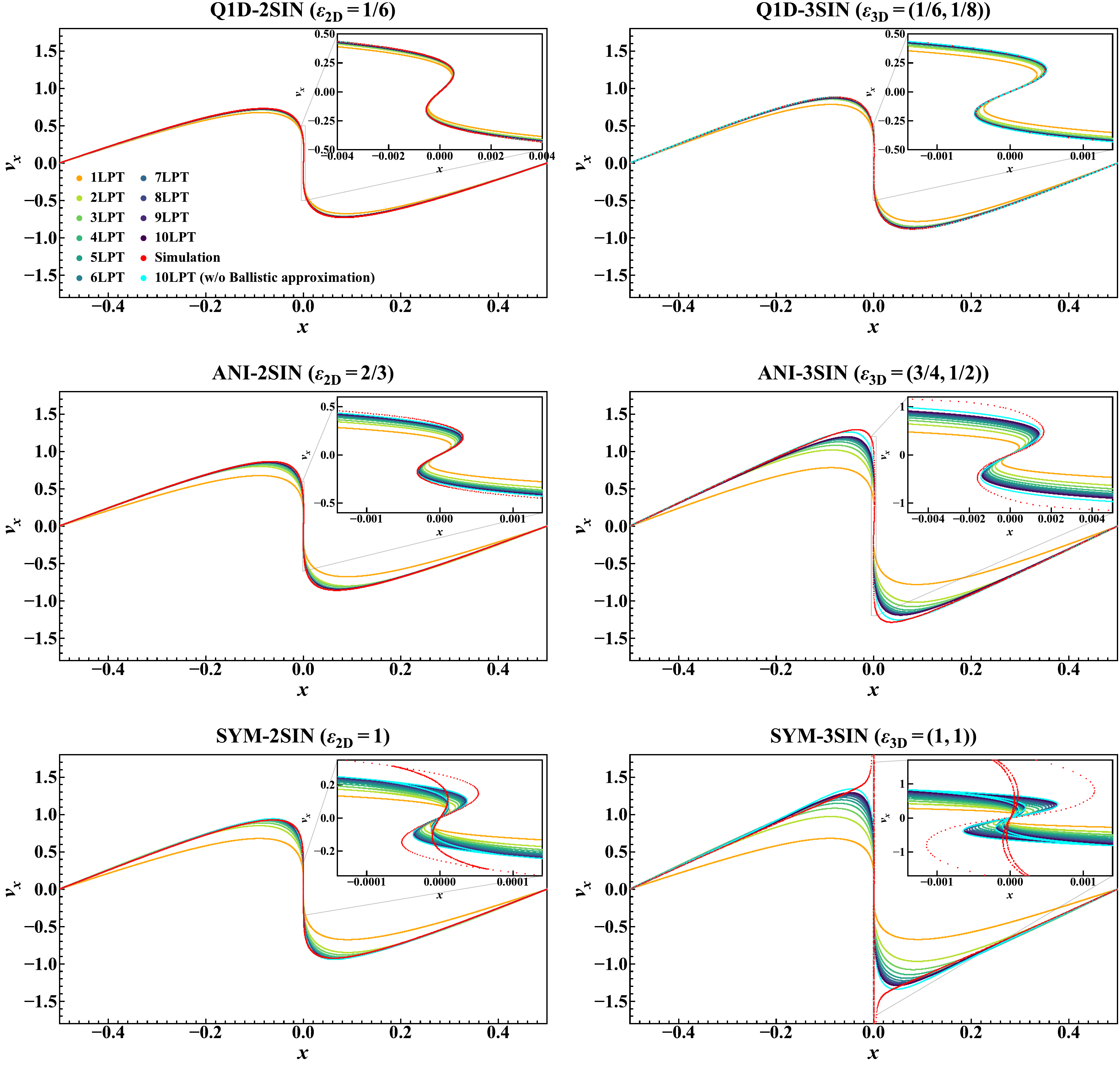}
\caption{Phase-space structure for two and three sine waves initial conditions shortly after collapse. This figure is analogous to Fig.~\ref{fig: phase precollapse}, except that we compare predictions of LPT at various orders in the ballistic approximation framework to measurements in Vlasov simulations shortly after collapse, that is for $a=a_{\rm sc}+\Delta a$ as listed in Table~\ref{tab: initial conditions} and discussed in the main text. Note that the additional cyan curve corresponds to the LPT prediction at 10th-order without using the ballistic approximation.
}
\label{fig: phase postcollapse}
\end{figure*}
%-------------------------------------------------
%-------------------------------------------------
\begin{figure*}[!htbp]
\centering
\includegraphics[width=0.75\textwidth]{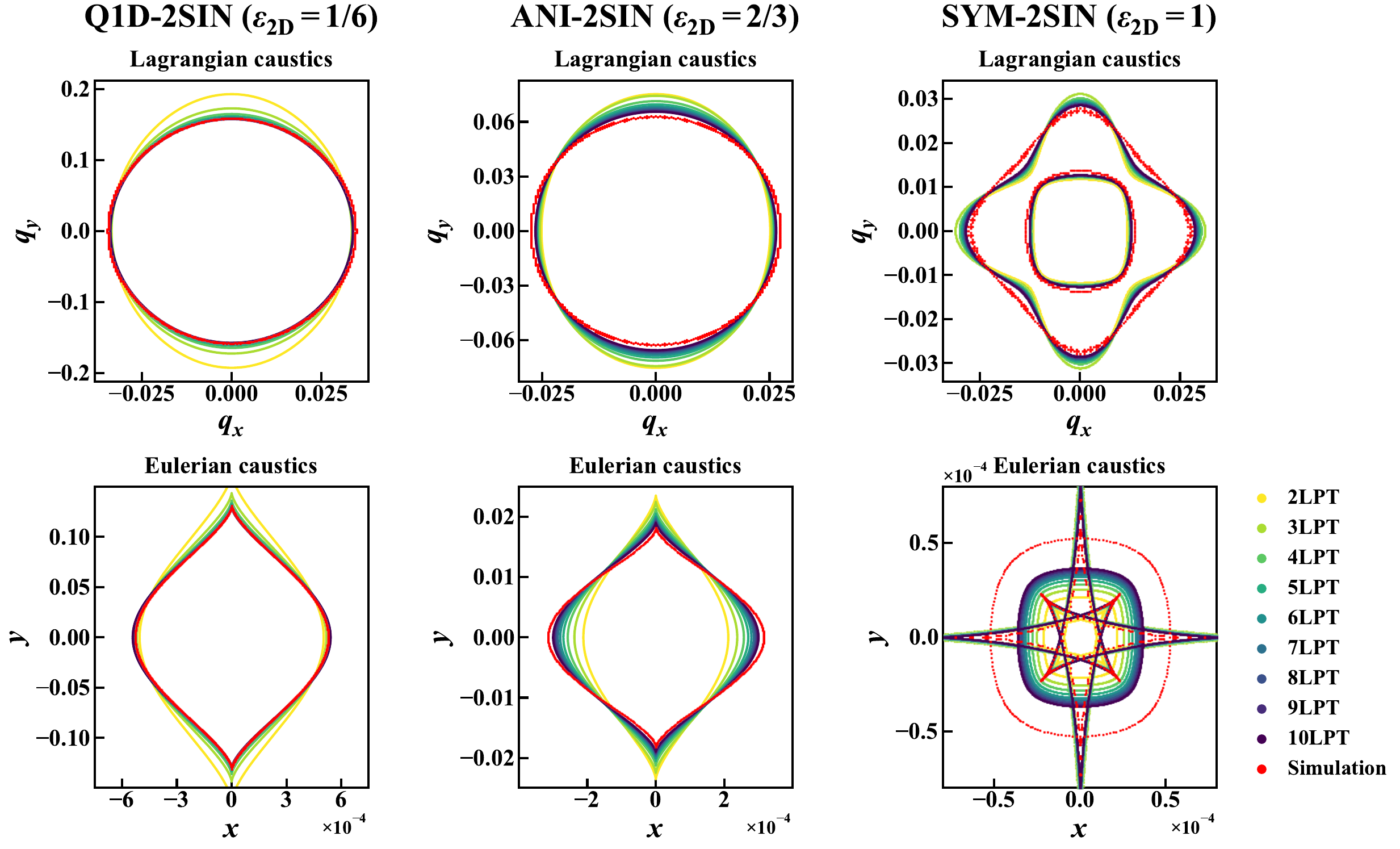}
\caption{The caustic pattern shortly after shell crossing in the two-dimensional case: comparison of LPT predictions using ballistic approximation to Vlasov runs. {\it Left, middle} and {\it right panels} correspond respectively to Q1D-2SIN, ANI-2SIN and SYM-2SIN configurations, while {\it top} and {\it bottom panels} show the caustics in Lagrangian and Eulerian spaces, respectively.}
\label{fig: 2SIN caustics}
\end{figure*}
%-------------------------------------------------
%-------------------------------------------------
\begin{figure*}[!htbp]
\centering
\includegraphics[width=0.75\textwidth]{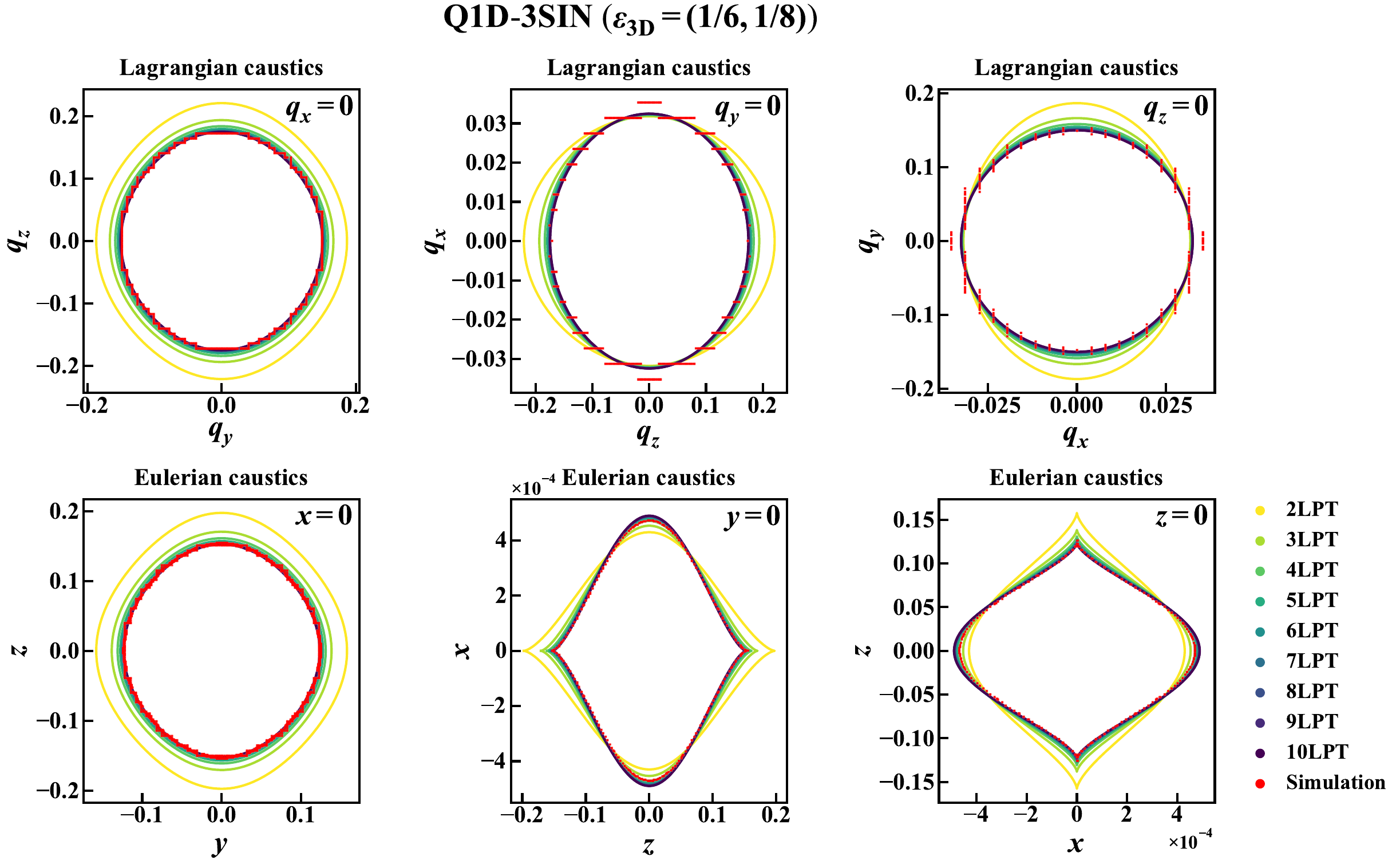}
\includegraphics[width=0.75\textwidth]{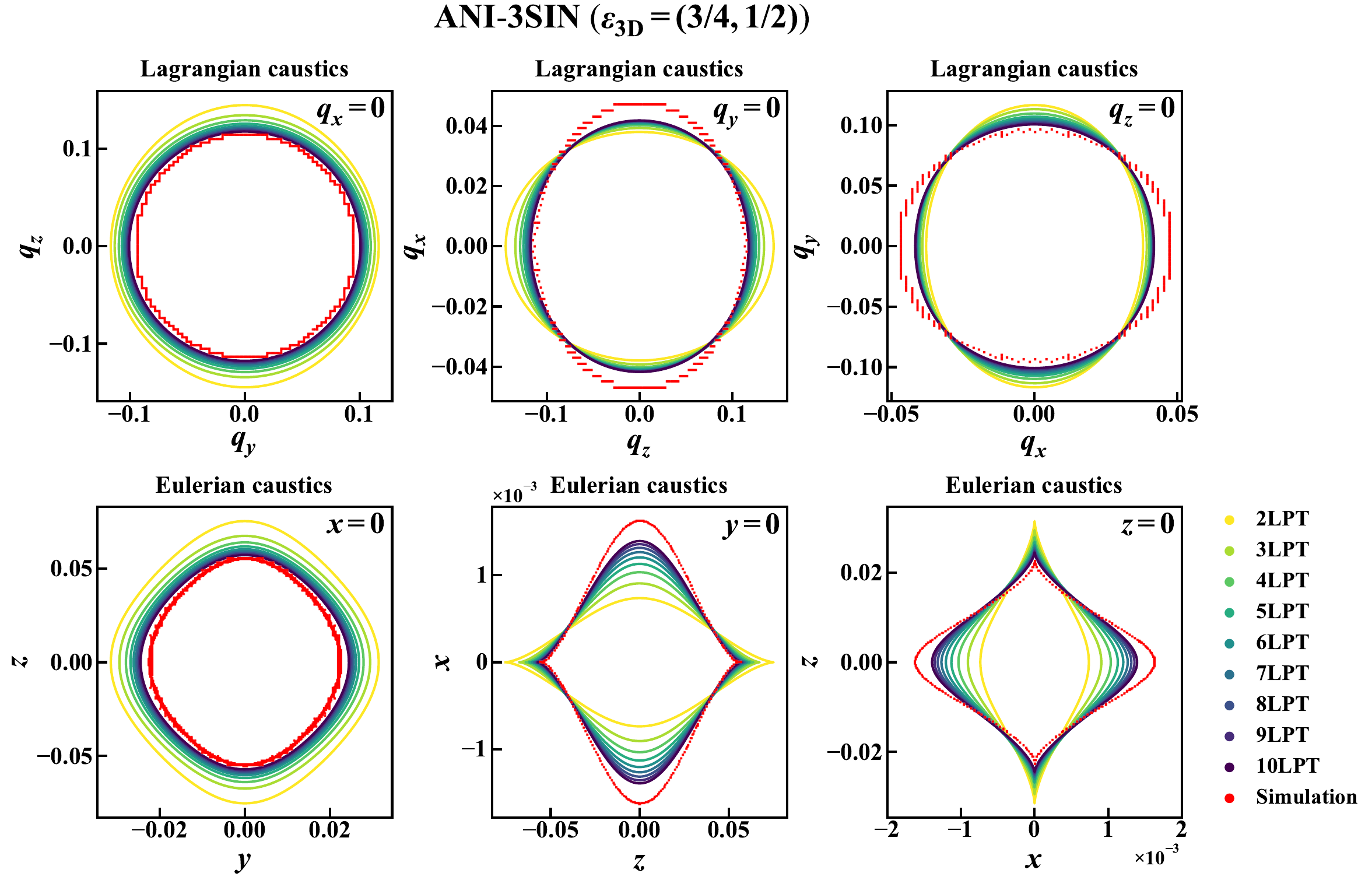}
\includegraphics[width=0.6\textwidth]{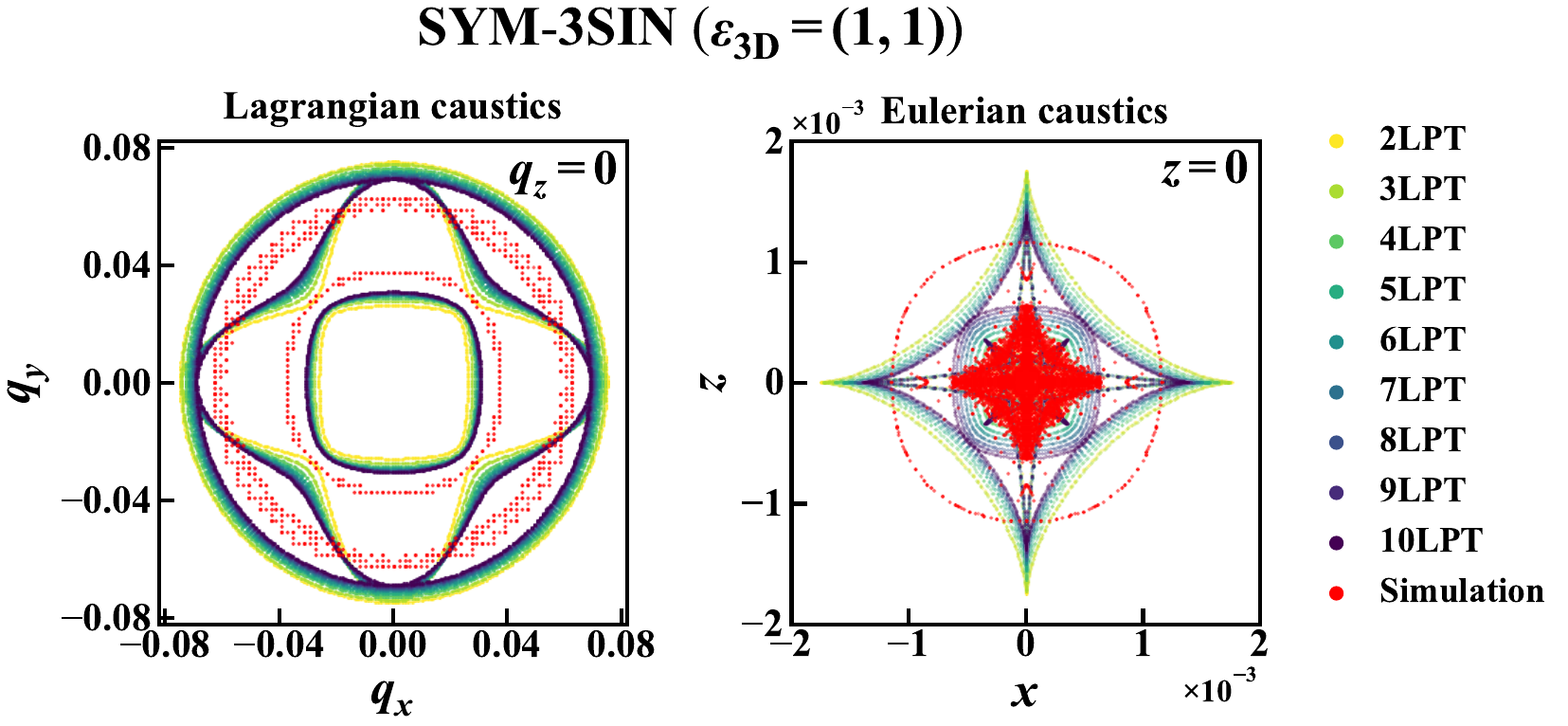}
\caption{The structure of caustics shortly after shell crossing for three sine waves initial conditions: comparison of LPT predictions using ballistic approximation to Vlasov runs. {\it Top six, middle six}, and {\it bottom two panels} respectively correspond to Q1D-3SIN, ANI-3SIN and SYM-3SIN. On each group of 6 panels, the {\it top} and {\it bottom lines} correspond to Lagrangian and Eulerian spaces, and the intersection of the caustic surfaces with the plane $q_x=0$, $q_y=0$ and $q_z=0$ respectively for the {\it first, second} and {\it third} column of each group. In the {\it bottom group of two panels,} the {\it left} and {\it right panels} correspond respectively to Lagrangian and Eulerian space and only show the intersection of the caustic surface network with the plane $q_z=0$ due to the symmetry of the system.}
\label{fig:3SINcaustics}
\end{figure*}
%-------------------------------------------------
\begin{figure}[!htbp]
\centering
\includegraphics[width=0.45\textwidth]{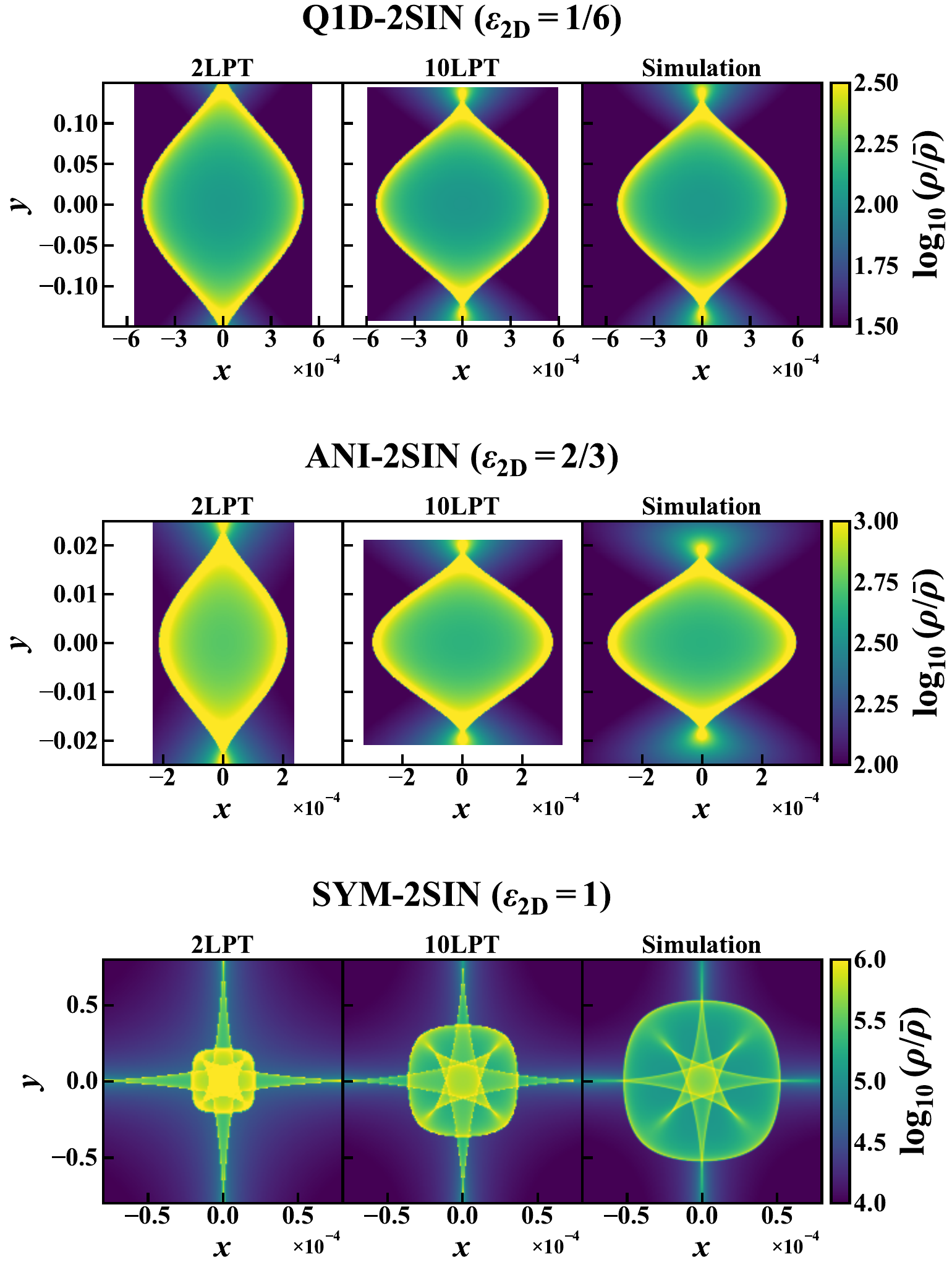}
\caption{Two-dimensional density shortly after shell-crossing: comparison of LPT predictions using ballistic approximation to Vlasov runs. From {\it top} to {\it bottom}, Q1D-2SIN, ANI-2SIN and SYM 2SIN. From {\it left} to {\it right}, 2LPT, 10LPT and measurements in Vlasov simulations.}
\label{fig: 2SIN density}
\end{figure}
%-------------------------------------------------
%-------------------------------------------------
\begin{figure*}[!htbp]
\centering
\includegraphics[width=0.45\textwidth]{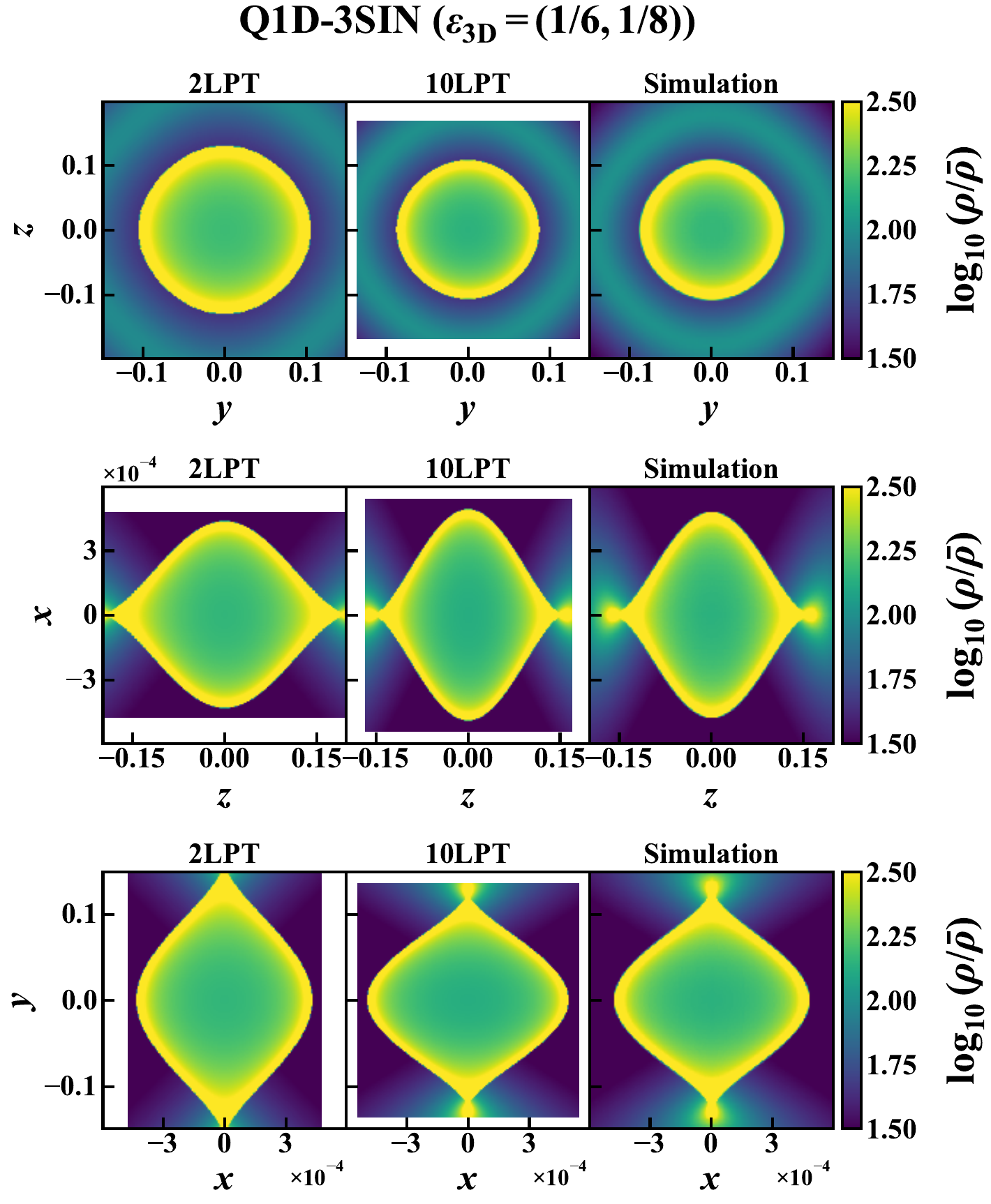}\includegraphics[width=0.45\textwidth]{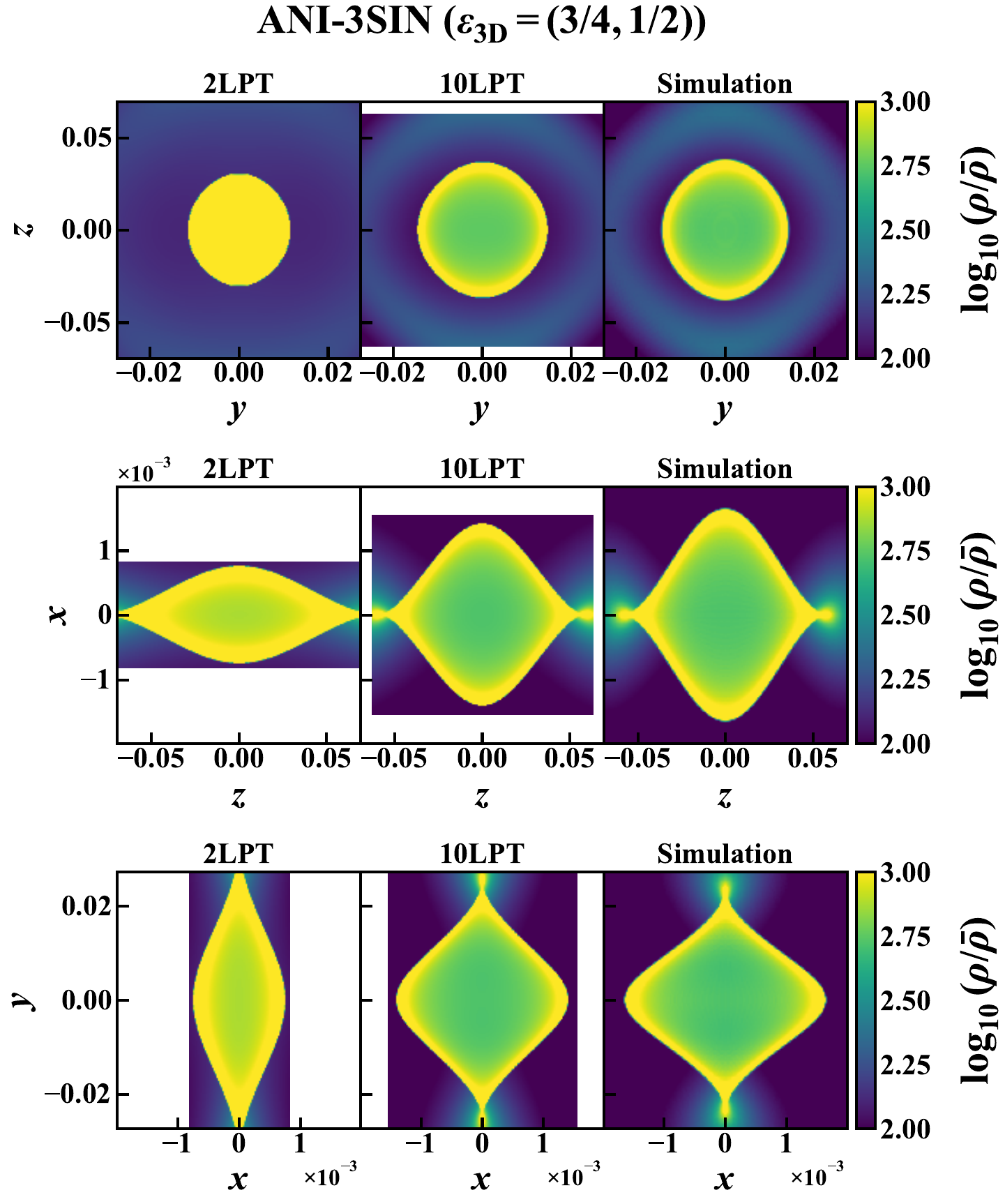}
\includegraphics[width=0.45\textwidth]{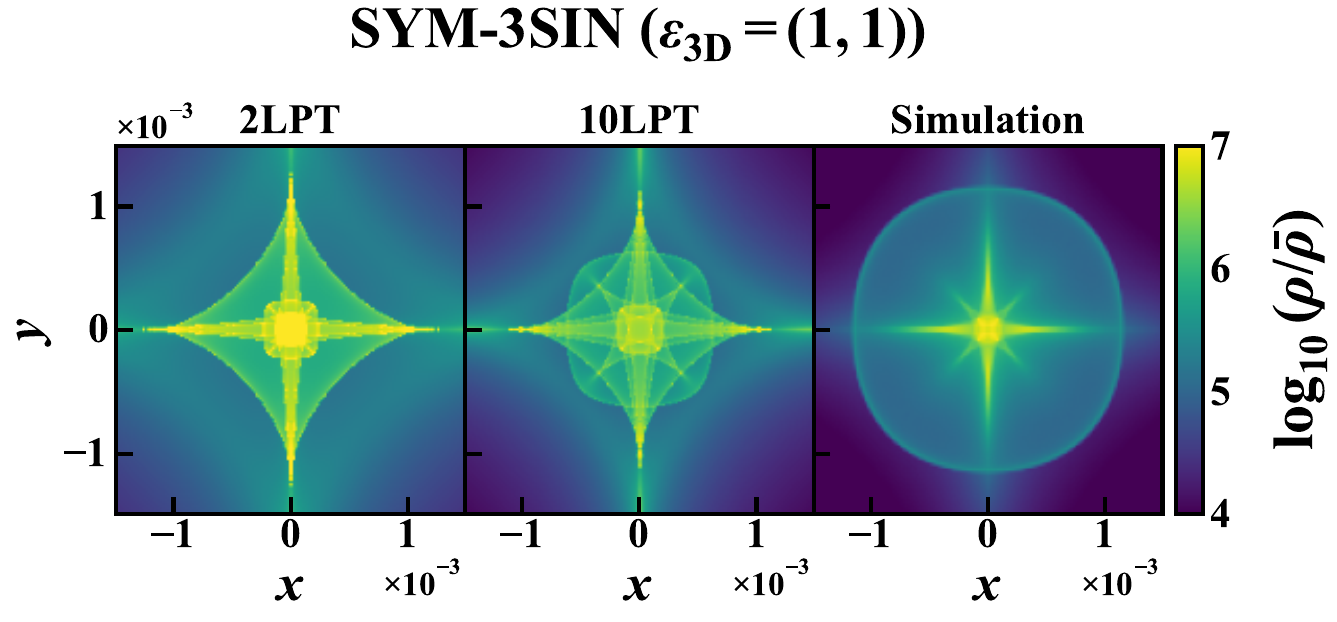}
\caption[]{
Slices of the projected density shortly after shell-crossing: comparison of LPT using ballistic approximation to Vlasov runs. The {\it left} and {\it right groups of nine panels} correspond respectively to Q1D-3SIN ({\it top, middle} and {\it bottom line of panels:} $x=-1.55\times 10^{-4}$, $y=-1.17\times 10^{-3}$ and $z=-1.56\times 10^{-3}$ slice) and ANI-3SIN ($x=-5.16\times10^{-4}$, $y=-2.15\times 10^{-4}$ and $z=-5.47\times10^{-4}$ slice), while the {\it bottom group of three panels} corresponds to SYM-3SIN ($z=-1.17\times 10^{-5}$ slice). On each group of panels, {\it left, middle} and {\it right columns} give respectively the 2nd-order LPT prediction, the 10th-order LPT prediction and the Vlasov code measurements. Due to the symmetry of the system for SYM-3SIN, only one slice is shown for the bottom panels.}
\label{fig:density}
\end{figure*}
%-------------------------------------------------
%-------------------------------------------------
\begin{figure}[!htbp]
\centering
\includegraphics[width=0.45\textwidth]{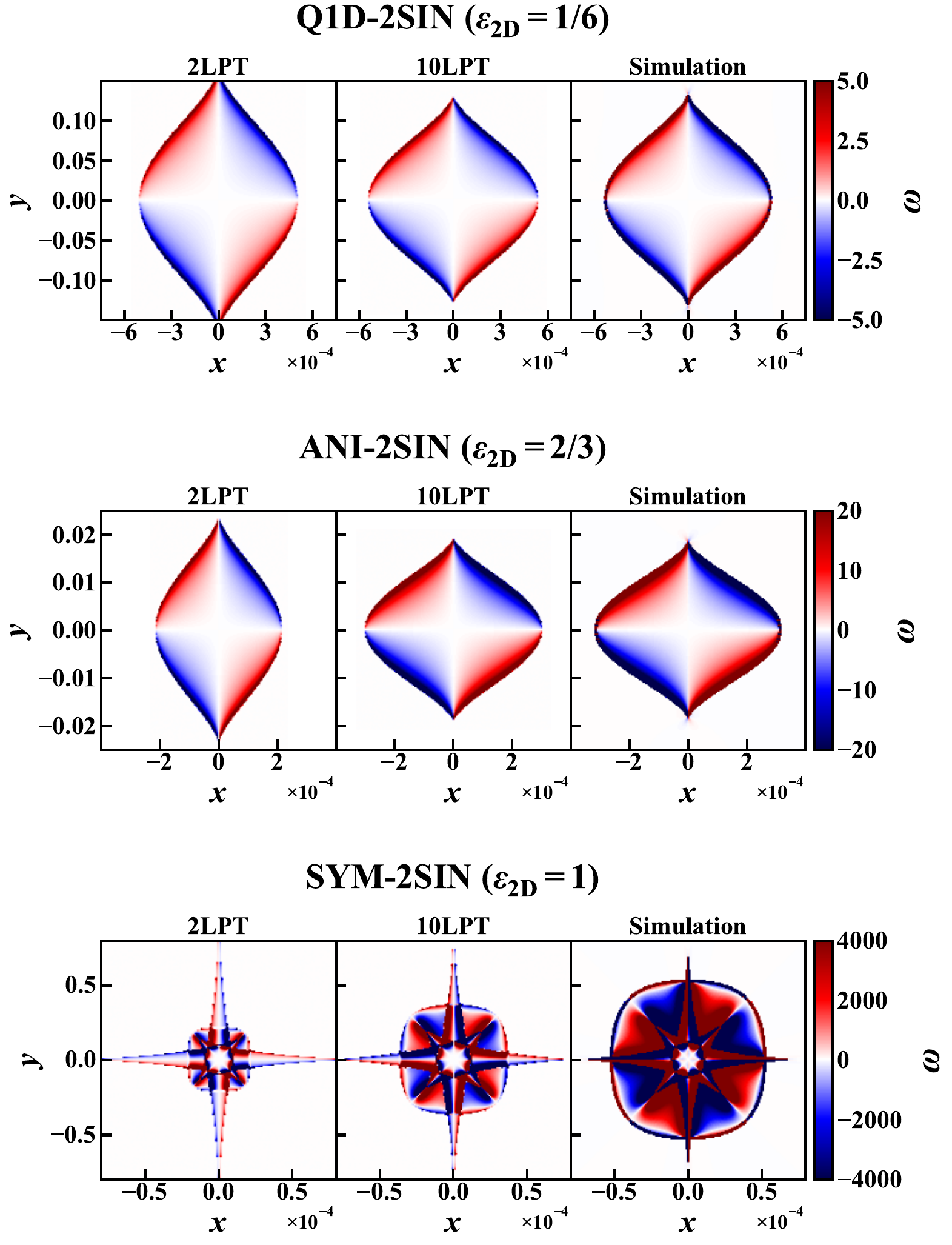}
\caption{Two-dimensional vorticity fields: comparison of LPT predictions to Vlasov runs.}
\label{fig: 2SIN vorticity}
\end{figure}
%-------------------------------------------------
\begin{figure*}[!htbp]
\centering
\includegraphics[width=\textwidth]{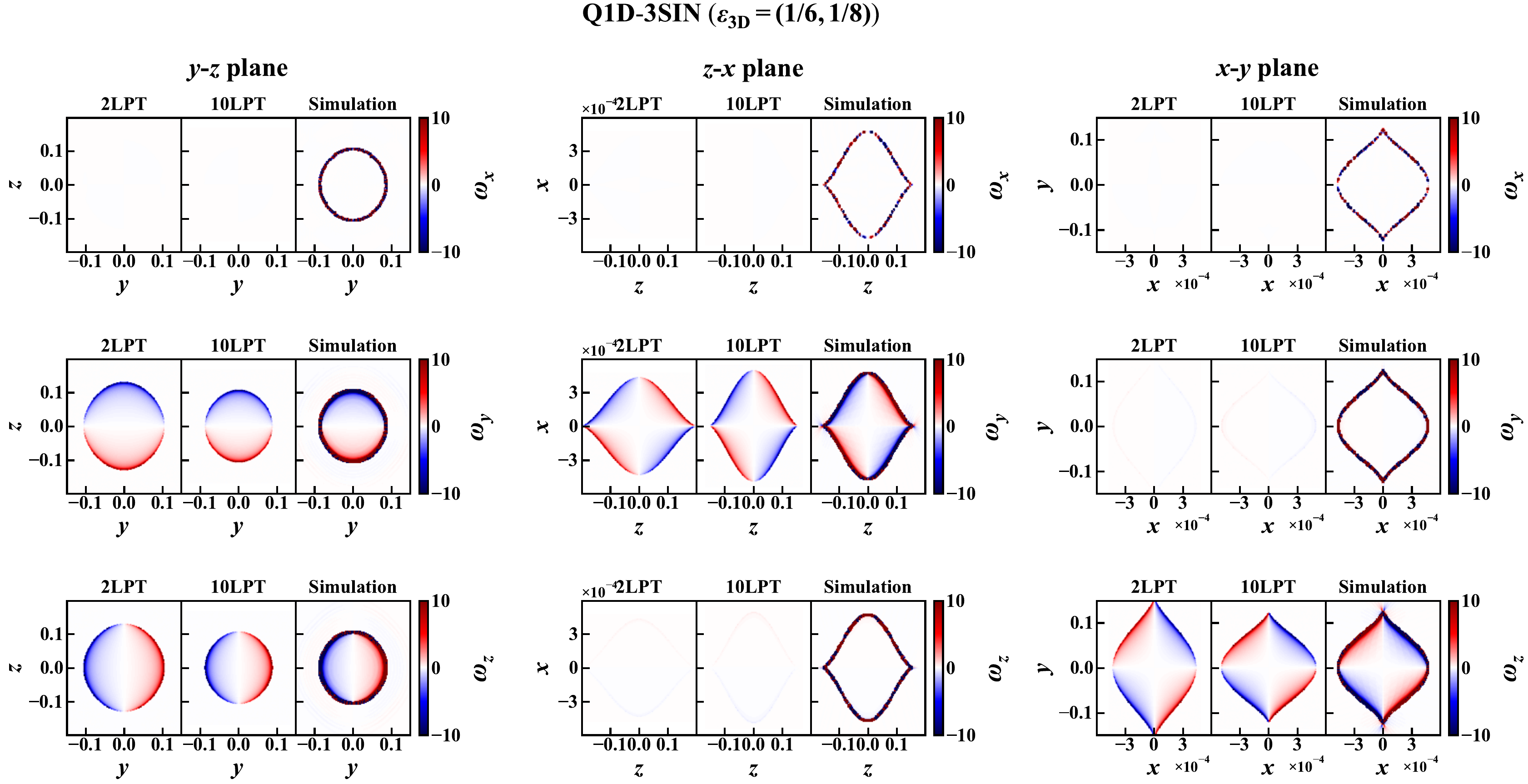}
\caption{Vorticity components shortly after shell-crossing: comparison of LPT with the ballistic approximation to Vlasov runs for Q1D-3SIN. The two-dimensional slices are the same as those shown in the top left group of nine panels of Fig.~\ref{fig:density}, except that the slice considered changes from {\it left} to {\it right}, while the vorticity component changes from {\it top} to {\it bottom}. Again, on each line of three panels, 2LPT ({\it left panel}) and 10LPT ({\it middle panel}) are compared to simulation measurements ({\it right panel}).}
\label{fig: 3SIN vorticity Q1D}
\end{figure*}
%-------------------------------------------------
\begin{figure*}[!htbp]
\centering
\includegraphics[width=\textwidth]{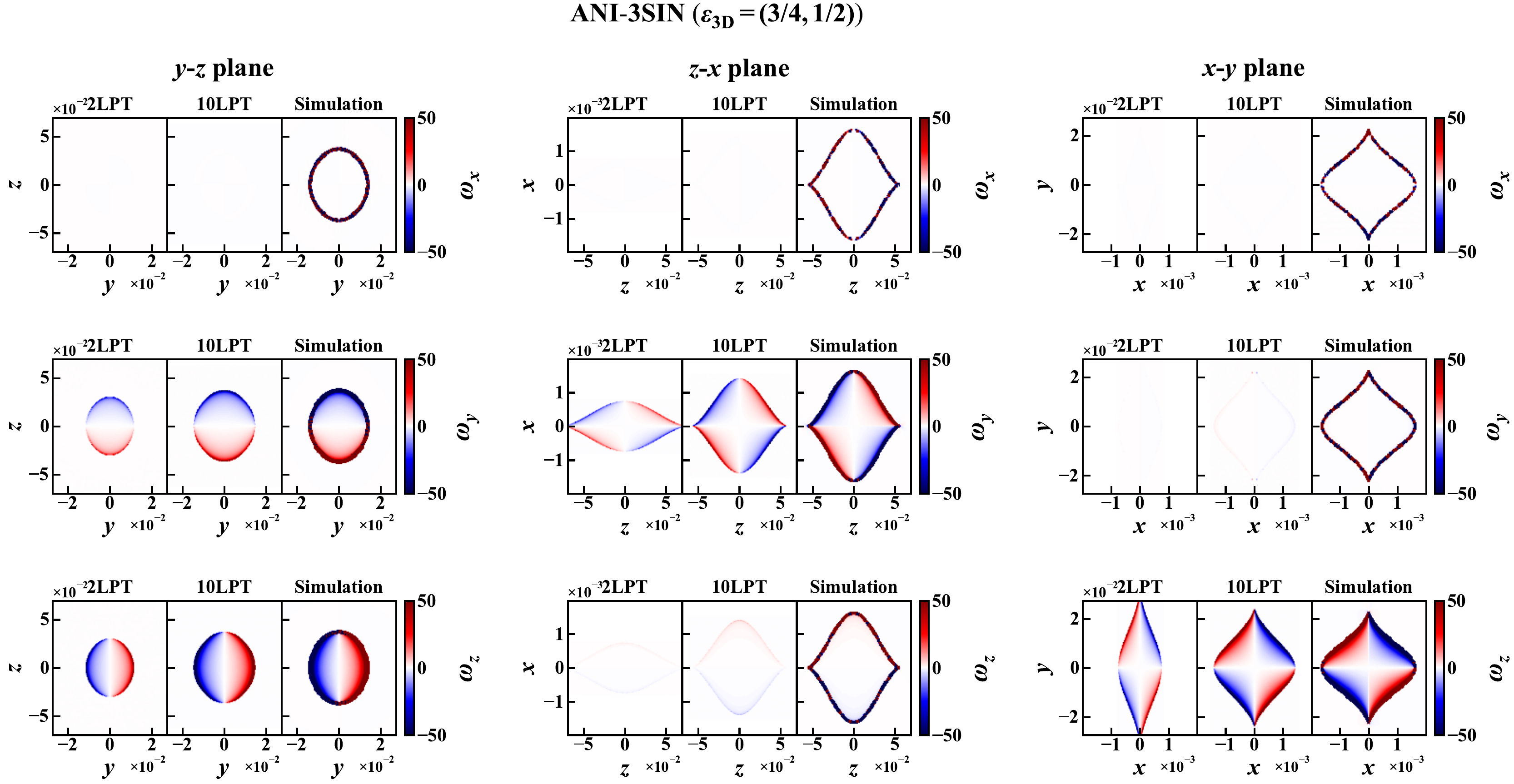}
\caption{Same as Fig.~\ref{fig: 3SIN vorticity Q1D}, but for ANI-3SIN. The two-dimensional slices considered are the same as in the top right group of nine panels of Fig.~\ref{fig:density}.}
\label{fig: 3SIN vorticity ANI}
\end{figure*}
%-------------------------------------------------
\begin{figure}[!htbp]
\centering
\includegraphics[width=0.45\textwidth]{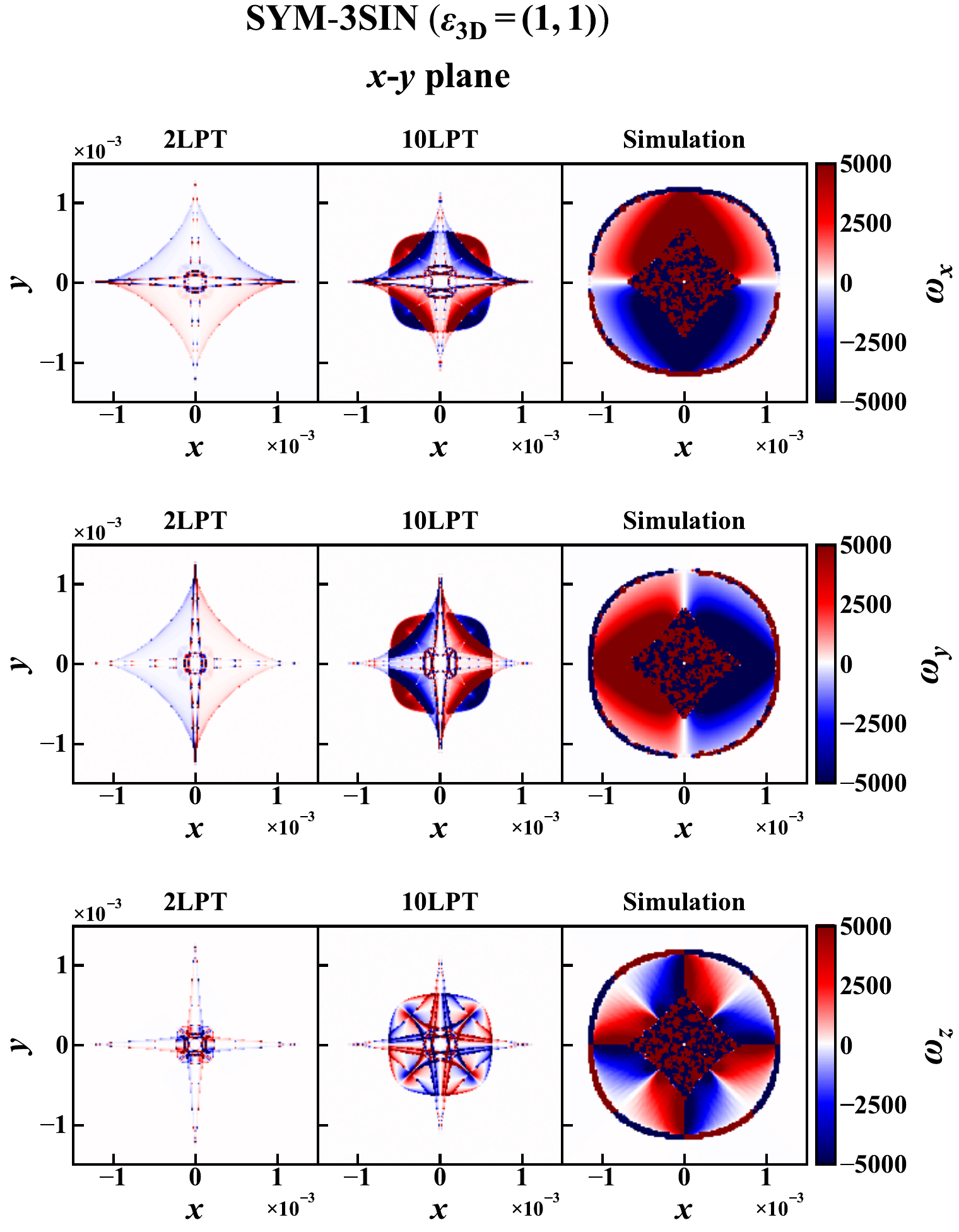}
\caption{Same as Fig.~\ref{fig: 3SIN vorticity Q1D}, but for SYM-3SIN.
Due to the symmetry of the initial conditions, only a $x$-$y$ slice is shown, which is the same as in bottom three panels of Fig.~\ref{fig:density}.}
\label{fig: 3SIN vorticity SYM}
\end{figure}
%-------------------------------------------------

Due to the degenerate nature of the sine waves initial conditions, it is therefore needed to go beyond the first order of the perturbative development of the dynamical equations to obtain a more realistic shape for the caustics. Indeed, for instance, 2LPT brings non-linear couplings between various axes of the dynamics:
\begin{align}
A^{(\rm 2LPT)}(\bm{q},a) &= q_A + \frac{L}{2\pi}D^{(\rm 2LPT)}_{+, {\rm sc}}\,\epsilon_{A}\,\left( 1 + f\frac{\Delta a}{a_{\rm sc}}\right) \sin\left( \frac{2\pi}{L}q_{A} \right)\notag \\
& \quad - \frac{3}{14} \frac{L}{2\pi}\left(D^{(\rm 2LPT)}_{+, {\rm sc}}\right)^2
\sum_{B}(1-\delta_{AB}) \,\epsilon_A\, \epsilon_B \notag \\
& \qquad \times \left( 1 + 2f\frac{\Delta a}{a_{\rm sc}}\right)\, \sin\left( \frac{2\pi}{L}q_{A} \right) \, \cos\left( \frac{2\pi}{L} q_B \right) ,\label{eq:bal2LPT}
\end{align}
where $D^{(\rm 2LPT)}_{+, {\rm sc}}$ stands for the growth factor at the shell-crossing time evaluated by using 2LPT, which can be obtained as a function of $\epsilon_{\rm 2D}$ or $\bm{\epsilon}_{\rm 3D}$ by solving the following second order polynomial:
\begin{align}
 1 + D^{(\rm 2LPT)}_{+, {\rm sc}} \epsilon_{x} - \frac{3}{14} \left( D^{(\rm 2LPT)}_{+, {\rm sc}} \right)^{2} \epsilon_{x} (\epsilon_{y} + \epsilon_{z}) = 0 . \label{eq: caustics 2LPT}
\end{align}
In Eq.~(\ref{eq:bal2LPT}) , the growth rate $f$ and the scale factor $a_{\rm sc}$ are evaluated at the same time as $D^{(\rm 2LPT)}_{+, {\rm sc}}$.
The additional terms compared to Eq.~(\ref{eq:zelbal}) imply, shortly after collapse, non-vanishing contributions from all axes at the quadratic level (and above) in the Lagrangian coordinate $\bm{q}$ in the equation $J^{(\rm 2LPT)}(\bm{q},a) = 0$ (as long as none of the $\epsilon_i$ cancels).
This leads to, in the non-axial-symmetric case (i.e., Q1D-2SIN, Q1D-3SIN, ANI-2SIN and ANI-3SIN), elliptic and ellipsoid shapes for the caustic curve/surface in Lagrangian space, respectively in 2D and 3D, as expected and as illustrated in the three-dimensional case by the top left panel of Fig~\ref{fig: 3SIN caustics}, while the bottom left one shows the expected typical pancake shape in Eulerian space.
Accordingly, in the analyses performed below, except for the phase-space diagrams that we examine first, LPT is examined from 2nd order and above. Note that, as will be studied below, the axial-symmetric cases $\epsilon_{\rm 2D}=1$ and $\bm{\epsilon}_{\rm 3D}=(1,1)$ are a bit more complex, as they require Taylor expansion of the motion beyond quadratic order in $\bm{q}$ to obtain a full description of the correct and more intricate topology of the caustic surfaces (e.g. 4th order in $\bm{q}$ instead of second for SYM-2SIN), as illustrated in 3D by right panels of Fig.~\ref{fig: 3SIN caustics}, due to simultaneous collapses along several axes. It would be beyond the scope of this article to go further in analytic details of catastrophe theory for these very particular configurations, so we shall be content with a descriptive analysis of the LPT results in this latter case.

%%%%%%%%%%%%%%%%%%%%%%%%%%%%%%%%%%%%%%%%%%%%%%%%%%
%%%%%%%%%%%%%%%%%%%%%%%%%%%%%%%%%%%%%%%%%%%%%%%%%%
\subsection{Phase-space structure}
\label{sec: postcollapse phase space}
%%%%%%%%%%%%%%%%%%%%%%%%%%%%%%%%%%%%%%%%%%%%%%%%%%
%%%%%%%%%%%%%%%%%%%%%%%%%%%%%%%%%%%%%%%%%%%%%%%%%%

Figs.~\ref{fig:sim_balval} and \ref{fig: phase postcollapse} display phase-space diagrams for our various sine-wave systems shortly after collapse, namely the intersection of the phase-space sheet with the hyperplanes $y=0$ and $y=z=0$, for 2D and 3D configurations, respectively. The first figure examines the validity of the ballistic approximation directly from simulations data, while the second one compares predictions from LPT at various orders to the simulations.

Interestingly, Fig.~\ref{fig:sim_balval} shows that the ballistic approximation can deviate significantly from the exact solution even quite shortly after collapse. While it remains precise for all configurations in 2D, with a slight worsening of the accuracy when going from the quasi-1D to the axial-symmetric set-up, as expected, significant or even dramatic deviations from the exact solution can be seen in 3D for the anisotropic configuration (ANI-3SIN) and the axial-symmetric configuration (SYM-3SIN). In the last case (bottom right panel), the intersection of the phase-space sheet with the hyperplane $y=z=0$ is composed, in the multi-stream region, of three curves with a ``S'' shape instead of a single one, as a result of simultaneous shell-crossing along all coordinates axes. The prediction from the ballistic approximation has then to be compared to the ``S'' with the largest amplitude along $x$-axis. Similar arguments apply to the 2D case (bottom left panel), but in this case, the intersection with the $y=0$ hyperplane is composed of two curves instead of three.\footnote{Note that when examining the bottom panels of Figs.~\ref{fig:sim_balval} (black curves) and \ref{fig: phase postcollapse}, solving the equation $x(q)=x_0$ in the multi-stream region does not seem to have the expected number of solutions, that is either 3, 9 or 27 (only in the bottom right panel for the latter case). But this is because we are sitting exactly at the centre of the system and symmetries imply superposition of solutions in the subspace $y=0$ in 2D and $y=z=0$ in 3D.
In particular, as discussed in Sec.~\ref{sec: ballistic approx}, first-order LPT has each coordinate axis totally decoupled from the others, which means that only one ``S'' shape, which corresponds respectively, in the $(x,v_x)$ subspace, to 3 and 9 superposed ``S'' shapes, is visible as the orange curve on bottom panels of Fig.~\ref{fig: phase postcollapse}.
On the other hand, higher-order LPT induces nonlinear couplings between various axes of the dynamics, which implies that we have, respectively, only 2 and 3 visible ``S'' shapes on the bottom left and right panels of Figs.~\ref{fig:sim_balval} (black curves) and \ref{fig: phase postcollapse}, both for the simulations and LPT predictions at order $n \geq 2$ among those the remaining $3-2=1$ and $9-3=6$ curves overlap with their visible ``S'' counterpart in $(x,v_x)$ subspace.
}

The results shown in Fig.~\ref{fig:sim_balval} must however be interpreted with caution. Indeed, strictly speaking, we aim to apply the ballistic approximation from collapse time, which is not exactly the case in this figure. The red curves, which correspond to the snapshots used to originate the ballistic motion, all correspond to a time $a_{\rm bc}$ slightly before actual collapse (namely $a_{\rm bc}=0.03103$ for SYM-3SIN and $a_{\rm bc}={\hat a}_{\rm sc}$ as given in Table~\ref{tab: initial conditions} for other cases), because we do not have exactly access to this latter. The level of proximity to collapse is, at least from the visual point of view, the lowest for the axial-symmetric cases, particularly SYM-3SIN. It is also for this latter case that the ratio $\Delta a_{\rm eff}/a_{\rm bc} \simeq 0.03$ is the largest, where $\Delta a_{\rm eff}$ is the difference between $a_{\rm pc}\equiv a_{\rm sc}+\Delta a$ and $a_{\rm bc}$ and is used to test the ballistic approximation. Note also that $\Delta a_{\rm eff}/a_{\rm bc} \simeq 0.03$ is also of the same order for ANI-3SIN. The greater the value of $\Delta a_{\rm eff}/a_{\rm bc}$, the greater the deviation due to nonlinear corrections of the motions to be expected. Furthermore, we also expect the ballistic approximation to worsen from quasi-1D cases to fully axial-symmetric. 

When examining Fig.~\ref{fig:sim_balval} more in detail, we also note that for ANI-3SIN, the central part of the ``S'' shape is well reproduced by the ballistic approximation, while the tails underestimate velocities. This last defect is in fact present to some extent in all configurations except for Q1D cases, given the uncertainties on the measurements. In the 3D axial-symmetric case, not only the velocities in the tails are strongly underestimated, but also the magnitude of the position of the caustics along $x$-axis (local extremum of $x$ coordinate). This suggests that the effects of acceleration in the vicinity of collapse are strong, which implies that the ballistic approximation can be applied only for a very short time (or it could be that it is not applicable, mathematically, due to the strength of the singularity).

Now we turn to the comparison of LPT predictions of various orders to the simulations, as shown in Fig.~\ref{fig: phase postcollapse}. Remind that the ballistic approximation is applied from the respective collapse times of each perturbation order, which obviously makes predictions of LPT artificially more realistic than it should be.

The conclusions of Sec.~\ref{sec: phase space shell-crossing}, where we compared LPT to simulations at collapse, still stand at the coarse level, obviously, since the time considered in Fig.~\ref{fig: phase postcollapse} is nearly equal to collapse time. When zooming on the central part of the system, we notice that the ballistic approximation applied to LPT works increasingly well with the order, as expected, especially when approaching quasi-1D dynamics. It can fail in the tails of the ``S'' shape of the phase-space sheet, as a combination of limits of LPT to describe the system in the vicinity of the velocity extrema and the limits of the ballistic approximation itself just discussed above.

It is important to note at this point that the ballistic approximation is not necessarily the best choice outside the multi-stream region, where LPT predictions can still perform very well. This is illustrated on Fig.~\ref{fig: phase postcollapse} by the cyan curves which show 10th-order LPT results without assuming ballistic approximation, to be compared to the dark purple curves which correspond to 10th-order LPT with ballistic approximation. While the differences are very small given the time considered and the accuracy of the measurements, it can be seen, when just examining middle right panel of Fig.~\ref{fig: phase postcollapse}, that pure 10th-order LPT seems to perform better than its ballistic counterpart in the tails of the ``S'' and nearby the local extrema of the velocity.

On the other hand, in bottom panels of Fig.~\ref{fig: phase postcollapse}, which correspond  to axial-symmetric configurations, it is not clear at all that the cyan curves bring any improvement over the dark purple ones outside the multi-stream region. Here, LPT performs significantly more poorly than for other configurations, with or without assuming ballistic approximation. In this case, the topology of the phase-space structure in the multi-stream region (two or three curves according to the number of dimensions) is correct and the very central part of the shell corresponding to $x$-axis motion remains synchronised with the simulation, but except for this, the structure of the system is reproduced only qualitatively. This will be confirmed by the analyses that follow next.

%%%%%%%%%%%%%%%%%%%%%%%%%%%%%%%%%%%%%%%%%%%%%%%%%%
%%%%%%%%%%%%%%%%%%%%%%%%%%%%%%%%%%%%%%%%%%%%%%%%%%
\subsection{Configuration space: caustics, density, vorticity}
\label{sec: postcollapse caustics}
%%%%%%%%%%%%%%%%%%%%%%%%%%%%%%%%%%%%%%%%%%%%%%%%%%
%%%%%%%%%%%%%%%%%%%%%%%%%%%%%%%%%%%%%%%%%%%%%%%%%%
We now enter into the heart of this work, which consists of examining in detail the structure of the system shortly after collapse in configuration space, both in Lagrangian and Eulerian coordinates. The multi-stream region is delimited by caustics, where density and vorticity are singular, so an accurate description of the caustic pattern represents the first test of LPT predictions. In most cases, the singularity corresponds to a simple fold of the phase-space sheet, with $\rho(r) \sim r^{-1/2}$ \cite[see, e.g.,][]{2018JCAP...05..027F} and likewise for the vorticity along the direction orthogonal to the caustic \citep[see, e.g., ][]{1999A&A...343..663P}. This means that a large part of the vorticity signal is generated in the vicinity of caustics. Having the correct shape of caustics is therefore fundamental because this information mostly determines the subsequent internal structure of the multi-stream region both for the density and the vorticity fields.
Hence, in what follows, we first comment on Figs.~\ref{fig: 2SIN caustics} and \ref{fig:3SINcaustics}, which compare, respectively in the 2D and 3D cases, the caustic pattern at various orders of LPT to the Vlasov code for our various sine waves initial conditions.
Then, we turn to the normalised density and the vorticity in Figs.~\ref{fig: 2SIN density} to \ref{fig: 3SIN vorticity SYM}, and compare LPT predictions at 2nd and 10th order to the simulations. We remind the reader again that LPT predictions with the ballistic approximation are computed by synchronising the respective collapse times obtained at each perturbation order, which obviously artificially improves the performances of LPT. 
Note cautiously that in plotting the results below in 3D cases, the caustics outputs (Figs.~\ref{fig: 2SIN caustics} and \ref{fig:3SINcaustics}) are shown at the intersections with the $x=0$, $y=0$ and $z=0$ planes, while the density and vorticity outputs (Figs.~\ref{fig:density} and \ref{fig: 3SIN vorticity Q1D}--\ref{fig: 3SIN vorticity SYM}) are shown in slices slightly away from the origin because the vorticity vanishes in the $x=0$, $y=0$, and $z=0$ planes due to the symmetries in the chosen setting.

The examination of left four panels of Fig.~\ref{fig: 2SIN caustics} and top twelve panels of Fig.~\ref{fig:3SINcaustics} provides us insights on the caustic pattern in the most typical configurations, $\epsilon_i \neq \epsilon_j$, $i\neq j$, that is the non-axial-symmetric case, both in 2D and in 3D. Note that in the 3D case, to have a more clear view of the caustic pattern, we consider the intersection of the caustic surfaces with the planes $x=0$, $y=0$ and $z=0$ (remind however that a 3D view of the caustic surfaces was given for 2LPT in Fig.~\ref{fig: 3SIN caustics}). In the case of the simulations, as explained in Sec.~\ref{sec:accuracy} and Appendix~\ref{sec:appcaus}, the caustic pattern corresponds to a tessellation, that is a set of ensembles of connected segments in 2D and of connected triangles in 3D. The intersection of a tessellation of triangles with a plane also gives a set of ensembles of connected segments. On the figures, only the vertices of these segments are represented, and the initial regular pattern of the tessellation used to represent the phase-space sheet is clearly visible in Lagrangian space.

Figs.~\ref{fig: 2SIN caustics} and \ref{fig:3SINcaustics} confirm the results of the previous section, namely that high order LPT provides a rather accurate description of the caustic pattern in the Q1D case, with a clear convergence to the simulation when the perturbation order increases. In the anisotropic cases, ANI-2SIN and ANI-3SIN, the match between LPT and simulations is not perfect, even at 10th order, but remains still reasonably good, especially in 2D. For the axial-symmetric configurations, the convergence of LPT with order seems much slower, particularly in Eulerian space and in 3D.

While the shape of the caustic pattern is very simple in the generic cases, it becomes significantly more intricate in the axial-symmetric cases, particularly in Eulerian space. Indeed, the caustics are composed of two connected curves instead of one for $\epsilon_{2{\rm D}}=1$, and three connected surfaces instead of one for $\bm{\epsilon}_{3{\rm D}}=(1,1)$. For $\epsilon_{2{\rm D}}=1$, the shapes of simulation caustics are approximately reproduced by LPT, but we already see that convergence to the exact solution is slow. In particular, the extension of the outer caustic in Eulerian space is quite underestimated by LPT and it seems that increasing perturbation order to arbitrary values is not going to improve the agreement with the simulations. The situation is much worse in 3D, particularly in Eulerian space, although the simulation measurements are very noisy, which makes comparison with theoretical predictions difficult. We know from the previous paragraph that these mismatches are at least partly attributable to limits of the ballistic approximation, along with limits of LPT to be able to describe the velocity field at collapse in the vicinity of its extrema, particularly the spike observed on the phase-space diagram in the 3D case.

Not surprisingly, the projected density measurements of Figs.~\ref{fig: 2SIN density} and \ref{fig:density} fully confirm the analyses of the caustic pattern. Putting aside the axial-symmetric cases, we can see that the agreement between simulations and LPT of high order is quite good, inside and outside the caustics. 
Note that the very high density contrasts observed in the axial-symmetric cases are due to the multiple foldings of the phase-space sheet, even in 2D, which explains the limits of the ballistic approximation, since feedback effects from the gravitational force field after collapse are expected to be orders of magnitude larger than in the generic case.

Turning to the vorticity, as shown in Figs.~\ref{fig: 2SIN vorticity} to \ref{fig: 3SIN vorticity SYM}, we obtain again a good agreement between theory and measurements in the non-axial-symmetric configurations, except that, at the exact location of the caustics, the simulation measurements are expected to be spurious. This is discussed in Appendix~\ref{app:quadfields} (see also end of Sec.~\ref{sec:accuracy}), which provides details on the measurements of various fields, in particular how we exploit a quadratic description of the phase-space sheet to achieve vorticity measurements with unprecedented accuracy, but yet still limited by strong variations of the fields in the vicinity of the caustics, in particular the strong discontinuous transition between the inner part and the outer part of the multi-stream regions at the precise locations of the caustics.

From the qualitative point of view, the topology of the vorticity field for the generic configurations, $\epsilon_i \neq \epsilon_j$, $i \neq j$, agrees perfectly with the predictions of \citet[][]{1999A&A...343..663P} based on Zel'dovich dynamics.\footnote{Remind however that the crossed sine waves configurations we consider are degenerate with respect to 1LPT, as discussed at the end of Sec.~\ref{sec: ballistic approx}, so the reasoning of \citet[][]{1999A&A...343..663P} requires 2LPT to be applicable in this particular case.} In 2D, the vorticity field is a scalar which can be decomposed into the four sectors inside the caustic, two of positive sign and two of negative one. In 3D, the vorticity field is a vector. Because shell crossing takes place along $x$-axis, each component of this vector field has specific properties, which are related to variations of the velocity and density field along with the phase-space sheet.
In particular, it is easy to convince oneself, that if collapse happens along $x$-axis, the strongest variations of all the fields are expected along this direction.
That means, since each coordinate of the vorticity vector depends on variations of the velocity field in other coordinates, that the magnitude of $\omega_{x}$ is expected to be small.
Due to symmetries, we also expect that $\omega_{y}$ and $\omega_{z}$ are, respectively, an odd function of $z$ and $y$, which means that $\omega_{y}$ approaches zero when approaching the $z=0$ plane, and similarly for $\omega_{z}$ when approaching the $y=0$ plane, which explains the pattern of the vorticity field on Figs.~\ref{fig: 3SIN vorticity Q1D} and \ref{fig: 3SIN vorticity ANI}. These symmetries impose us to perform measurements on slices slightly shifted from the centre of the system, as detailed in the caption of Fig.~\ref{fig:density}. Note as well, in these figures, that the pattern is analogous to 2D for the $y$ and $z$ coordinate of the vorticity fields in the $x-z$ and $x-y$ subspace, respectively.

Turning to the axial-symmetric case, the agreement between theory and measurements is again only partial. Yet the high magnitude of vorticity is of the same order for theory and measurements. The structure of the vorticity field, significantly more complex than in the generic case due to multiple foldings of the phase-space sheet, is qualitatively in agreement between LPT and Vlasov code in 2D, while simulations measurements are too noisy in the 3D case to make definitive conclusions. What is clear, however, is that the outer part of the multi-stream region seems to be qualitatively reproduced by the theory, but its size is totally wrong.

%%%%%%%%%%%%%%%%%%%%%%%%%%%%%%%%%%%%%%%%%%%%%%%%%%
%%%%%%%%%%%%%%%%%%%%%%%%%%%%%%%%%%%%%%%%%%%%%%%%%%
%%%%%%%%%%%%%%%%%%%%%%%%%%%%%%%%%%%%%%%%%%%%%%%%%%
\section{Summary}
\label{sec: summary}
%%%%%%%%%%%%%%%%%%%%%%%%%%%%%%%%%%%%%%%%%%%%%%%%%%
%%%%%%%%%%%%%%%%%%%%%%%%%%%%%%%%%%%%%%%%%%%%%%%%%%
%%%%%%%%%%%%%%%%%%%%%%%%%%%%%%%%%%%%%%%%%%%%%%%%%%
In this paper, following the footsteps of \citet{2018PhRvL.121x1302S}, we have investigated the structure of primordial cold dark matter (CDM) haloes seeded by two or three crossed sine waves of various amplitudes at and shortly after shell-crossing, by comparing thoroughly up to 10th-order Lagrangian perturbation theory (LPT) to high-resolution Vlasov-Poisson simulations performed with the public Vlasov solver {\tt ColDICE}. We devoted our attention first to the phase-space structure and radial profiles of the density and velocities at shell-crossing, and, second, to the phase-space structure, caustics, density and vorticity fields shortly after shell-crossing. In particular, measurements of unprecedented accuracy of the vorticity in the simulations were made possible by exploiting the fact that {\tt ColDICE} is able to follow locally the phase-space sheet structure at the quadratic level.

We studied three qualitatively different initial conditions characterised by the amplitude of three crossed sine waves, as summarised in Table~\ref{tab: initial conditions}: quasi one-dimensional (Q1D-2SIN, Q1D-3SIN), where one amplitude of the sine waves dominates over the other one(s), anisotropic (ANI-2SIN, ANI-3SIN), where the amplitude of each wave is different but remains of the same order, and axial-symmetric (SYM-2SIN, SYM-3SIN), where all amplitudes are the same. In order to predict the protohalo structure shortly after shell-crossing in an analytical way, we used the ballistic approximation, where the acceleration is neglected after the shell-crossing time.

Our main findings can be summarised as follows:
%----------------------------------
\begin{itemize}
\item {\it Phase-space diagrams at collapse:} except for SYM-3SIN, one expects the system to build up a classic pancake singularity at shell-crossing, with a phase-space structure along $x$-axis analogous to what is obtained in one dimension. This pancake is well reproduced by LPT which converges increasingly well to the exact solution when approaching quasi one-dimensional dynamics, as expected. The local extrema of the velocity field around the singularity are the locations where LPT differs most from the exact solution, by underestimating the magnitude of the velocity. Convergence with perturbation order becomes very slow when approaching three-dimensional axial-symmetry, where spikes appear on the velocity field on each side of the singularity.

\item {\it Radial profiles at collapse:} with a sufficiently high order of the perturbation, LPT can reproduce arbitrarily high density contrasts at collapse, but we note a slower convergence when turning to velocity profiles. Still, the convergence of LPT is sufficiently good to probe the asymptotic logarithmic slope at small radii expected at collapse for various profiles from singularity theory, as summarised in Table~\ref{tab: summary at shell-crossing}, and confirmed by simulation measurements. These profiles are $\rho(r)\propto r^{-2/3}$ for generic initial conditions, that is Q1D and ANI, and $\rho(r)\propto r^{-4/3}$ for SYM-2SIN, $\rho(r) \propto r^{-2}$ for SYM-3SIN, while velocities profiles always present the same power-law behaviour, e.g. $v^{2}(r) \propto r^{2/3}$. Confirming predictions of singularity theory at collapse and explicit convergence to asymptotic profiles at small radii is expected but non-trivial. 

\item {\it Synchronisation:} agreement with singularity theory predictions is made even more explicit by computing LPT profiles at the respective collapse times computed at each perturbation order. In this case, convergence to the power-law profile predicted at small radii by singularity theory is explicit. In fact, the radial profiles obtained from LPT with such synchronisation are strikingly alike, particularly for the density, which motivates the use of a ballistic approximation to study the motion slightly beyond collapse. 

\item {\it Ballistic approximation:} measurements in simulations show that the ballistic approximation provides an accurate description of the phase-space structure of the system slightly beyond collapse, except in the axial-symmetric case SYM-3SIN, where the very highly contrasted nature of the system due to multiple superpositions of phase-space sheet folds introduces strong force feedback effects. We notice however that even in the generic pancake case, the ballistic approximation can still slightly underestimate velocities in the vicinity of the singularity. Obviously, these conclusions depend on how long this approximation is used. In this work, we considered maximum relative variations of the expansion factor of the order of three percent. For these values, the deviations from pure ballistic motion were the most significant, as expected. 

\item{\it Structure beyond collapse:} given the limits discussed above, we find, when considering generic, non-axial-symmetric configurations, that LPT of sufficient order combined with the ballistic approximation provides a very good description of the structure of the system beyond collapse time, with excellent agreement between the predicted density and the vorticity fields inside the multi-stream region and those measured in the Vlasov simulations. Turning to axial-symmetric cases, this agreement is only obtained at the qualitative level, even for 10th-order LPT, particularly for SYM-3SIN, where the size of the outer caustics is strongly underestimated.
\end{itemize}
%----------------------------------

Obviously, the ballistic approximation is only the first step for a more complete calculation taking into account the feedback due to the force field, as first proposed in the one dimensional case by \citet[][]{2015MNRAS.446.2902C,2017MNRAS.470.4858T} \citep[see also the recent work of][]{2019arXiv191200868R}, and formulated in terms of ``post-collapse perturbation theory''.
The next step is indeed to implement a Lagrangian perturbation theory approach where the small parameter is the interval of time $\Delta a=a-a_{\rm sc}$ from collapse time and where a Taylor expansion of the phase-space sheet is performed around the singularity in terms of Lagrangian coordinates. Shortly after collapse, in the generic, non-axial-symmetric case, the system presents a ``S'' shape in $x-v_x$ subspace, similar to the 1D case where position and velocities can be approximated as third-order polynomials of the Lagrangian coordinate $\bm{q}$. The calculation of the force field requires then to solve a 3 value problem in the multi-stream region, similarly as in the 1D case. While technically challenging, generalisation of post-collapse perturbation theory to 2D and 3D seems possible. It might bring strong insights on nonlinear corrections due to multi-stream dynamics on large scale structure statistics, e.g. predictions of the power spectrum of the large-scale matter distribution from higher order perturbation theory.

%%%%%%%%%%%%%%%%%%%%%%%%%%%%%%%%%%%%%%%%%%%%%%%%%%
%%%%%%%%%%%%%%%%%%%%%%%%%%%%%%%%%%%%%%%%%%%%%%%%%%
%%%%%%%%%%%%%%%%%%%%%%%%%%%%%%%%%%%%%%%%%%%%%%%%%%
\section*{Acknowledgements}
%%%%%%%%%%%%%%%%%%%%%%%%%%%%%%%%%%%%%%%%%%%%%%%%%%
%%%%%%%%%%%%%%%%%%%%%%%%%%%%%%%%%%%%%%%%%%%%%%%%%%
%%%%%%%%%%%%%%%%%%%%%%%%%%%%%%%%%%%%%%%%%%%%%%%%%%
This work was supported in part by JSPS Research Fellow Grant-in-Aid Number 17J10553 (SS), MEXT/JSPS KAKENHI Grants Numbers JP17H06359, JP20H05861, and JP21H01081, JST AIP Acceleration Research grant JP20317829 (AT), ANR grant ANR-13-MONU-0003 (SC), as well as Programme National Cosmology et Galaxies (PNCG) of CNRS/INSU with INP and IN2P3, co-funded by CEA and CNES (SC). Numerical computation with {\tt ColDICE} was carried out using the HPC resources of CINES (Occigen supercomputer) under the GENCI allocations c2016047568, 2017-A0040407568 and 2018-A0040407568. Post-treatment of {\tt ColDICE} data were performed on HORIZON cluster of Institut d'Astrophysique de Paris. Calculations of theoretical predictions have made use of the Yukawa Institute Computer Facility.

\bibliography{ref}
\bibliographystyle{aa}

\begin{appendix}

%%%%%%%%%%%%%%%%%%%%%%%%%%%%%%%%%%%%%%%%%%%%%%%%%%%%%%%%%%%%%%%%%%%%%%%%%%%%%%%%%%
%%%%%%%%%%%%%%%%%%%%%%%%%%%%%%%%%%%%%%%%%%%%%%%%%%%%%%%%%%%%%%%%%%%%%%%%%%%%%%%%%%
%%%%%%%%%%%%%%%%%%%%%%%%%%%%%%%%%%%%%%%%%%%%%%%%%%%%%%%%%%%%%%%%%%%%%%%%%%%%%%%%%%
\section{Expressions of the LPT solutions}
\label{app: solutions}
%%%%%%%%%%%%%%%%%%%%%%%%%%%%%%%%%%%%%%%%%%%%%%%%%%%%%%%%%%%%%%%%%%%%%%%%%%%%%%%%%%
%%%%%%%%%%%%%%%%%%%%%%%%%%%%%%%%%%%%%%%%%%%%%%%%%%%%%%%%%%%%%%%%%%%%%%%%%%%%%%%%%%
%%%%%%%%%%%%%%%%%%%%%%%%%%%%%%%%%%%%%%%%%%%%%%%%%%%%%%%%%%%%%%%%%%%%%%%%%%%%%%%%%%

In this appendix, we present the LPT solutions up to 5th order, which are obtained by solving the recursion relations given in Eqs.~(\ref{eq:longitudinal3}) and (\ref{eq:transverse3}). Since higher-order solutions are straightforwardly derived in the same way, we do not explicitly show them here.

The results are partly presented in \citet{1991ApJ...382..377M} up to 3rd order, taking into account the contributions other than the fastest growth mode \citep[see also][for the solutions including decaying modes]{1997A&A...318....1B}. Here, we first explicitly show the analytical solutions of the sine waves initial conditions up to 5th order.

As shown in Sec.~\ref{sec:recur}, the fastest growing mode can be expanded as follows:
\begin{align}
\bm{\Psi}(\bm{q},t) = \sum^{\infty}_{n=1}D^{n}_{+}(t)\, \bm{\Psi}^{(n)}(\bm{q})~.
\end{align}
For presentation purposes, we only show the $x$-components of the displacement field, i.e., $\Psi^{(n)}_{x}(q)$ with the definition $(\epsilon_{1},\, \epsilon_{2}) \equiv (\epsilon_{y}/\epsilon_{x},\, \epsilon_{z}/\epsilon_{x})$. Given the $x$-components, the $y$- and $z$-components can be derived by permutating all $x$, $y$, and $z$.

\begin{widetext}

\begin{align}
\Psi^{(1)}_{x} &= \frac{\epsilon_{x}}{2\pi}\sin(2\pi\, q_{x})~, \\
\Psi^{(2)}_{x} &= - \frac{3\epsilon_{x}^{2}}{28\pi}
\left[ \epsilon_{1}\cos(2\pi\, q_{y}) + \epsilon_{2}\cos(2\pi\, q_{z})\right] \sin(2\pi\, q_{x})~, \\
\Psi^{(3)}_{x} &= \frac{\epsilon_{x}^{3}}{2520\pi}
\Bigl[
78 \cos (2 \pi q_{x}) (\epsilon_{1} \cos (2 \pi q_{y})+\epsilon_{2} \cos (2 \pi q_{z}))
\notag \\
& \qquad
+160 \epsilon_{1} \epsilon_{2} \cos (2 \pi q_{y}) \cos (2 \pi q_{z})-3 \epsilon_{1}^2 \cos (4 \pi q_{y})
-3 \epsilon_{2}^2 \cos (4 \pi q_{z})
+75 \left(\epsilon_{1}^2+\epsilon_{2}^2\right)
\Bigr]
\sin(2\pi\, q_{x})~, \\
\Psi^{(4)}_{x} &= - \frac{\epsilon_{x}^{4}}{7761600\pi}
\Bigl[
\epsilon_{2} \cos (2 \pi q_{z}) \left(4242 \cos (4 \pi q_{x})+28550 \epsilon_{1}^2 \cos (4 \pi q_{y})+208850 \epsilon_{1}^2+57015 \epsilon_{2}^2+89166\right)
\notag \\
& \qquad
+ 60 \cos (2 \pi q_{x}) \left(6010 \epsilon_{1} \epsilon_{2} \cos (2 \pi q_{y}) \cos (2 \pi q_{z})+1274 \epsilon_{1}^2 \cos (4 \pi q_{y})+1274 \epsilon_{2}^2 \cos (4 \pi q_{z})+2039 \left(\epsilon_{1}^2+\epsilon_{2}^2\right)\right)
\notag \\
& \qquad
+2 \epsilon_{1} \cos (2 \pi q_{y}) \left(2121 \cos (4 \pi q_{x})-9303 \epsilon_{1}^2 \cos (4 \pi q_{y})+14275 \epsilon_{2}^2 \cos (4 \pi q_{z})+33159 \epsilon_{1}^2+104425 \epsilon_{2}^2+44583\right)\notag \\
& \qquad
-9303 \epsilon_{2}^3 \cos (6 \pi q_{z})
\Bigr]
\sin(2\pi\, q_{x})~,\\
\Psi^{(5)}_{x} &= \frac{\epsilon_{x}^{5}}{36793476720000\pi}
\Bigl[
895050 \cos (4 \pi q_{x}) \left(181296 \epsilon_{1} \epsilon_{2} \cos (2 \pi q_{y}) \cos (2 \pi q_{z})+41657 \left(\epsilon_{1}^2+\epsilon_{2}^2\right)\right)
\notag \\
&\qquad
+560226137900 \epsilon_{1}^2 \epsilon_{2} \cos (2 \pi q_{x}) \cos (4 \pi q_{y}) \cos (2 \pi q_{z})+560226137900 \epsilon_{1} \epsilon_{2}^2 \cos (2 \pi q_{x}) \cos (2 \pi q_{y}) \cos (4 \pi q_{z})
\notag \\
&\qquad
+594423768510 \epsilon_{1}^3 \cos (2 \pi q_{x}) \cos (2 \pi q_{y})-18642271230 \epsilon_{1}^3 \cos (2 \pi q_{x}) \cos (6 \pi q_{y})\notag \\
&\qquad
+57401651835 \epsilon_{1}^2 \cos (4 \pi (q_{x}-q_{y}))+57401651835 \epsilon_{1}^2 \cos (4 \pi (q_{x}+q_{y}))+1006179766020 \epsilon_{1} \epsilon_{2}^2 \cos (2 \pi q_{x}) \cos (2 \pi q_{y})
\notag \\
&\qquad
+106526974554 \epsilon_{1} \cos (2 \pi q_{x}) \cos (2 \pi q_{y})-6828912090 \epsilon_{1} \cos (6 \pi q_{x}) \cos (2 \pi q_{y})+1006179766020 \epsilon_{1}^2 \epsilon_{2} \cos (2 \pi q_{x}) \cos (2 \pi q_{z})
\notag \\
&\qquad
+57401651835 \epsilon_{2}^2 \cos (4 \pi (q_{x}-q_{z}))+594423768510 \epsilon_{2}^3 \cos (2 \pi q_{x}) \cos (2 \pi q_{z})-18642271230 \epsilon_{2}^3 \cos (2 \pi q_{x}) \cos (6 \pi q_{z})
\notag \\
&\qquad
+57401651835 \epsilon_{2}^2 \cos (4 \pi (q_{x}+q_{z}))+106526974554 \epsilon_{2} \cos (2 \pi q_{x}) \cos (2 \pi q_{z})-6828912090 \epsilon_{2} \cos (6 \pi q_{x}) \cos (2 \pi q_{z})
\notag \\
&\qquad
+46659033850 \epsilon_{1}^2 \epsilon_{2}^2 \cos (4 \pi (q_{y}-q_{z}))+46659033850 \epsilon_{1}^2 \epsilon_{2}^2 \cos (4 \pi (q_{y}+q_{z}))+503168866200 \epsilon_{1}^3 \epsilon_{2} \cos (2 \pi q_{y}) \cos (2 \pi q_{z})
\notag \\
&\qquad
-54358176600 \epsilon_{1}^3 \epsilon_{2} \cos (6 \pi q_{y}) \cos (2 \pi q_{z})+503168866200 \epsilon_{1} \epsilon_{2}^3 \cos (2 \pi q_{y}) \cos (2 \pi q_{z})-54358176600 \epsilon_{1} \epsilon_{2}^3 \cos (2 \pi q_{y}) \cos (6 \pi q_{z})
\notag \\
&\qquad
+877320054000 \epsilon_{1} \epsilon_{2} \cos (2 \pi q_{y}) \cos (2 \pi q_{z})+115709716980 \epsilon_{1}^2 \epsilon_{2}^2 \cos (4 \pi q_{y})-35064073044 \epsilon_{1}^4 \cos (4 \pi q_{y})
\notag \\
&\qquad
-1607701095 \epsilon_{1}^4 \cos (8 \pi q_{y})+241408089450 \epsilon_{1}^2 \cos (4 \pi q_{y})+115709716980 \epsilon_{1}^2 \epsilon_{2}^2 \cos (4 \pi q_{z})-35064073044 \epsilon_{2}^4 \cos (4 \pi q_{z})\notag \\
&\qquad
+241408089450 \epsilon_{2}^2 \cos (4 \pi q_{z})-1607701095 \epsilon_{2}^4 \cos (8 \pi q_{z})
\notag \\
&\qquad
+546975 \left(\epsilon_{1}^2 \left(940700 \epsilon_{2}^2+443058\right)+171045 \epsilon_{1}^4+21 \epsilon_{2}^2 \left(8145 \epsilon_{2}^2+21098\right)\right)
\Bigr]
\sin(2\pi\, q_{x})~.
\end{align}
\end{widetext}
Note that because of the underlying symmetry of the initial conditions, the $x$-components of the LPT solutions are symmetric under the exchange of $y \leftrightarrow z$ and $\epsilon_{1} \leftrightarrow \epsilon_{2}$.

%%%%%%%%%%%%%%%%%%%%%%%%%%%%%%%%%%%%%%%%%%%%%%%%%%%%%%%%%%%%%%%%%%%%%%%%%%%%%%%%%%
%%%%%%%%%%%%%%%%%%%%%%%%%%%%%%%%%%%%%%%%%%%%%%%%%%%%%%%%%%%%%%%%%%%%%%%%%%%%%%%%%%
%%%%%%%%%%%%%%%%%%%%%%%%%%%%%%%%%%%%%%%%%%%%%%%%%%%%%%%%%%%%%%%%%%%%%%%%%%%%%%%%%%
\section{Measurements on the tessellation of Vlasov simulations}
\label{app: coldice}
%%%%%%%%%%%%%%%%%%%%%%%%%%%%%%%%%%%%%%%%%%%%%%%%%%%%%%%%%%%%%%%%%%%%%%%%%%%%%%%%%%
%%%%%%%%%%%%%%%%%%%%%%%%%%%%%%%%%%%%%%%%%%%%%%%%%%%%%%%%%%%%%%%%%%%%%%%%%%%%%%%%%%
%%%%%%%%%%%%%%%%%%%%%%%%%%%%%%%%%%%%%%%%%%%%%%%%%%%%%%%%%%%%%%%%%%%%%%%%%%%%%%%%%%
This appendix explains in detail how measurements of various quantities studied in this article are performed on the output of {\tt ColDICE}, which consists of a tessellation of the phase-space sheet with simplices, respectively triangles and tetrahedra in 4D and 6D phase-space. 
\subsection{Phase-space diagrams}
\label{app:A1}
The intersection of a hypersurface of dimension $D_1=D$ with a hyperplane $P$ of dimension $D_2=D+1$ inside a phase-space of dimension $D_3=2D$ is of dimension $D_1+D_2-D_3=1$. In other words, this intersection corresponds in the non-trivial case to a set of curves. Therefore, in 2D, the intersection of the phase-space sheet with the hyperplane $x=0$ is a set of curves, and likewise in 3D, the intersection of the phase-space sheet with the hyperplane $x=y=0$. Furthermore, as discussed more in detail in C21, because the phase-space sheet remains at all times a fully connected hypersurface, this set of curves should also be fully connected, with no hanging point.

To produce a phase-space diagram, we employ an approach valid at linear order, consisting in simply computing, for each simplex, its geometric intersection with the hyperplane $P$, which is either empty, a point or a segment. As a result, phase-space diagrams extracted from the Vlasov runs consist of sets of connected segments of which the ends are plotted in Figs.~\ref{fig: phase precollapse} and \ref{fig: phase postcollapse}.
\subsection{Expansion factor at collapse: $a_{\rm sc}$}
\label{app:colla}
An accurate estimate of the value of the expansion factor corresponding to collapse time is essential when studying the properties of the system at shell crossing and shortly after. The curves generated in Appendix~\ref{app:A1} can be used for this purpose, at two expansion times $a_i$, $i=1,2$, supposed to be just before and just after collapse. At both these times, we consider the unique phase-space diagram segment portion $S$ of which one of the ends is as close as possible to the origin, while the other one, with abscissa $x_i$, has $v_x > 0$ and the magnitude of other coordinate(s) of the velocity, in theory null, as small as possible. The latter condition excludes, in the axial-symmetric case, the component(s) of the flow corresponding to simultaneous collapse along the $y$ or/and $z$ direction, which has/have non-zero value of $v_{y}$ or/and $v_z$. Once this segment is identified at both times corresponding to $a_1$ and $a_2$, a simple linear interpolation is used to estimate expansion factor at collapse:
\begin{equation}
 a_{\rm sc} \simeq \frac{a_1x_2-a_2 x_1}{x_2-x_1}.
 \label{eq:asc}
\end{equation}
This formula does not require to be just before and after collapse time to estimate $a_{\rm sc}$, but, to provide sufficiently accurate results, it is obviously needed to consider times sufficiently close to actual collapse time. A first guess of collapse time is estimated by using perturbation theory predictions extrapolated to infinite order (STC18), as listed in 5th column of Table~\ref{tab: initial conditions}. Examination of Table~\ref{tab: initial conditions} shows that the value $a_{\rm sc}^{\infty}$ predicted this way agrees extremely well with the measurements in the Vlasov runs provided by equation (\ref{eq:asc}), and shown in 6th column of the table. 

Note however that, beyond the approximate nature of the linear approximation underlying equation (\ref{eq:asc}), collapse time estimate can, among others, as discussed in the main text in Sec.~\ref{sec:accuracy}, be significantly influenced by force resolution, that is the value of $n_{\rm g}$. As discussed in detail in C21, decreasing force resolution delays collapse time, hence, to have an accurate estimate of this latter, it is required to have a sufficiently large value of $n_{\rm g}$. We performed extensive force resolution tests for the 3D simulation with $\bm{\epsilon}_{\rm 3D}=(3/4,1/2)$. Our measurements of $a_{\rm sc}$ with the above method provide $a_{\rm sc}=0.02912$ (in excellent agreement with the predicted value $a_{\rm sc}^{\infty}=0.02911$), $0.0292$ and $0.0293$ respectively for $n_{\rm g}=1024$, $512$ and $256$. Our $n_{\rm g}=512$ simulation thus provides, for this value of $\bm{\epsilon}_{\rm 3D}$, an estimate of collapse expansion factor accurate at approximately the $10^{-4}$ level and we expect this to apply as well to the quasi one-dimensional case $\bm{\epsilon}_{\rm 3D}=(1/6,1/8)$. However, for the axial-symmetric case, $\bm{\epsilon}_{\rm 3D}=(1,1)$, the measured value might still overestimate the actual one by an amount larger than $\sim 10^{-4}$ due to the high strength of the singularity building up at the center of the system. On the contrary, in the 2D case, given the high value of $n_{\rm g}=2048$ used to perform the simulations, we expect our estimates of collapse time to be very accurate, at an order better than the $10^{-4}$ level, but we did not test this explicitly.
\subsection{Caustics}
\label{sec:appcaus}
Caustics are regions where the determinant $J$ of the Jacobian matrix changes sign. At linear order in the local description of the phase-space sheet, geometrically this means that the orientation of the simplex changes in configuration space, which allows one to define unambiguously regions corresponding to the intersections of simplices with $J \geq 0$ and simplices with $J < 0$, where the sign of $J$ is directly estimated from the current orientation of the simplex with respect to the original one. This is actually performed during runtime by {\tt ColDICE} which can output caustics directly when needed. In 2D, the phase-space sheet is composed of a tessellation of triangles, hence the caustics estimated this way are given by sets of segments of which the ends are shown in Fig.~\ref{fig: 2SIN caustics}. In 3D, the caustics are given by sets of triangles, of which we compute the intersection with the $y=z=0$ plane to get again a set of segments of which the extremities are shown in Fig.~\ref{fig:3SINcaustics}. Because we are using a leading order approach, the caustic lines or surfaces are not necessarily smooth but should trace accurately enough the actual caustics for the purpose of this work (see also Sec.~\ref{sec:accuracy}).
\subsection{Radial profiles}
\label{app:radp}
To measure radial profiles in logarithmic bins, each simplex is replaced with a large number of particles as explained for the 3D case in Appendix A2 of C21. We refer to this work for the reader interested in the details of this procedure, that we straightforwardly generalised to the 2D case. 
\subsection{Density field and vorticity: from linear to quadratic order}
\label{app:quadfields}
In this section, we aim to compute the following quantities:
\begin{itemize}
\item the jacobian of the transformation between initial and final positions, $J(\bm{q})$ as defined in equation (\ref{eq: def Jij});
\item the Lagrangian projected density $\rho_{\rm L}(\bm{q})$ defined in equation (\ref{eq:rhoLdef});
\item the total Eulerian projected density, which stems from the superposition of one or more folds of the phase-space sheet, as described by equation (\ref{eq:rhoE});
\item the Eulerian velocity field, given by equation (\ref{eq:vE});
\item the vorticity, $\omega^{2{\rm D}}$ and $\omega^{3{\rm D}}$, defined in equations (\ref{eq: vorticity 2D}) and (\ref{eq: vorticity 3D}). Again, because the acceleration derives from a potential, local vorticity on
each phase-space sheet fold cancels. Only the nonlinear superposition of
phase-space sheet folds contained in the first term of these equations
induces non-zero vorticity, while the second term
should not contribute, although we shall take it into account in our
numerical calculations. 
\end{itemize}
In what follows, we explain how to compute the various quantities
defined above by exploiting the decomposition in simplices of the
phase-space sheet by {\tt ColDICE}. In \S~\ref{sec:bary}, we introduce barycentric coordinates,
which are useful to define the position of any point inside each simplex. In \S~\ref{sec:lin},
we show how the barycentric coordinates can be used to perform calculations at the linear
level inside each simplex, in particular partial derivatives of a function, as
already discussed for instance by \citet{2015MNRAS.454.3920H}. In \S~\ref{sec:quad}, we generalize the
procedure to the case when a quadratic description of the phase-space
sheet is available. Finally, in \S~\ref{sec:proj}, we describe the
way we sample various quantities described above on an high resolution cartesian grid, 
by using proper sets of sampling particles associated to each
simplex.

%%%%%%%%%%%%%%%%%%%%%%%%%%%%%%%%%%%%%%%%%%%%%%%%%%
%%%%%%%%%%%%%%%%%%%%%%%%%%%%%%%%%%%%%%%%%%%%%%%%%%
\subsubsection{Barycentric coordinates}
\label{sec:bary}
%%%%%%%%%%%%%%%%%%%%%%%%%%%%%%%%%%%%%%%%%%%%%%%%%%
%%%%%%%%%%%%%%%%%%%%%%%%%%%%%%%%%%%%%%%%%%%%%%%%%%
Since the phase-space sheet is sampled with simplices, it is useful to
define a well known local system of coordinates on each simplex. Given
the positions $\bm{X}_k$, $k=1,\cdots,N_{\rm s}$ of the simplex vertices,
with $N_{\rm s}=D+1$ where $D$ is dimension of configuration space, and a function $g(\bm{X})$, 
one can define the following linear interpolation
\begin{equation}
g_{\rm linear}(\bm{X})=\sum_{k=1}^{N_{\rm s}} \xi_k g_k,
\label{eq:linint}
\end{equation}
for 
\begin{equation}
\bm{X} = \bm{X}_{\rm linear} \equiv \sum_{k=1}^{N_{\rm s}} \xi_k
\bm{X}_k,\label{eq:lineco}
\end{equation}
where the {\it barycentric coordinates} $\xi_k$ are positive quantities verifying $\xi_k=1$,
$\xi_{k'\neq k}=0$, for
$\bm{X}=\bm{X}_k$ and $\sum_k \xi_k=1$. When working in Lagrangian
space, the space of initial positions, the phase-space sheet is flat
and the linear interpolation
\begin{equation}
\bm{Q}_{\rm linear} =\sum_{k=1}^{N_{\rm s}} \xi_k \bm{Q}_k
\label{eq:linecoq}
\end{equation}
is exact. In what follows, we shall therefore use equation
(\ref{eq:linecoq}) to define 
barycentric coordinates. With this definition of $\xi_k$, this means that at
given time $t$, equation (\ref{eq:lineco}) remains valid, but only at
the linear level, since dynamical evolution of the phase-space sheet
produces curvature.

%%%%%%%%%%%%%%%%%%%%%%%%%%%%%%%%%%%%%%%%%%%%%%%%%%
%%%%%%%%%%%%%%%%%%%%%%%%%%%%%%%%%%%%%%%%%%%%%%%%%%
\subsubsection{Calculations of various quantities at the linear level}
\label{sec:lin}
%%%%%%%%%%%%%%%%%%%%%%%%%%%%%%%%%%%%%%%%%%%%%%%%%%
%%%%%%%%%%%%%%%%%%%%%%%%%%%%%%%%%%%%%%%%%%%%%%%%%%

With the barycentric coordinate representation, any function
defined at vertices positions can be estimated
locally at the linear level using equation (\ref{eq:linint}). The
derivative of the function inside each simplex can thus be written
\begin{equation}
\frac{\partial g}{\partial x_\beta}=\sum_{k=1}^{N_{\rm s}}
g_k \frac{\partial \xi_k}{\partial x_\beta}=\sum_{k=1}^{N_{\rm s}}
g_k\, M^{-1}_{k,\beta+1},
\label{eq:derivlin}
\end{equation}
where $M$ is the matrix of the partial derivatives of the vector
$\bm{P}\equiv (\sum_k \xi_k, X_1, X_2, X_3)$ with respect to $\xi_k$, with $\bm{X}$
given by equation (\ref{eq:lineco}). In other words,
\begin{align}
M_{1,k} &= 1, \label{eq:Mmat1} \\
M_{\alpha+1,k}& = \frac{\partial X_\alpha}{\partial \xi_k}=
X_{k,\alpha}, \label{eq:Mmat2}
\end{align}
with $X_{k,\alpha}$ the $\alpha^{\rm th}$ coordinate of $k^{\rm th}$
vertex with position $\bm{X}_k$. At the linear level, the derivative defined by
equation (\ref{eq:derivlin}) is therefore simply constant inside each
simplex. 

This calculation can be performed in Lagrangian space
to obtain directly the Jacobian of the transformation between initial
and present position, equation (\ref{eq: def Jij}), from the jacobian matrix
\begin{equation}
J_{\alpha,\beta}\equiv \frac{\partial X_\alpha}{\partial
Q_\beta}=\sum_{k=1}^{N_{\rm s}} M_{\alpha+1,k}\, W^{-1}_{k,\beta+1},
\label{eq:Tcalc}
\end{equation}
where $W$ is defined similarly as $M$ but in Lagrangian space,
\begin{align}
W_{1,k} &= 1, \\
W_{\alpha+1,k}& = \frac{\partial Q_\alpha}{\partial \xi_k}=
Q_{k,\alpha},
\end{align}
with $Q_{k,\alpha}$ the $\alpha^{\rm th}$ Lagrangian coordinate
of $k^{\rm th}$ vertex with Lagrangian position $\bm{Q}_k$.

This means that the linear description allows one to obtain the
Jacobian $J(\bm{q})$ and the corresponding Lagrangian density
$\rho_{\rm L}(\bm{q})=1/|J(\bm{q})|$ as constant quantities inside each
simplex. Unfortunately, this is not sufficient, since equations (\ref{eq: vorticity 2D}) and (\ref{eq: vorticity 3D}) involve partial derivatives of the projected density. Rigorously
speaking, this means that a representation of the simplices at the
quadratic level is required.

%%%%%%%%%%%%%%%%%%%%%%%%%%%%%%%%%%%%%%%%%%%%%%%%%%
%%%%%%%%%%%%%%%%%%%%%%%%%%%%%%%%%%%%%%%%%%%%%%%%%%
\subsubsection{Calculations of various quantities at the quadratic level}
\label{sec:quad}
%%%%%%%%%%%%%%%%%%%%%%%%%%%%%%%%%%%%%%%%%%%%%%%%%%
%%%%%%%%%%%%%%%%%%%%%%%%%%%%%%%%%%%%%%%%%%%%%%%%%%
Fortunately, {\tt ColDICE} employs a quadratic description
of the phase-space sheet for performing local refinement
of the tessellation, by using tracers defined as mid-points of edges
of the simplices in Lagrangian space. For instance, the tracer
associated to vertices $(k,k')$ corresponds to barycentric coordinates $\xi_k=\xi_{k'}=1/2$,
$\xi_{j}=0$ for $j\neq k$ and $j\neq k'$. 

This allows one to define in a unique way a quadratic hypersurface inside each simplex
coinciding with vertices and tracers belonging to it. Note that the second order
representation preserves the continuity of the phase-space sheet but does
not warrant its smoothness at the differential level. Exploiting this representation
will nevertheless provide, in practice, sufficiently accurate measurements of the vorticity. 

In the quadratic representation, one defines the following conventional
shape functions $N_k$,
\begin{align}
N_k(\bm{\xi}) & \equiv \xi_k (2 \xi_k-1), \quad k \leq N_{\rm s},\\
 & \equiv 4 \xi_{K(k)}\, \xi_{L(k)}, \quad N_{\rm s} < k \leq
 N_{\rm s}+N_{\rm t}, 
\end{align}
where functions $K(k)$ and $L(k)$ are appropriately chosen to cover
all the combinations $1 \leq K(k) < L(k) \leq N_{\rm s}$, and where
$N_{\rm s}=D+1$ is the number of simplices and $N_{\rm t}=D(D+1)/2$ is
the number of tracers, given $D$ the dimension of space. 

Then equation (\ref{eq:linint}) becomes
\begin{equation}
g_{\rm quad}(\bm{X})=\sum_{k=1}^{N_{\rm s}+N_{\rm t}}
N_k(\bm{\xi})\, g_k,
\label{eq:quadint}
\end{equation}
where $g_k$ is now also defined over tracers. This is true in
particular for quadratically interpolated positions
\begin{equation}
\bm{X}=\bm{X}_{\rm quad}=\sum_{k=1}^{N_{\rm s}+N_{\rm t}}
N_k(\bm{\xi})\, \bm{X}_k,
\label{eq:coord2nd}
\end{equation}
while corresponding Lagrangian coordinates $\bm{Q}$ are still given by the linear
representation (\ref{eq:linecoq}) since the phase-space sheet is
initially flat, which means that matrix $W$ does not change. 

From equation (\ref{eq:coord2nd}), we obtain the matrix $M^{\rm quad}$ of partial derivatives of the vector 
$\bm{P}\equiv (\sum_k \xi_k, X_1, X_2, X_3)$ with respect to $\xi_k$,
which replaces equations (\ref{eq:Mmat1}) and (\ref{eq:Mmat2}): 
\begin{align}
M^{\rm quad}_{1,k} &=1, \\
M^{\rm quad}_{\alpha+1,k}& = \frac{\partial X_\alpha}{\partial \xi_k}=
\sum_{k'=1}^{N_{\rm s}+N_{\rm t}}
\frac{\partial N_{k'}}{\partial \xi_k} X_{k',\alpha},
\end{align}
with $X_{k',\alpha}$ the $\alpha^{\rm th}$ coordinate of $k'^{\rm th}$
vertex/tracer with position $\bm{X}_{k'}$. While, at the linear
level, this matrix was constant inside each simplex, it now depends 
linearly on barycentric coordinates, given the quadratic
nature of the shape functions $N_k$. 

The calculation of the jacobian matrix exploiting
the quadratic representation then just consists in replacing matrix $M$ by
$M^{\rm quad}$ in equation (\ref{eq:Tcalc}), since matrix $W$
remains unchanged, which allows us to compute $J(\bm{q})$ in the
quadratic representation. In particular, we can compute $J_k=J(\bm{Q}_k)$ at each
vertex and tracer position to prepare it for the quadratic interpolation
procedure given by equation (\ref{eq:quadint}) (but staying aware of
the fact that this calculation is accurate only up to linear order), which, in addition to
phase-space coordinates themselves, is enough to compute all the
quantities we are interested in. 

Once we have a scalar function $g(\bm{x})$ defined at vertices and
tracers positions, its derivative is given from equation
(\ref{eq:quadint}) by 
\begin{align}
\left[\frac{\partial g}{\partial x_{\beta}}\right]_{\rm quad} & = \sum_{k=1}^{N_{\rm s}+N_{\rm t}}
g_k \sum_{k'=1}^{N_{\rm s}}\frac{\partial N_k}{\partial \xi_{k'}}
\frac{\partial \xi_{k'}}{\partial x_\beta} \\
& = \sum_{k=1}^{N_{\rm s}+N_{\rm t}}
g_k \sum_{k'=1}^{N_{\rm s}}\frac{\partial N_k}{\partial \xi_{k'}}
[M^{\rm quad}(\bm{\xi})]^{-1}_{k',\beta+1}.
\label{eq:derivquad}
\end{align}
This quantity depends in a non-linear way on barycentric position
and requires matrix $M^{\rm quad}$ to be invertible, so to be able to have a
finite estimate of the partial derivative, it
is necessary to avoid the subspace occupied by the caustics (or, more exactly the element 
of surface/curve approximating the actual caustics). 

In particular, we estimate the partial derivatives of the projected
density as follows
\begin{equation}
\frac{\partial \rho}{\partial x_\beta} = - \frac{{\rm sign}(J)}{J^2} \frac{\partial
 J}{\partial x_\beta},
\label{eq:rhoderiv}
\end{equation}
using equation (\ref{eq:derivquad}) to estimate ${\partial J}/{\partial x_{\beta}}$.
Indeed, the jacobian is a smooth function of the Lagrangian coordinate. This is not the case for function $\rho_{\rm L}(\bm{q})=1/|J(\bm{q})|$, which presents strong variations in the vicinity of caustics. Using the Jacobian instead of the density to estimate derivatives allows us to capture better asymptotic behaviours in the vicinity of the caustics. 

%%%%%%%%%%%%%%%%%%%%%%%%%%%%%%%%%%%%%%%%%%%%%%%%%%
%%%%%%%%%%%%%%%%%%%%%%%%%%%%%%%%%%%%%%%%%%%%%%%%%%
\subsubsection{Practical estimate: particle representation and local
 coarse graining}
\label{sec:proj}
%%%%%%%%%%%%%%%%%%%%%%%%%%%%%%%%%%%%%%%%%%%%%%%%%%
%%%%%%%%%%%%%%%%%%%%%%%%%%%%%%%%%%%%%%%%%%%%%%%%%%

In order to compute any quantity at position $\bm{x}_0$, one has to
compute the intersection of the phase space sheet with the position
$\bm{x}=\bm{x}_0$, or to resolve inside each simplex the equation
$\bm{X}(\bm{\xi})=\bm{x}_0$. While this is a straightforward
linear problem in the linear representation of the phase-space sheet
(equation \ref{eq:lineco}), it becomes non-trivial in the quadratic case
(equation \ref{eq:coord2nd}), for which some iterative procedure seems
necessary. Here, to simplify the approach (and also to reduce effects
of aliasing and divergences near caustics), we adopt a forward point 
of view consisting in projecting the tessellation on
a cartesian mesh of resolution $n_{\rm ana}$ in configuration space, 
similarly as it is performed by {\tt ColDICE} to
compute the projected density. This means that we apply
additional coarse graining over each pixel/voxel of
the cartesian mesh: Integrals of the form
\begin{align}
\int {\rm d}^D v \, g(\bm{x},\bm{v})\, f(\bm{x},\bm{v})
\end{align}
become
\begin{align} \frac{1}{v_{\rm pixel/voxel}}
 \int_{{\rm
 pixel/voxel}} {\rm d}^D x \int {\rm d}^D v \, g(\bm{x},\bm{v})\,
f(\bm{x},\bm{v}),
\label{eq:coarsegrain}
\end{align}
with $v_{\rm pixel/voxel}$ the area/volume of the pixel/voxel under consideration.

While \citet{2016JCoPh.321..644S} use actual ray-tracing exact to linear
order to compute such an integral, we replace each simplex with a large
number of particles, which allows us to give an account of the
quadratic shape of the simplices in a very simple way. However, employing particles
introduces some discrete noise. To reduce the noise, we use
cloud-in-cell (CIC) interpolation procedure on the target cartesian
mesh \citep{1988csup.book.....H}, that is each particle
is replaced by a voxel/pixel of the same (small) size as the voxels/pixels of the cartesian
mesh, and associate a weight $W_{\rm inter}$ to the particle proportional to the
volume of intersection between the voxel/pixel associated with the particle and the
voxel/pixel of the mesh. In addition, we do not place
particles at random within the simplex, but instead refine the
simplex hierarchically in an homogeneous fashion $\ell_{\rm max}$ times, 
and then replace each sub-simplex obtained this way with a
particle corresponding to the centre of the simplex in the barycentric
representation.
The calculation of $\ell_{\rm max}$ is such that
discreteness effects due to the projection inside each voxel/pixel of the
cartesian mesh are kept under control:
\begin{equation}
\ell_{\rm max}={\rm max} \left\{ \lfloor {\rm log}_2( S/\Delta)+4-D \rfloor,0\right\}
\end{equation}
where $D$ is the dimension of space, $\Delta$ is the step of the cartesian mesh and
\begin{equation}
S={\rm max}_{k \neq k',\alpha}|X_{k,\alpha}-X_{k',\alpha}|,
\end{equation}
with $X_{k,\alpha}$ the $\alpha^{\rm th}$ coordinate of $k^{\rm th}$
vertex with position $\bm{X}_{k}$ and likewise for
$X_{k',\alpha}$. 

In practice, prior to CIC interpolation, each particle $p$
samples a small element of volume $\delta V_p$ in the spatial part of the integral
of right hand side of equation (\ref{eq:coarsegrain}),
\begin{equation}
\delta V_p=V_{\rm L}\, 2^{-D \ell_{\rm max}}\, J_p,
\end{equation}
where $V_{\rm L}$ is the Lagrangian volume of the simplex and $J_p$ the
jacobian (quadratically) interpolated at position of the particle. So,
at the end, the contribution of each particle in right hand 
integral of equation (\ref{eq:coarsegrain}) is equal to $\delta V_p \,
W_{\rm inter} \, g(\bm {X}_p,\bm{V}_p)$, where we remind that $W_{\rm
 inter}$ is the CIC weight defined previously and where function $g$
is estimated at particle phase-space position $(\bm
{X}_p,\bm{V}_p)$ using the methods described in previous
paragraphs.

Note that coarse graining due to the CIC interpolation is
expected to introduce defects/biases in regions where large variations
are expected, that is in the vicinity of caustics. Obviously these
effects become more dramatic when differentiating quantities,
so one expects the vorticity to be affected numerically nearby the
caustics, as can be noted in Figs.~\ref{fig: 2SIN vorticity}, \ref{fig: 3SIN vorticity Q1D}, \ref{fig: 3SIN vorticity ANI} and \ref{fig: 3SIN vorticity SYM}.

%%%%%%%%%%%%%%%%%%%%%%%%%%%%%%%%%%%%%%%%%%%%%%%%%%%%%%%%%%%%%%%%%%%%%%%%%%%%%%%%%%
%%%%%%%%%%%%%%%%%%%%%%%%%%%%%%%%%%%%%%%%%%%%%%%%%%%%%%%%%%%%%%%%%%%%%%%%%%%%%%%%%%
%%%%%%%%%%%%%%%%%%%%%%%%%%%%%%%%%%%%%%%%%%%%%%%%%%%%%%%%%%%%%%%%%%%%%%%%%%%%%%%%%%
\section{Perturbative treatment of quasi one-dimensional flow}
\label{app: Q1DPT}
%%%%%%%%%%%%%%%%%%%%%%%%%%%%%%%%%%%%%%%%%%%%%%%%%%%%%%%%%%%%%%%%%%%%%%%%%%%%%%%%%%
%%%%%%%%%%%%%%%%%%%%%%%%%%%%%%%%%%%%%%%%%%%%%%%%%%%%%%%%%%%%%%%%%%%%%%%%%%%%%%%%%%
%%%%%%%%%%%%%%%%%%%%%%%%%%%%%%%%%%%%%%%%%%%%%%%%%%%%%%%%%%%%%%%%%%%%%%%%%%%%%%%%%%
In Sec.~\ref{sec: phase space shell-crossing}, we test theoretical predictions relying on the perturbative treatment of a quasi one-dimensional flow as proposed by RF17. In this appendix, we briefly present this approach, focusing mainly on three sine waves initial conditions configurations. Generalisation to the two sine waves case is straightforward.

When the system is exactly one dimensional, the first-order LPT solution is exact before shell-crossing. 
This fact leads to an approximate treatment in three-dimensional space. We exploit the exact solution of the one-dimensional problem along $x$-axis given by Zel’dovich approximation as the unperturbed zeroth-order state:
\begin{align}
\Psi^{(0)}_{i} (\bm{q},t) = D_{+}(t)\,\Psi^{\rm ini}_{x}(q_{x})\,\delta_{i,x},
\end{align}
with the subscript $i=x$, $y$ for two sine waves and $i=x$, $y$, $z$ for three sine waves. 
The transverse fluctuations are considered as small first-order perturbations to this set-up:
\begin{align}
\bm{\Psi}^{(1)}(\bm{q},t) = \sum_{n=1}^{\infty}D^{n}_{+}(t)\,\bm{\Psi}^{(1,n)}(\bm{q}).
\end{align}
In the case of the three sine waves, we assume the explicit form of the functions $\Psi^{\rm ini}_{x}$ and $\bm{\Psi}^{(1,1)}$ to be
\begin{align}
\Psi^{\rm ini}_{x} &= \frac{L}{2\pi}\,\epsilon_{x}\,\sin{\left( \frac{2\pi}{L}q_{x} \right)} ,~\\
\bm{\Psi}^{(1,1)} &= \left(
\begin{array}{c}
0\\
\frac{L}{2\pi}\,\epsilon_{y}\,\sin{\left( \frac{2\pi}{L}\,q_{y} \right)} \\
\frac{L}{2\pi}\,\epsilon_{z}\,\sin{\left( \frac{2\pi}{L}\,q_{z} \right)}
\end{array}
\right) .
\end{align}
Then, substituting these initial conditions into the recursion relations in Eqs.~(\ref{eq:longitudinal3}) and (\ref{eq:transverse3}), we obtain
\begin{align}
\bm{\nabla}_{q}\cdot\bm{\Psi}^{(1,n)} &= -\Psi^{\rm ini}_{x,x}\,\frac{2n-1}{2n+3}\,\left( \Psi^{(1,n-1)}_{y,y} + \Psi^{(1,n-1)}_{z,z}\right) , \label{eq: longitudinal Q1D}\\
 \bm{\nabla}_{q}\times\bm{\Psi}^{(1,n)}&= -\Psi^{\rm ini}_{x,x} \frac{n-2}{n}\,
\left(
\begin{array}{c}
0\\
\Psi^{(1,n-1)}_{x,z} \\
- \Psi^{(1,n-1)}_{x,y}
\end{array}
\right) . \label{eq: transverse Q1D}
\end{align}
By solving Eqs.~(\ref{eq: longitudinal Q1D}) and (\ref{eq: transverse Q1D}), one can construct the Q1D first-order solutions.

We further extend the Q1D treatment proposed by RF17 up to second order in the transverse fluctuations as described below. 
On top of the zeroth and first order perturbations, we define the second order perturbation by
\begin{equation}
\bm{\Psi}^{(2)}(\bm{q},t) = \sum_{n=2}^{\infty}D^{n}_{+}(t)\,\bm{\Psi}^{(2,n)} .
\end{equation}
Again, using Eqs.~(\ref{eq:longitudinal3}) and (\ref{eq:transverse3}), we obtain the following recursion relation:
\begin{align}
\bm{\nabla}_{q}\cdot\bm{\Psi}^{(2,n)} &= -\Psi^{\rm ini}_{x,x} \,\frac{2n^{2}-3n+1}{2n^{2}+n-3} \left( \Psi^{(2,n-1)}_{y,y} + \Psi^{(2,n-1)}_{z,z} \right)
\notag \\
& - \varepsilon_{ijk}\,\varepsilon_{ipq}\sum_{m=1}^{n-1}\frac{2m^{2}+m-3/2}{2n^{2}+n-3}\, \Psi^{(1,n-m)}_{j,p}\,\Psi^{(1,m)}_{k,q}
\notag \\
&
 -\Psi^{\rm ini}_{x,x}\, \varepsilon_{xjk}\,\varepsilon_{xqr}\notag \\
& \quad \times \sum_{m=1}^{n-2}\frac{2m^{2}+m}{2n^{2}+n-3}\, \Psi^{(1,n-m-1)}_{j,q}\,\Psi^{(1,m)}_{k,r} , \\
\left[ \bm{\nabla}_{q}\times \bm{\Psi}^{(2,n)}\right]_{i} &= \Psi^{\rm ini}_{x,x}\,\frac{2n^{2}-3n-2}{2n^{2}+n}\, \varepsilon_{ixj}\,\Psi^{(2,n-1)}_{x,j}
\notag \\
&
+ \sum^{n-1}_{m=1} \frac{2m^{2}+m}{2n^{2}+n}\,\varepsilon_{ijk}\,\Psi^{(1,n-m)}_{l,j}\,\Psi^{(1,m)}_{l,k} .
\end{align}
The recursion is initialised by 
\begin{align}
\bm{\nabla}_{q}\cdot \bm{\Psi}^{(2,2)} &= -\frac{3}{14}\,\varepsilon_{ijk}\,\varepsilon_{ipq}\,\Psi^{(1,1)}_{j,p}\,\Psi^{(1,1)}_{k,q} ,\\
\bm{\nabla}_{q}\times \bm{\Psi}^{(2,2)} &= 0 .
\end{align}
Using the above recursion relations, the perturbative expansion is then performed by keeping terms proportional to $\epsilon_{y}^{n}$ and $\epsilon_z^{m}$ up to second order, $n+m=2$, while keeping terms proportional to $\epsilon_{x}^{k}$ up to tenth order, $k=10$, as shown in Fig.~\ref{fig: phase precollapse}. This approach provides a very accurate description of the dynamics when the system is initially quasi one-dimensional, i.e., $|\epsilon_x| \gg |\epsilon_{y,z}|$.

The results of the Q1D solution at first order in the transverse fluctuations are presented in RF17. Here, we show (for the first time) the Q1D solution at second second-order in the transverse fluctuations and up to 5th-order in the longitudinal direction in terms of the growth factor, i.e., $\bm{\Psi}^{(2,5)}$ as follows. We use the same notations as in Appendix~\ref{app: solutions}.

\begin{widetext}
For the $x$-components of the Q1D solutions, we have
\begin{align}
\Psi^{(2,2)}_{x} &= -\frac{3 \epsilon^{2}_{x}}{28\pi}
\sin (2 \pi q_{x}) \Bigl[ \epsilon_{1} \cos (2 \pi q_{y})+\epsilon_{2} \cos (2 \pi q_{z})
\Bigr]~,\\
\Psi^{(2,3)}_{x} &= \frac{\epsilon_{x}^3}{2520 \pi}
\sin (2 \pi q_{x})
\Bigl[
39 \epsilon_{1} \cos (2 \pi (q_{x}-q_{y}))+39 \epsilon_{1} \cos (2 \pi (q_{x}+q_{y}))+39 \epsilon_{2} \cos (2 \pi (q_{x}-q_{z}))+39 \epsilon_{2} \cos (2 \pi (q_{x}+q_{z}))
\notag \\
& \qquad 
+80 \epsilon_{1} \epsilon_{2} \cos (2 \pi (q_{y}-q_{z}))+80 \epsilon_{1} \epsilon_{2} \cos (2 \pi (q_{y}+q_{z}))-3 \epsilon_{1}^2 \cos (4 \pi q_{y})-3 \epsilon_{2}^2 \cos (4 \pi q_{z})+75 \left(\epsilon_{1}^2+\epsilon_{2}^2\right)
\Bigr] ~,\\
\Psi^{(2,4)}_{x} &= \frac{\epsilon_{x}^4}{2587200 \pi}
\Bigl[
-10 \sin (4 \pi q_{x}) \left(6010 \epsilon_{1} \epsilon_{2} \cos (2 \pi q_{y}) \cos (2 \pi q_{z})+1274 \epsilon_{1}^2 \cos (4 \pi q_{y})+1274 \epsilon_{2}^2 \cos (4 \pi q_{z})+2039 \left(\epsilon_{1}^2+\epsilon_{2}^2\right)\right)
\notag \\
& \qquad 
-29015 \sin (2 \pi q_{x}) (\epsilon_{1} \cos (2 \pi q_{y})+\epsilon_{2} \cos (2 \pi q_{z}))-707 \sin (6 \pi q_{x}) (\epsilon_{1} \cos (2 \pi q_{y})+\epsilon_{2} \cos (2 \pi q_{z}))
\Bigr] ~,\\
\Psi^{(2,5)}_{x} &= \frac{\epsilon_{x}^5}{4088164080000 \pi}
\Bigl[
4507471800 \epsilon_{1} \epsilon_{2} \cos (2 \pi (2 q_{x}-q_{y}-q_{z})) +4507471800 \epsilon_{1} \epsilon_{2} \cos (2 \pi (2 q_{x}+q_{y}-q_{z}))
\notag \\
& \qquad 
+4507471800 \epsilon_{1} \epsilon_{2} \cos (2 \pi (2 q_{x}-q_{y}+q_{z}))
+4507471800 \epsilon_{1} \epsilon_{2} \cos (2 \pi (2 q_{x}+q_{y}+q_{z}))
+6377961315 \epsilon_{1}^2 \cos (4 \pi (q_{x}-q_{y}))
\notag \\
& \qquad 
+6377961315 \epsilon_{1}^2 \cos (4 \pi (q_{x}+q_{y}))
+5918165253 \epsilon_{1} \cos (2 \pi (q_{x}-q_{y}))
+5918165253 \epsilon_{1} \cos (2 \pi (q_{x}+q_{y}))
\notag \\
& \qquad 
-379384005 \epsilon_{1} \cos (2 \pi (3 q_{x}+q_{y}))
-379384005 \epsilon_{1} \cos (6 \pi q_{x}-2 \pi q_{y})
\notag \\
& \qquad 
+33 \left(193271555 \epsilon_{2}^2 \cos (4 \pi (q_{x}-q_{z}))+179338341 \epsilon_{2} \cos (2 \pi (q_{x}-q_{z}))+815965150 \left(\epsilon_{1}^2+\epsilon_{2}^2\right)\right)
\notag \\
& \qquad 
+6377961315 \epsilon_{2}^2 \cos (4 \pi (q_{x}+q_{z}))
+5918165253 \epsilon_{2} \cos (2 \pi (q_{x}+q_{z}))
-379384005 \epsilon_{2} \cos (2 \pi (3 q_{x}+q_{z}))
\notag \\
& \qquad 
-379384005 \epsilon_{2} \cos (6 \pi q_{x}-2 \pi q_{z})
+4142788650 \cos (4 \pi q_{x}) \left(\epsilon_{1}^2+\epsilon_{2}^2\right)
+48740003000 \epsilon_{1} \epsilon_{2} \cos (2 \pi (q_{y}-q_{z}))
\notag \\
& \qquad 
+48740003000 \epsilon_{1} \epsilon_{2} \cos (2 \pi (q_{y}+q_{z}))+26823121050 \epsilon_{1}^2 \cos (4 \pi q_{y})+26823121050 \epsilon_{2}^2 \cos (4 \pi q_{z})
\Bigr] \sin{(2\pi q_{x})} ~.
\end{align}

For the $y$-components of the Q1D solutions, we derive
\begin{align}
\Psi^{(2,2)}_{y} &=
- \frac{3\epsilon_{1}\epsilon^{2}_{x}}{28\pi}
\Bigl[
\sin (2 \pi q_{y}) (\cos (2 \pi q_{x})+\epsilon_{2} \cos (2 \pi q_{z}))
\Bigr]~, \\
\Psi^{(2,3)}_{y} &=
\frac{\epsilon_{1}\epsilon^{3}_{x}}{2520\pi}
\Bigl[
\sin (2 \pi q_{y}) (2 \cos (2 \pi q_{x}) (39 \epsilon_{1} \cos (2 \pi q_{y})+80 \epsilon_{2} \cos (2 \pi q_{z}))-3 \cos (4 \pi q_{x})+75)
\Bigr]~, \\
\Psi^{(2,4)}_{y} &=
- \frac{\epsilon_{1}\epsilon^{4}_{x}}{7761600\pi}
\Bigl[
\sin (2 \pi q_{y}) (60 \epsilon_{1} (1274 \cos (4 \pi q_{x})+2039) \cos (2 \pi q_{y})
\notag \\
& \qquad 
+50 \epsilon_{2} (571 \cos (4 \pi q_{x})+4177) \cos (2 \pi q_{z})+57015 \cos (2 \pi q_{x})-9303 \cos (6 \pi q_{x}))
\Bigr]~, \\
\Psi^{(2,5)}_{y} &=
\frac{\epsilon_{1}\epsilon^{5}_{x}}{4088164080000\pi}
\Bigl[
\sin (2 \pi q_{y}) (33 (-340 \epsilon_{1} \cos (2 \pi q_{x}) (369227 \cos (4 \pi q_{x})-6071163) \cos (2 \pi q_{y})
\notag \\
& \qquad 
-91 (1297372 \cos (4 \pi q_{x})+59485 \cos (8 \pi q_{x})-3461625))-618800 \epsilon_{2} \cos (2 \pi q_{x}) (19521 \cos (4 \pi q_{x})-100109) \cos (2 \pi q_{z}))
\Bigr]~.
\end{align}
\end{widetext}
Note that the $y$-components and $z$-components are symmetric under the exchange of $q_{y} \leftrightarrow q_{z}$ and $\epsilon_{y} \leftrightarrow \epsilon_{z}$ in the above expressions.

%%%%%%%%%%%%%%%%%%%%%%%%%%%%%%%%%%%%%%%%%%%%%%%%%%
%%%%%%%%%%%%%%%%%%%%%%%%%%%%%%%%%%%%%%%%%%%%%%%%%%
\section{Asymptotic structure of the singularity at collapse}
\label{sec: analytic log slope}
%%%%%%%%%%%%%%%%%%%%%%%%%%%%%%%%%%%%%%%%%%%%%%%%%%
%%%%%%%%%%%%%%%%%%%%%%%%%%%%%%%%%%%%%%%%%%%%%%%%%%

In this appendix, we analytically investigate the dependence on initial conditions of the logarithmic slopes of various profiles expected at shell-crossing and small radii, as discussed in Sec.~\ref{sec:radialprof}. The main results of this investigation are summarised in Table~\ref{tab: summary at shell-crossing}. Since the extension to the two-dimensional case is obvious by simply setting $\epsilon_z=0$, we present here the structure of the singularity obtained from LPT to some order for the three sine waves case. Note that the calculations presented in this appendix are conceptually not new since singularity theory applied to cosmology is already well known \citep[see, e.g.][]{1982GApFD..20..111A,1989RvMP...61..185S,2014MNRAS.437.3442H,2018JCAP...05..027F}.

%%%%%%%%%%%%%%%%%%%%%%%%%%%%%%%%%%%%%%%%%%%%%%%%%%
\subsection{Taylor expansion around the singularity}
%%%%%%%%%%%%%%%%%%%%%%%%%%%%%%%%%%%%%%%%%%%%%%%%%%

To facilitate the analysis, we focus on the relation between Eulerian and Lagrangian coordinates around the origin.
Expanding the relation (\ref{eq: def Psi}) between the Eulerian coordinate $\bm{x}$ and the Lagrangian coordinate $\bm{q}$ around the origin, we obtain, after neglecting $O(q^{4})$ and higher order terms, 
\begin{align}
x_{i}(\bm{q},t) = \tilde{A}_{ia}(t) q_{a} + C_{iabc}(t)\,q_{a}\,q_{b}\,q_{c} , \label{eq: Psi expanded}
\end{align}
with $\tilde{A}_{ij}(t) \equiv \delta_{ij} + A_{ij}(t)$, where $A_{ij}(t)$ is some function of time.
In these equations, we have exploited the symmetric nature of the three sine waves initial conditions, which imply that Eulerian coordinates around the shell-crossing point can be expanded in terms of odd third-order polynomials of the Lagrangian coordinates, that is polynomial forms $P$ verifying $P(-\bm{q})=-P(\bm{q})$ \citep[see, e.g.,][for the one-dimensional case]{2015MNRAS.446.2902C,2017MNRAS.470.4858T}. 

The coefficients $A_{ij}$ and $C_{ijkl}$ are expressed in terms of partial derivatives of the displacement field with respect to the Lagrangian coordinates at the origin as follows,
\begin{align}
A_{ij}(t) = \Psi_{i,j}(\bm{0},t) , \quad
C_{ijkl}(t) = \frac{1}{3!}\Psi_{i,jkl}(\bm{0}, t) .
\end{align}
Using Eq.~(\ref{eq: Psi expanded}), the Jacobi matrix is given by
\begin{align}
J_{ij}(\bm{q},t) = \tilde{A}_{ij}(t) + 3 C_{ijab}(t)\,q_{a}\,q_{b} . \label{eq: def Jij expanded}
\end{align}
The shell-crossing time $t_{\rm sc}$ is obtained by solving the equation $J(\bm{0},t_{\rm sc}) = 0$, that is
\begin{align}
\det\tilde{A}(t_{\rm sc}) = 0 . \label{eq: sc condition}
\end{align}
Hereafter, the dependence on collapse time $t_{\rm sc}$ will be omitted in the notations.

Using Eq.~(\ref{eq: def Jij expanded}), the Jacobian is given by
\begin{align}
J(\bm{q}) & =
\frac{3}{2}\varepsilon_{ijk}\,\varepsilon_{abc}
\Biggl[
\tilde{A}_{ia} \, \tilde{A}_{jb} \, C_{kcde} \, q_{d}\,q_{e}
\notag \\
& \qquad
+ 3 \tilde{A}_{ia}\, C_{jbde}\, C_{kcfg} \, q_{d}\,q_{e}\, q_{f}\,q_{g}
\notag \\
& \qquad
+ 3 C_{iade}\, C_{jbfg} \, C_{kcmn} \, q_{d}\,q_{e}\, q_{f}\,q_{g} \,q_{m}\,q_{n}
\Biggr] .
\end{align}

To simplify furthermore the calculations, we noticed, up to 10th order of the LPT development, that for an initial displacement field given by three orthogonal sine waves aligned with each coordinate axis, the matrix $\tilde{A}_{ij}$ is diagonal and the matrix $C_{ijkl}$ verifies
\begin{align}
 C_{iiii} & \neq 0 , \notag \\
 C_{iijj} &=C_{ijij}= C_{ijji} \neq 0\,, \quad i \neq j , \notag \\
 C_{ijkl} &=0 \,, \quad {\rm otherwise} . \label{eq:Csym}
\end{align}
We did not find any simple way to demonstrate that these properties stand at any order, but the fact that they are verified up to 10th-order LPT is a really convincing clue. In this case, the shell-crossing condition, Eq.~(\ref{eq: sc condition}), is reduced to $\tilde{A}_{xx} \,\tilde{A}_{yy}\, \tilde{A}_{zz} = 0$. Prior to shell-crossing, the Eulerian coordinates read
\begin{align}
A(\bm{q}) &=\left( \tilde{A}_{xx} \, q_{x} \,\delta_{Ax} + \tilde{A}_{yy} \, q_{y}\, \delta_{Ay} + \tilde{A}_{zz}\, q_{z} \,\delta_{Az}\, \right) \notag \\
 &+ C_{AAAA} q_I^3 + \sum_{B}3 \left(1-\delta_{AB}\right) C_{AABB}\, q_{A}\,q_{B}^2 ,
\end{align}
for $A,B = x,y,z$, and the Jacobian
\begin{align}
J(\bm{q}) &=
3
\left(
\tilde{A}_{xx} \, \tilde{A}_{yy} \, C_{zzde} 
+ \tilde{A}_{yy} \, \tilde{A}_{zz} \, C_{xxde} 
+ \tilde{A}_{zz} \,\tilde{A}_{xx}\, C_{yyde} 
\right) \notag \\
& \qquad \qquad \times q_{d}\,q_{e}
\notag \\
&
+
\frac{9}{2}
\left( 
\varepsilon_{xjk}\,\varepsilon_{xbc}\,\tilde{A}_{xx} 
+ \varepsilon_{yjk}\,\varepsilon_{ybc}\,\tilde{A}_{yy} 
+ \varepsilon_{zjk}\,\varepsilon_{zbc}\,\tilde{A}_{zz} 
\right)
\notag \\
& \qquad\qquad \times C_{jbde}\, C_{kcfg} \, q_{d}\,q_{e}\, q_{f}\,q_{g}
\notag \\
&
+
\frac{9}{2}\varepsilon_{ijk}\,\varepsilon_{abc}\,
C_{iade} \, C_{jbfg}\, C_{kcmn}\, q_{d}\,q_{e} \,q_{f}\,q_{g}\, q_{m}\,q_{n} . \label{eq: expanded J}
\end{align}
In this last expression, we did not exploit yet explicitly the symmetries on $C_{ijkl}$ matrix. 
On the basis of these equations, we can now investigate the slope of the density profile for the three types of singularities we consider, depending on the values of the eigenvalues of matrix $\tilde{A}_{ij}$.

%%%%%%%%%%%%%%%%%%%%%%%%%%%%%%%%%%%%%%%%%%%%%%%%%%
\subsection{Asymptotic behaviour of the profiles at collapse}
%%%%%%%%%%%%%%%%%%%%%%%%%%%%%%%%%%%%%%%%%%%%%%%%%%

We now analyse the asymptotic behaviour of the profiles around the origin.
To this end, when we consider the following scaling, $\bm{x}\to s\, \bm{x}$, implying $r\to s\, r$, we examine how the Lagrangian coordinate changes. By doing so, we can understand the behaviour of the scaling of the Jacobian in terms of the Lagrangian coordinates, and thus reveal the behaviour of the density profile at the origin.

It is important to note that this proof is somewhat simplified and ignores details on the angular dependence when performing integrals over spherical shells to obtain the radial profiles. Therefore, the proof given in this appendix is not mathematically rigorous, but it leads to the same conclusions as the exact calculations in which proper form factors are estimated. The purpose of this section is indeed to provide a simplified rephrasing of singularity theory already presented in a more rigorous fashion in other works~\citep[see, e.g.][]{1982GApFD..20..111A,1989RvMP...61..185S,2014MNRAS.437.3442H,2018JCAP...05..027F}.

First, we examine the case $0 = \tilde{A}_{xx}\neq \tilde{A}_{yy}\neq \tilde{A}_{zz}$, which corresponds to the following sine waves initial conditions: Q1D-2SIN, Q1D-3SIN, ANI-2SIN, and ANI-3SIN, i.e., $0 \leq \epsilon_{\rm 2D} < 1$ or $0 \leq \epsilon_{{\rm 3D},i}<1$ for $i=1, 2$, as summarised in first line of Table~\ref{tab: summary at shell-crossing}. This configuration corresponds to shell-crossing along $x$-axis.

When computing the radial profiles close to the center of the system, we consider spherical shells of radius $r$ with $r^2=x^2+y^2+z^2 \ll 1$. At leading order in $\bm{q}$ we have, at shell-crossing, 
\begin{align}
x & \simeq C_{xxxx} \,q_x^3+ 3 C_{xxyy}\, q_{x}\,q_{y}^2 + 3 C_{xxzz}\, q_{x}\,q_{z}^2 , \label{eq: x} \\
y & \simeq \tilde{A}_{yy}\,q_{y} \label{eq: Q1D y} ,\\
z & \simeq \tilde{A}_{zz}\,q_{z} \label{eq: Q1D z} .
\end{align}
Applying the scaling $\bm{x} \to s \bm{x}$ implies
\begin{align}
 q_{y} \to s \,q_{y} ,\quad q_{z} \to s\, q_{z},
\label{eq: scale Q1Da}
\end{align}
then, applying the same scaling in Eq.~(\ref{eq: x}) and taking the limit $s \ll 1$, we simply obtain
\begin{align}
q_{x} \to s^{1/3}q_{x} .
\label{eq: scale Q1Db}
\end{align}
After exploiting the symmetries of matrix $C_{ijkl}$ (Eq.~\ref{eq:Csym}), the Jacobian up at leading order in $\bm{q}$ is given by
\begin{align}
J(\bm{q}) &\simeq
3 \tilde{A}_{yy}\, \tilde{A}_{zz} \left( C_{xxxx}\, q^{2}_{x} + C_{xxyy}\, q^{2}_{y} + C_{xxzz}\, q^{2}_{z} \right) . \label{eq: jacob Q1D}
\end{align}
Using the scalings in Eqs.~(\ref{eq: scale Q1Da}) and (\ref{eq: scale Q1Db}), we have
\begin{align}
J(\bm{q}) &\to
3 \tilde{A}_{yy}\, \tilde{A}_{zz} \left( C_{xxxx}\, s^{2/3} q^{2}_{x} + C_{xxyy}\, s^{2} q^{2}_{y} + C_{xxzz}\, s^{2} q^{2}_{z} \right) .
\end{align}
Then, we have $J\propto s^{2/3}$ when $s \ll 1$, resulting in $\rho \propto r^{-2/3}$.

Second, we consider the case, $0 = \tilde{A}_{xx}=\tilde{A}_{yy}\neq \tilde{A}_{zz}$ which corresponds to the initial condition SYM-2SIN, i.e., $\epsilon_{\rm 2D} = 1$ or $\epsilon_{{\rm 3D},1}=1$ and $\epsilon_{{\rm 3D},2}<1$, as summarised in second line of Table~\ref{tab: summary at shell-crossing}. This configuration corresponds to simultaneous shell-crossings along $x$ and $y$ axes.
In this case, the main contribution to Eulerian coordinates reads, at shell-crossing,
\begin{align}
 x & \simeq C_{1} q^{3}_x+ 3 C_{2}\, q_{x}\,q_{y}^2 + 3 C_{3}\, q_{x}\,q_{z}^2 \label{eq: SYM2 x} ,\\
 y & \simeq C_{1} q^{3}_y+ 3 C_{2}\, q_{y}\,q_{x}^2 + 3 C_{3}\, q_{y}\,q_{z}^2 \label{eq: SYM2 y} ,\\
 z & \simeq \tilde{A}_{zz}\,q_{z} , \label{eq: SYM2 z}
\end{align}
where we have exploited symmetry between $x$ and $y$ which implies $C_{xxxx}=C_{yyyy}\equiv C_1$, $C_{xxyy}=C_{yyxx}\equiv C_2$ and $C_{xxzz}=C_{yyzz} \equiv C_3$. Applying the same reasoning as above, the scaling $\bm{x} \to s \bm{x}$ implies
\begin{align}
 q_{z} \to s \, q_{z}, \label{eq:SYM2 scalea}
\end{align}
which leads us, when applying the scaling to equations (\ref{eq: SYM2 x}) and (\ref{eq: SYM2 y}) and assuming $s \ll 1$ to neglect the last term in these equations, hence
\begin{align}
 s\,x & \simeq C_{1} q^{3}_x+ 3 C_{2}\, q_{x}\,q_{y}^2 \label{eq: SYM2 x 2} ,\\
 s\,y & \simeq C_{1} q^{3}_y+ 3 C_{2}\, q_{y}\,q_{x}^2 \label{eq: SYM2 y 2} .
\end{align}
which leads to the scalings
\begin{align}
q_{x} \to s^{1/3} q_{x} ,\quad 
q_{y} \to s^{1/3} q_{y} . \label{eq: SYM2 scaleb}
\end{align}
With $\tilde{A}_{xx}=\tilde{A}_{yy}=0$, the first term in Eq.~(\ref{eq: expanded J}) vanishes, and the second term becomes the leading contribution:
\begin{align}
J(\bm{q})& =\frac{9}{2}
\varepsilon_{zjk}\,\varepsilon_{zbc}\,\tilde{A}_{zz} 
C_{jbde}\, C_{kcfg} \, q_{d}\,q_{e}\, q_{f}\,q_{g} .
\end{align}
Exploiting equation (\ref{eq:Csym}) then allows us to write this term as 
\begin{align}
J(\bm{q}) &\simeq \sum_{i+j+k=2} b_{ijk}\,q^{2i}_{x}q^{2j}_{y}q^{2k}_{z} ,
\end{align}
where $b_{ijk}$ are constant coefficients.
Using the scalings given by Eqs.~(\ref{eq:SYM2 scalea}) and (\ref{eq: SYM2 scaleb}), we have
\begin{align}
J(\bm{q}) &\to
\sum_{i+j+k=2} s^{2(i+j)/3+k}\,b_{ijk}\,q^{2i}_{x}q^{2j}_{y}q^{2k}_{z} . \label{eq: J SYM2}
\end{align}
The lowest power of $s$ in Eq.~(\ref{eq: J SYM2}) is $s^{4/3}$ with $i=j=1$ and $k=0$. Thus we have $\rho \propto r^{-4/3}$.

Third, we consider the case $\tilde{A}_{xx}=\tilde{A}_{yy}=\tilde{A}_{zz}=0$, which corresponds in the three sine waves case to the SYM-3SIN, i.e., $\epsilon_{{\rm 3D}}=(1,1)$, as summarised in third line of Table~\ref{tab: summary at shell-crossing}. This configuration corresponds to simultaneous shell-crossings along all the axes. The approach is analogous to the previous case. 
The Eulerian coordinates can be written as follows:
\begin{align}
 x & \simeq C_{1} q^{3}_x+ 3 C_{2}\, q_{x}\,q_{y}^2 + 3 C_{3}\, q_{x}\,q_{z}^2 ,\label{eq: SYM3 x}\\
 y & \simeq C_{1} q^{3}_y+ 3 C_{2}\, q_{y}\,q_{z}^2 + 3 C_{3}\, q_{y}\,q_{x}^2 ,\\
 z & \simeq C_{1} q^{3}_z+ 3 C_{2}\, q_{z}\,q_{x}^2 + 3 C_{3}\, q_{z}\,q_{y}^2 , \label{eq: SYM3 z}
\end{align}
with, again $C_1=C_{xxxx}=C_{yyyy}=C_{zzzz}$, $C_2=C_{iijj}$ for $i \neq j$.
Eqs.~(\ref{eq: SYM3 x})--(\ref{eq: SYM3 z}) directly lead to the scaling of $\bm{q}$:
\begin{align}
q_{x} \to s^{1/3}q_{x},\quad
q_{y} \to s^{1/3}q_{y},\quad
q_{z} \to s^{1/3}q_{z}, \label{eq: SYM3 scale}
\end{align}
Now, the first and second terms in Eq.~(\ref{eq: expanded J}) vanish, and we have
\begin{align}
J(\bm{q}) =
\frac{9}{2}\varepsilon_{ijk}\,\varepsilon_{abc}\,
C_{iade}\, C_{jbfg} \, C_{kcmn}\, q_{d}\,q_{e}\, q_{f}\,q_{g} \,q_{m}\,q_{n} ,
\end{align}
which reads, after exploitation of the symmetries (\ref{eq:Csym}),
\begin{align}
J(\bm{q}) \simeq \sum_{i+j+k=3}b_{ijk}\,q^{2i}_{x}\,q^{2j}_{y}\,q^{2k}_{z} ,
\end{align}
Using the scaling given by Eq.~(\ref{eq: SYM3 scale}), we have
\begin{align}
J(\bm{q}) \to
\sum_{i+j+k=3} s^{2(i+j+k)/3}b_{ijk}q^{2i}_{x}q^{2j}_{y}q^{2k}_{z} . \label{eq: J SYM3}
\end{align}
The lowest power of $s$ in Eq.~(\ref{eq: J SYM3}) is $s^{2}$. Thus we have $\rho \propto r^{-2}$.

Finally, we focus on the velocity and pseudo phase-space density profiles.
Using Eq.~(\ref{eq: Psi expanded}), the velocity field up to leading order in the Lagrangian coordinate is given by
\begin{align}
v_{i}(\bm{q},t_{\rm sc}) &= a \,\dot{\tilde{A}}_{ij}(t_{\rm sc})\,q_{j} ,
\end{align}
where the dot denotes derivative with respect to the cosmic time $t$.
As can be seen from this relation, the velocity field is proportional to the Lagrangian coordinate, irrespective of initial conditions, because the shell-crossing condition $\tilde{A}_{ij}(t_{\rm sc})=0$ does not imply $\dot{\tilde{A}}_{ij}(t_{\rm sc})=0$. Since, according to the calculations performed above, the leading contribution from the Lagrangian vector always come from the scaling $q_i\to s^{1/3} q_i$ when $s \ll 1$, it is fairly easy to convince oneself that the velocity profiles are given, at small radii, by $v^{2}\propto r^{2/3}$, $v^{2}_{r}\propto r^{2/3}$, and $-v_{r}\propto r^{1/3}$, from which we can infer as well the logarithmic slope of the pseudo phase-space density $Q(r)$ through Eq.~(\ref{eq: pseudo Q}).

\end{appendix}

\end{document}